\documentclass[twocolumn]{aastex631}

\newcommand\N[1]{{\color{blue} \bf #1}} 
\usepackage{soul}

\shorttitle{An Extreme Cool Supergiant Variable in M51}
\shortauthors{Jencson et al.}
\graphicspath{{./}{figures/}}

\begin{document}

\title{An Exceptional Dimming Event for a Massive, Cool Supergiant in M51}

\correspondingauthor{Jacob E.\ Jencson}
\email{jjencson@email.arizona.edu}

\author[0000-0001-5754-4007]{Jacob E. Jencson}
\affil{Steward Observatory, University of Arizona, 933 North Cherry Avenue, Tucson, AZ 85721-0065, USA}

\author[0000-0003-4102-380X]{David J. Sand}
\affil{Steward Observatory, University of Arizona, 933 North Cherry Avenue, Tucson, AZ 85721-0065, USA}

\author[0000-0003-0123-0062]{Jennifer E. Andrews}
\affil{Gemini Observatory/NSF's NOIRLab, 670 N. A'ohoku Place, Hilo, HI, 96720, USA}

\author[0000-0001-5510-2424]{Nathan Smith}
\affil{Steward Observatory, University of Arizona, 933 North Cherry Avenue, Tucson, AZ 85721-0065, USA}

\author[0000-0002-0744-0047]{Jeniveve Pearson}
\affil{Steward Observatory, University of Arizona, 933 North Cherry Avenue, Tucson, AZ 85721-0065, USA}

\author[0000-0002-1468-9668]{Jay Strader}
\affil{Center for Data Intensive and Time Domain Astronomy, Department of Physics and Astronomy, Michigan State University, East Lansing, MI 48824, USA}

\author[0000-0001-8818-0795]{Stefano Valenti}
\affil{Department of Physics and Astronomy, University of California, 1 Shields Avenue, Davis, CA 95616-5270, USA}

\author[0000-0003-4666-4606]{Emma R.\ Beasor}\altaffiliation{Hubble Fellow}
\affil{NSF's NOIR Lab, 950 N.\ Cherry Avenue, Tucson, AZ 85721, USA}

\author[0000-0003-2283-2185]{Barry Rothberg}
\affil{LBT Observatory, University of Arizona, 933 N.\ Cherry Ave,Tucson AZ 85721, USA}
\affil{Department of Physics \& Astronomy, MS 3F3, George Mason University, 4400 University Drive, Fairfax, VA 22030, USA}



\begin{abstract}

We present the discovery of an exceptional dimming event in a cool supergiant star in the Local Volume spiral M51. The star, dubbed M51-DS1, was found as part of a \textit{Hubble Space Telescope} (\textit{HST}) search for failed supernovae (SNe). The supergiant, which is plausibly associated with a very young ($\lesssim6$~Myr) stellar population, showed clear variability (amplitude $\Delta F814W\approx0.7$\,mag) in numerous \textit{HST} images obtained between 1995 and 2016, before suddenly dimming by $>$2\,mag in $F814W$ sometime between late 2017 and mid-2019. In follow-up data from 2021, the star rebrightened, ruling out a failed supernova. Prior to its near-disappearance, the star was luminous and red ($M_{F814W}\lesssim-7.6$\,mag, $F606W-F814W=1.9$--2.2\,mag). Modeling of the pre-dimming spectral energy distribution of the star favors a highly reddened, very luminous ($\log[L/L_{\odot}] = 5.4$--5.7) star with $T_{\mathrm{eff}}\approx$3700--4700~K, indicative of a cool yellow or post-red supergiant (RSG) with an initial mass of $\approx26$--40\,$M_{\odot}$. However, the local interstellar extinction and circumstellar extinction are uncertain, and could be lower: the near-IR colors are consistent with an RSG, which would be cooler ($T_{\mathrm{eff}}\lesssim3700$~K) and slightly less luminous ($\log[L/L_{\odot}] = 5.2$--5.3),  giving an inferred initial mass of $\approx19$--22\,$M_{\odot}$. In either case, the dimming may be explained by a rare episode of enhanced mass loss that temporarily obscures the star, potentially a more extreme counterpart to the 2019--2020 ``Great Dimming'' of Betelgeuse. Given the emerging evidence that massive evolved stars commonly exhibit variability that can mimic a disappearing star, our work highlights a substantial challenge in identifying true failed SNe.

\end{abstract}



\section{Introduction} \label{sec:intro}
Massive stars ($\gtrsim$8~$M_{\odot}$) end their life in core collapse after nuclear burning is no longer sustainable.  Often the collapse of a massive star results in the ejection of the stellar envelope and a luminous core-collapse supernova (SN), but this may not always be the case. Indeed, one surprising claim is that standard type IIP SNe, for which progenitor stars have been identified in deep pre-explosion data, arise from red supergiant (RSG) stars with only modest initial masses $\lesssim$18~$M_{\odot}$ \citep[e.g.,][]{smartt09,smartt15},  even though RSGs as a population extend to $\gtrsim$25~$M_{\odot}$ \citep{humphreys79}. The apparent lack of high-mass RSG progenitors to type IIP SNe suggests that either they are systematically missed by transient surveys (e.g., they explode preferentially in heavily obscured regions) or that the highest-mass RSGs do not end their life as SNe. However, the value of the mass cutoff at $\sim$18\,$M_{\odot}$, its statistical validity, and its interpretation are controversial due to the complexities of connecting the limited observations to uncertain models \citep[e.g.,][]{davies07,smith11,walmswell12,kochanek12,davies18a,davies20a,davies20b,kochanek20}.

The putative lack of type II SN progenitors with initial masses $\gtrsim$18~$M_{\odot}$ has led to the hypothesis that more massive stars cannot yield successful SN explosions, instead ending their lives as ``failed'' SNe that collapse directly into black holes \citep[e.g.,][]{kochanek08}, possibly with an associated fast, blue \citep{kashiyama15} or faint, $\approx$year-long transient \citep{lovegrove13,piro13}. Producing simulations that yield successful high-energy SN explosions has been a long-standing difficulty.  Modern simulations and theoretical work have found general agreement with this picture of a large fraction of failed SNe from high-mass RSGs, in that a majority of lower-mass RSG stars have compact cores that are more likely to produce successful explosions, while more massive ($\gtrsim$16--$20\,M_{\odot}$) stars may be more likely to collapse into black holes \citep[e.g.,][]{oconnor11,ugliano12,sukhbold14,pejcha15,ertl16,sukhbold16}. Though not strictly monotonic in mass, the general trend may be related to a transition in the star's core between convective and radiative carbon burning near the end of its life \citep{sukhbold19}.

Verification that massive RSG stars can directly collapse into black holes without an accompanying SN would shed light on the observed compact remnant population \citep{kochanek14} and point to the stellar-mass progenitors of the black hole mergers observed by gravitational-wave experiments \citep{GWTC-1,GWTC-2}.  The ideal observational signature of a failed SN would manifest as a bright RSG star that winks out of existence (or nearly so) from one epoch to the next, and stays in that state permanently.  With no accompanying luminous SN, the only explanation for the disappearance of such a massive star would be a direct collapse into a black hole, although some fading, optical/near-IR emission may remain from a weak envelope ejection or fallback accretion onto the newly formed black hole \citep{lovegrove13,perna14}.  

Multiple projects have searched for disappearing massive stars.  The first and most comprehensive of these has utilized the Large Binocular Telescope (LBT) to reimage nearby galaxy fields ($\lesssim$10\,Mpc) with a cadence of $\approx$months over the last decade \citep{gerke15,adams17a,neustadt21}. This search has identified the best-known candidate for a failed SN, dubbed NG6946-BH1 \citep{adams17b}, which was an apparent $\sim$25~$M_{\odot}$ ($\sim$10$^{5.5}$~$L_{\odot}$) RSG which brightened on a year timescale before nearly vanishing in the optical and IR. Follow-up observations show some fading remnant emission at the position, consistent with expectations of fallback accretion \citep{adams17b,basinger21}. The ongoing LBT survey has identified other  potential failed SN candidates, including a blue supergiant star in M101 \citep{neustadt21}. A separate, multiepoch archival \textit{Hubble Space Telescope} (\textit{HST}) search also identified a yellow supergiant (YSG) star that apparently vanished \citep{reynolds15}. 

A primary challenge in identifying true failed SNe is that massive, evolved stars are subject to several forms of large-amplitude variability on essentially all timescales, some which may mimic a disappearing star at optical wavelengths. Eruptive variability, like that of the luminous blue variables (LBVs; e.g., \citealp{humphreys99,smith01b,smith17}) or massive stellar mergers can form dust in their dense outflows, obscuring the surviving star in the optical, though producing bright mid-IR emission (e.g., \citealp{smith16,smith18,blagorodnova17,pejcha16a,pejcha16b}; see also IR-dominated transients in \citealp{jencson19}). 

Massive, cool supergiants, while subject to pulsational instabilities \citep{heger97,yoon10}, manifesting as periodic or semiregular variations on timescales of months to years \citep[e.g.,][]{kiss06,yang11,yang12,soraisam18}, can also undergo episodes of enhanced mass loss accompanied by observed LBV-like color changes \citep{smith04a}. These episodes may produce complex circumstellar material (CSM) like that observed for nearby, Galactic Y/RSGs that suggests prior, distinct mass ejections \citep[e.g.,][]{humphreys97,smith01a,smith04b,monnier04}. During such an event, increased opacity from a dense molecular wind (as in Mira variables, \citealp{reid02}; and for recent work on RSGs see, e.g., \citealp{davies21}) or the formation of dust can obscure a star causing a dramatic visual dimming. Such a scenario has, for example, been discussed at length to explain the recent historic minimum of Betelgeuse that occurred in 2019--2020, known as the ``Great Dimming'' \citep[e.g.,][]{levesque20,montarges21}. Modern extragalactic transient searches are designed to find bright, explosive outbursts that can be seen at large distances. Dramatic fading events may also signify important instabilities in massive evolved stars; however, as with failed SNe, they are more challenging to find and our knowledge of their impact on stellar evolution is limited. 

In this paper, we describe the first results from a program to search for failed SNe using a set of new \textit{HST} data directed at nearby galaxies that have both a long history of archival \textit{HST} imaging and have hosted multiple core-collapse SNe. We have identified a massive, cool supergiant star in M51 which underwent a significant dimming in 2019, preceded by a roughly year-long brightening phase, both of which pointed to a potential failed SN. As the first compelling candidate from our search, we named the object M51-DS1 (i.e., M51 ``Dimming Star'' 1) and began a dedicated follow-up campaign. Our subsequent \textit{HST} and ground-based near-IR imaging showed that the star had rebrightened in 2021, ruling out a failed SN explanation for the dimming event. Nevertheless, we take a detailed look at the available data for the object as both a false-positive in our failed SN search and as a prime example of exceptional variability for a massive, evolved supergiant. 

\section{A New Hubble Space Telescope Search for Failed Supernovae}\label{sec:FSN_search}
As described above, a clear observational signature of a failed SN is the permanent disappearance of a massive star. In order to observe more examples of the failed SN phenomenon, we chose nearby galaxies that are both proven core-collapse SN producers and have multiple epochs of \textit{HST} imaging over the last $\sim$30~yr.  To do this, we first identified galaxies that have hosted two or more core-collapse SNe in the modern era using the Asiago Supernova Catalogue \citep{asiago}, supplemented by SNe found in the last several years using a query of the Transient Name Server.\footnote{https://www.wis-tns.org/} We cross-matched this list of galaxies with those that had two or more existing epochs of \textit{HST} data in the $F814W$ filter, and we reobserved the resulting 31 galaxies (all with $D\lesssim$45 Mpc) over 41 orbits with the Advanced Camera for Surveys (ACS) in Cycle 26 (PI: D.\ Sand; PID: 15645), matching the footprint of previous \textit{HST} imaging as best as possible.  The $F814W$ filter was chosen as our discovery filter because it is among the most sensitive to the supergiant progenitors to failed SNe, and has been commonly used throughout the \textit{HST}'s lifetime with ACS, the Wide Field Planetary Camera 2 (WFPC2), and the Wide Field Camera 3 (WFC3); $F814W$ is also less sensitive to extinction in regions of star formation than bluer bands.  Each newly acquired $F814W$ image is a full orbit with a standard subpixel dither sequence, and has a sufficient depth ($F814W\approx27.2$\,mag with a signal-to-noise ratio, S/N, of $\approx$5) to observe massive stars with initial stellar masses $\gtrsim$10~$M_{\odot}$ throughout the distance range we cover, although some of the archival imaging we use is only sensitive to initial stellar masses of $\gtrsim$15~$M_{\odot}$.
 
We identify massive star candidates that have vanished using difference imaging, subtracting older \textit{HST} epochs with our newly acquired data. We outline the general procedure in Section~\ref{sec:obs} as we describe the identification of M51-DS1.  Key to our strategy is the wide range of data available in both the \textit{HST} and ground-based data archives for each of our target galaxies, which allows us to establish a baseline of existence for the progenitor star of any given failed SN candidate stretching back years prior to disappearance.  We will provide further details about our strategy and expected number of failed SNe in a future work, as analysis is ongoing.

\section{Observations and Data Processing} \label{sec:obs}

Here we describe the observations and data processing associated with the identification of M51-DS1 as a failed SN candidate, along with subsequent follow-up observations. 

\subsection{HST Image Processing and Subtraction}
To search for large-amplitude variables and ``disappearing'' luminous stars, we performed image subtraction with the available archival WFC3/UVIS and ACS/WFC images in the $F814W$ filter that overlap with the footprint of our Cycle 26 program observations. For M51, this includes ACS/WFC frames taken from 2005 January 12--22 (PI: S.\ Beckwith; PID: 10452) and a large campaign consisting of 34 epochs taken between 2016 October 5 and 2017 February 17 to study massive star variability (PI: C.\ Conroy; PID: 14704; see \citealp{conroy18}, hereafter C18). 

We used the Cycle 26 ACS/WFC $F814W$ observations of M51 as template images for subtraction with each of the archival frames. These include two spatially overlapping visits taken on 2019 May 24 and 31, respectively, designed to maximize coverage of M51 and overlap with the archival images. Each visit consisted of four dithered 564~s exposures. We first downloaded the \texttt{CALACS}-calibrated and charge-transfer-efficiency (CTE)-corrected \texttt{flc} frames for both visits from the Mikulski Archive for Space Telescopes. We processed the images using the \texttt{AstroDrizzle} software package, including automated cosmic-ray rejection, subpixel alignments with \texttt{TweakReg}, and final combination into a single, drizzled mosaic at the native WFC pixel scale of $0\farcs05$ with an effective point-spread function (ePSF) of $\approx2$\,pixels. 

We then processed the available archival $F814W$ frames from ACS/WFC in a similar manner. The 2005 observations consisted of a six-pointing, mosaicked mapping of M51, with each pointing composed of four dithered, individual 340\,s exposures. We aligned each of the 24 individual frames to the drizzled Cycle 26 template mosaic using \texttt{TweakReg}, typically using between $\approx$300 and $4000$ common stars depending on the size of the overlap region with the template. For the 2016--2017 data, each of the 34 epochs consists of four individual exposures totalling 2200\,s. We again aligned each frame to the template using \texttt{TweakReg}, where the number of common stars used for alignment varied between $\approx$100--500 stars. We achieved ($\mathrm{rms} \lesssim 0.1$ WFC pixels in the $x$- and $y$-directions) for every frame. We then ran the aligned images through \texttt{AstroDrizzle}. For the purposes of image subtraction, we produced one large drizzled mosaic of the 2005 images and split the 2016--2017 into three batches consisting of observations taken between 2016 October -- 2017 January, 2017 February--May, and 2017 June--September. We used the 2019 template image as a reference in \texttt{Astrodrizzle} so that all the output images were drizzled onto the same pixel grid. 

To prepare the drizzled images for subtraction, we ran each through the Astromatic packages \texttt{SExtractor} and \texttt{PSFEx} \citep{bertin96,bertin11} to generate source catalogs and models of the point-spread function (PSF) for each image. The images and corresponding noise maps generated from the inverse-variance-map output of \texttt{AstroDrizzle} were then converted to units of electrons on the scale of the 2019 template using the zero-point and exposure time information in the image headers. The background level of the images were also estimated and removed using the minimum value from a median filter on a 15$\times$15 grid of the region of overlap with the template. 

We then performed image subtraction using the ZOGY algorithm \citep{zackay16} on the background-subtracted images, with the noise maps and PSF model postage stamps as additional inputs. The subtraction produces a difference image, the difference image PSF, and a Scorr image (Equation 25 in \citealp{zackay16}), which is a match-filtered S/N image optimized for point-source detection. In order to use the Scorr image for the detection of variable objects, we normalize the Scorr values down to have a standard deviation of 1 over the image. This is necessary to account for the contribution of correlated pixel noise inherent in our drizzled images, which is not included in the calculation of the Scorr image with ZOGY. Finally, \texttt{SExtractor} is run on both the ``positive'' (archival-minus-template) and ``negative'' (template-minus-archival) difference images to build source catalogs of both fading and brightening objects. These are cross-matched with the corresponding, normalized Scorr image to select significant detections of $\mathrm{S/N} \geq 5$ (those objects with normalized Scorr values $\geq$5 in at least one pixel). 

\subsection{Candidate Identification}\label{sec:candidates}
To select initial candidates, we require that a source be detected at $|\Delta \lambda L_{\lambda}| \gtrsim 10^4~L_{\odot}$ in the ``positive'' difference images from at least two archival $F814W$ frames compared to the 2019 template, corresponding to an absolute magnitude in the difference image photometry of $M_{F814W} \leq -6.38$\,mag (Vega system) and indicating a luminous object that has faded significantly. This cutoff in the change in luminosity is the same as that used for the LBT search \citep{gerke15,adams17a,neustadt21}, which is sufficient to exclude lower-mass, large-amplitude variables including R CrB stars and Miras \citep[$M_I>-6$~mag;][]{tisserand09,soszynski09}. We also require that the two detections be separated in time by at least 1\,month. This establishes an archival baseline for a given source as a persistent object, i.e., to reject real astrophysical transient events such as novae as well as chance cosmic-ray hits and other image processing and subtraction artifacts.

We then visually inspect the position of the object in all available $F814W$ \textit{HST} imaging to further reject any obvious image artifacts and examine the full history of the source in that band. A strong failed SN candidate will appear as a star-like source that is present in all archival frames of sufficient depth prior to its first significant fading, after which the source will not appear to rebrighten. We note that we do not require that a source actually completely ``disappear,'' because a possible companion or unrelated, nearby star in a crowded region may continue to be detectable at the location even after a true failed SN. Any source that passes this visual inspection will be assigned a ``DS'' name (short for ``Dimming Star'') and flagged for detailed analysis, including PSF-fitting photometry of the available \textit{HST} imaging in other filters (see Section~\ref{sec:dolphot} for details). 

\begin{figure*}
\centering
\includegraphics[width=0.7\textwidth]{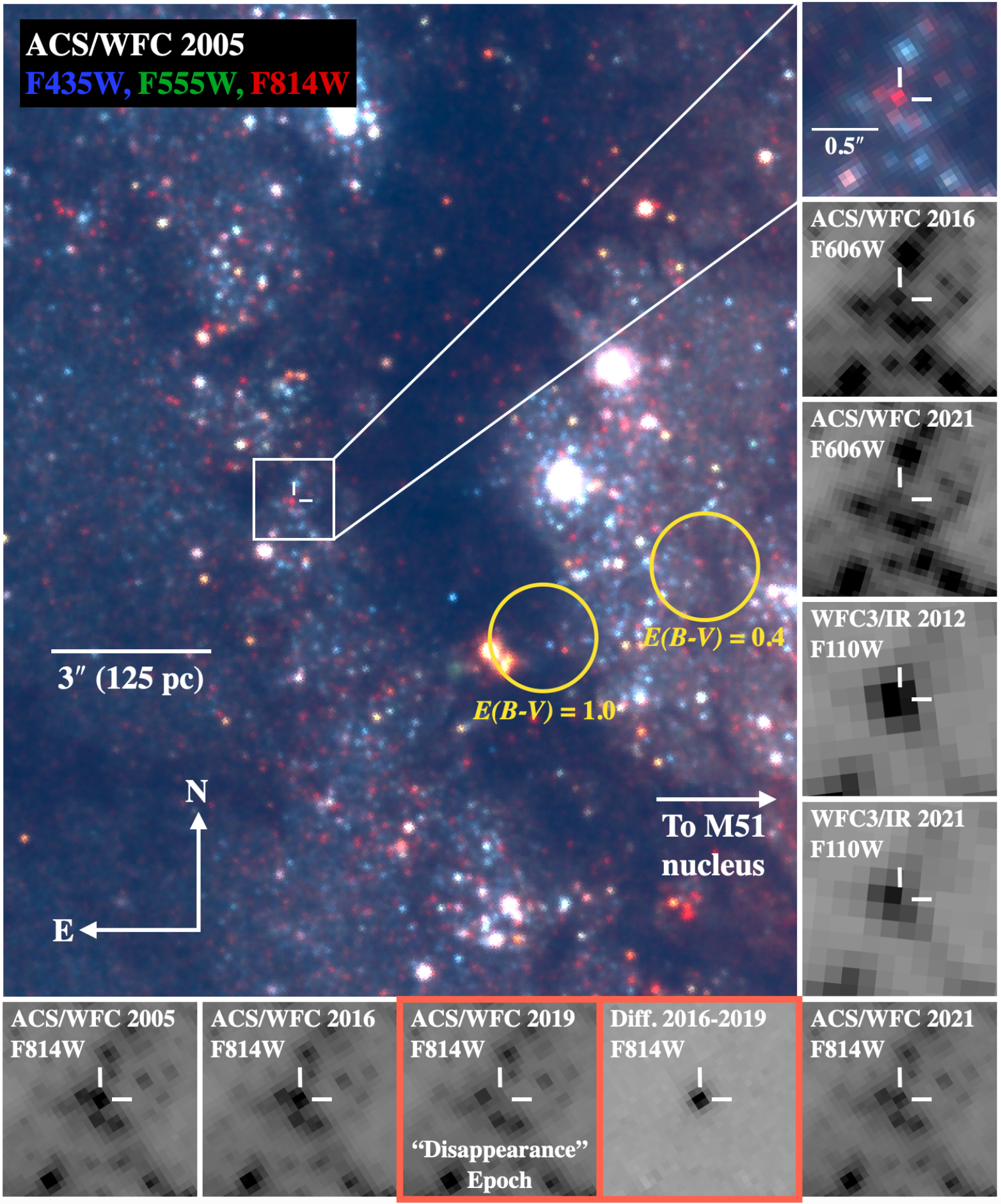}
\caption{\label{fig:hst_im}
HST imaging of M51-DS1 and the surrounding region. From left to right along the bottom row of panels, we show the sequence of ACS/WFC $F814W$ images from 2005, 2016, 2019, the 2016--2019 difference image, and the most recent 2021 image. The source underwent a dramatic fading, nearly disappearing in the 2019 frame (highlighted in the red-bordered panels), before rebrightening in 2021. The large, upper-left panel shows the color-composite ACS/WFC images from 2005 ($F435W$ in blue, $F555W$ in green, and $F814W$ in red) of the location of M51-DS1 along an inner spiral arm and prominent dust lane $\approx$36$\arcsec$ east of the galaxy nucleus. The positions of the nearest two \ion{H}{2} regions from \citet{croxall15} are indicated with yellow circles of 1$\arcsec$ radius and labeled with the derived $E(B-V)$ values for each one. A zoomed-in view of the immediate $1\farcs5 \times 1\farcs5$ (62.4~pc on a side) region around M51-DS1 is shown in the upper-rightmost panel. Below this, down the right side of the figure, we then show the same region in the ACS/WFC $F606W$ images from 2016 and 2021, and the WFC3/IR $F110W$ images from 2012 and 2021, demonstrating the dimming of the source across visible and near-IR wavelengths.
}
\end{figure*}

A subset of the sequence of available \textit{HST} $F814W$ imaging for M51-DS1, our first compelling candidate (as defined by the criteria outlined above), is shown in the bottom row of panels in Figure~\ref{fig:hst_im}. The star was clearly detected in the 2005 and 2016 ACS/WFC images (also in 1995 and 2008 WFPC2 images, not shown). In the 2019 image obtained as part of our search the star has seemingly vanished and is clearly detected as a fading candidate in our difference images. 

\subsection{Archival Imaging, Follow-up Observations, and Photometry}\label{sec:imaging}


Following our identification of M51-DS1 as a disappearing star and promising candidate for a failed SN, we began examining the wealth of multiband, archival \textit{HST} imaging, including the ACS/WFC $F814W$ images described above and the complimentary $F435W$, $F555W$, and $F658N$ (2005) and F606W (2016--2017) from the same programs. There is also coverage with WFC3/IR ($0\farcs13$ pixels) in $F110W$ and $F128N$ (Pa$\beta$) taken on 2012 September 4 (PI: J.\ Koda; PID: 12490). As shown in Figure~\ref{fig:hst_im}, M51-DS1 is notably red in appearance in the available archival imaging, and the optical source appears to be associated with a bright, relatively isolated star in the near-IR frames. Of the available WFC3/UVIS observations, we use only the 2012 observations in the redder $F673N$ and $F689M$ filters (PI: K.\ Kuntz; PID: 12762) where the star is clearly identifiable in the drizzled images. Similarly for WFPC2, while there are numerous prior observations in many filters that cover the location on the WF cameras, we consider only the $F814W$ images from 1995 (PI: R.\ Kirshner; PID: 5777) and 2008 (PI: M.\ Meixner; PID:  11229) given the red color of the star and the increasing likelihood of contamination in bluer filters for the larger pixels of the WF chips ($0\farcs0996$).

We were also awarded an \textit{HST} Cycle 28 mid-cycle proposal for additional deep, optical, and near-IR imaging of the field (PI: J.\ Jencson; PID: 16508) in order to verify the extreme fading of this source. The images for this program were obtained with ACS/WFC in $F606W$ and $F814W$ (total exposure time of 2208\,s for each filter) on 2021 April 28--29 and with WFC3/IR in $F110W$ and $F160W$ on 2021 June 6 (1200\,s exposure per filter). These new images revealed a surviving star at the location of M51-DS1 that had clearly rebrightened at $F814W$ from the 2019 minimum (see the bottom-right corner of Figure~\ref{fig:hst_im}). While this rules out a failed SN scenario for this object, the star is notably fainter at each of $F606W$, $F814W$, and $F110W$ compared to the corresponding pre-disappearance imaging. 

\subsubsection{\texttt{DOLPHOT} Photometry}\label{sec:dolphot}
We processed the available \textit{HST} imaging mentioned above with \texttt{DOLPHOT} \citep{dolphin00,dolphin16} to obtain PSF-fitting photometry of M51-DS1 and the surrounding stars in its vicinity. As inputs to \texttt{DOLPHOT}, we use the CTE-corrected \texttt{flc} frames for ACS/WFC and WFC3/UVIS observations, \texttt{flt} frames for WFC3/IR observations, and the standard-exposure-level \texttt{c0f} and data-quality \texttt{c1f} frames for WFPC2. Each input image was also run first through \texttt{Astrodrizzle} to flag cosmic-ray hits. We employ the parameter settings used for the \textit{HST} PHAT survey \citep{dalcanton12,williams14}. These settings were also used by C18 to build their photometry catalogs of M51, to which we can directly compare our results. 

For most of the data, we ran \texttt{DOLPHOT} separately for each instrument and filter combination over the available epochs, using the corresponding drizzled images as references for alignment. Generally, \texttt{DOLPHOT} produced excellent alignments better than $\approx$0.2\,pixels rms for ACS/WFC and WFC3/UVIS, and $\approx$0.3\,pixels rms for WFC3/IR and WFPC2. Owing to the large quantity of ACS/WFC $F814W$ and $F606W$ data (primarily from the C18 program), we had to run them in batches. In order to obtain consistent results across all runs for a given filter, we use the ``warmstart'' option in \texttt{DOLPHOT}, in which we use the output star catalog from an initial batch where M51-DS1 is well detected to fix the positions of stars in the subsequent runs for a given filter.  \texttt{DOLPHOT} computes and applies aperture corrections to a radius of $0\farcs5$ for the reported photometry. We then applied the appropriate corrections to infinite apertures for each instrument and filter combination.\footnote{See \citet{bohlin16} for ACS; for WFC3, see \url{https://www.stsci.edu/hst/instrumentation/wfc3/data-analysis/photometric-calibration}}
To estimate the photometric uncertainties, we compute the rms deviations of the measurements for the individual frames that comprise a given \textit{HST} observing visit of all stars within a 100\,pixel radius of M51-DS1 as a function of magnitude, added in quadrature with the nominal statistical uncertainty reported by \texttt{DOLPHOT}. 

For all the data that we ran, a ``good'' star (``object type''$=$1) was detected by \texttt{DOLPHOT} at the location of M51-DS1 as measured in the 2016 $F814W$ ACS frames within $\approx$0.5 ACS pixels, with the exception of the 2005 $F435W$ and $F555W$ ACS images. In these bluer filters, photometry of M51-DS1 is limited by crowding of nearby sources. We therefore adopt the magnitudes of the nearest detected star (1.3 and 0.8 pixels away for $F435W$ and $F555W$, respectively) as limits. Our final \texttt{DOLPHOT} photometry is presented in Table~\ref{table:HST_phot}. Throughout this work, all photometry is reported on the Vega magnitude system.  

Our measurements of M51-DS1 at $F814W$ from the ACS/WFC 2016--2017 data are typically about 0.05--0.1\,mag fainter than those reported for the star in the C18 catalog, though the offset is comparable with our measurement uncertainties. The offset is somewhat larger at $F606W$, with our measurements systematically at $\approx$0.3\,mag fainter that those of C18. In experimenting with \texttt{DOLPHOT} runs, we found that the choice of the alignment reference image as well as the specific images included in a given batch affected the final star list. In our final catalogs, there were stars immediately adjacent to M51-DS1 that were not in the C18 catalog. The simultaneous fitting of these additional stars by \texttt{DOLPHOT} may explain our lower flux measurements. 

\begin{deluxetable*}{ccccc}
\tablecaption{\textit{HST} \texttt{DOLPHOT} Photometry \label{table:HST_phot}}
\tablehead{\colhead{UT Date} & \colhead{MJD} & \colhead{Inst.} & \colhead{Band} & \colhead{App.\ Magnitude\tablenotemark{a}} \\ 
\colhead{} & \colhead{} & \colhead{} & \colhead{} & \colhead{(mag)} }
\startdata
1995 Jan 15.76 & 49732.76 & WFPC2   & $F814W$ & 22.03 (0.20) \\
2005 Jan 21.36 & 53391.36 & ACS/WFC & $F435W$ & $>24.8$      \\
2005 Jan 21.37 & 53391.37 & ACS/WFC & $F555W$ & $>25.1$      \\
2005 Jan 21.38 & 53391.38 & ACS/WFC & $F814W$ & 22.25 (0.11) \\
2008 Mar 31.14 & 54556.14 & WFPC2   & $F814W$ & 22.79 (0.20) \\
2012 Sep 04.82 & 56174.82 & WFC3/IR & $F110W$ & 20.09 (0.17) \\
2012 Sep 04.83 & 56174.83 & WFC3/IR & $F128N$ & 19.54 (0.09) \\
2016 Oct 05.25 & 57666.25 & ACS/WFC & $F606W$ & 24.51 (0.10) \\
2016 Oct 05.31 & 57666.31 & ACS/WFC & $F814W$ & 22.44 (0.05) \\
2016 Oct 14.25 & 57675.25 & ACS/WFC & $F606W$ & 24.54 (0.09) \\
\enddata
\tablenotetext{a}{Vega magnitudes. 1$\sigma$ uncertainties are given in parentheses.}
\tablecomments{Table~\ref{table:HST_phot} will be published in its entirety in the machine-readable format. A portion is shown here for guidance regarding its form and content.}
\end{deluxetable*}

\subsubsection{Ground-based Imaging and Photometry}
We obtained follow-up imaging in the near-IR with the MMT and Magellan Infrared Spectrograph (MMIRS, $0\farcs2$ pixels; \citealp{mcleod12}) on the 6.5~m MMT Observatory telescope on Mt.\ Hopkins at the Smithsonian's Fred Lawrence Whipple Observatory. We obtained $JHK_\mathrm{s}$ imaging over a two-night run on UT 2021 February 23 and 24, consisting of dithered sequences alternating between the target position in the central regions of M51 and an offset blank-sky field every few minutes to allow for accurate subtraction of the bright near-IR sky background. We reduced the images using a custom pipeline\footnote{Adapted from the MMIRS imaging pipeline developed by K.\ Paterson, available here: \url{https://github.com/CIERA-Transients/Imaging_pipelines}} that performs standard dark-current subtraction, flat-fielding, sky background estimation and subtraction, astrometric alignments, and final stacking of the individual exposures. 

We also obtained $J$- and $K_\mathrm{s}$-band imaging with the Near-Infrared Imager\footnote{\url{http://www.gemini.edu/instrumentation/niri}} (NIRI) on the 8~m Gemini-N Telescope on Maunakea through a Directors Discretionary Time program (PI: J.\ Jencson; PID: GN-2021A-DD-101). The images were obtained with the f/6 camera ($0\farcs117$ pixels) over multiple nights from 2021 April 1--6, again using sequences of dithered images alternating between the target position and an offset sky position. We also downloaded NIRI $JHK$ images from the Gemini Observatory archive covering the location of M51-DS1 and taken with in a similar manner with the same camera setup on 2005 June 27 (PI: S.\ Smartt; PID: GN-2005A-Q-49). We reduced all the images using \texttt{DRAGONS}\footnote{\url{https://dragons.readthedocs.io/en/v2.1.1/index.html}}, a Python-based platform for reducing Gemini data, and following the procedures for extended sources outlined in the NIRI imaging-reduction tutorial\footnote{\url{https://dragons.readthedocs.io/projects/niriimg-drtutorial/en/stable/}}. 

Lastly, near-IR $K_{\rm s}$-band observations were obtained on 2022 January 22 at the LBT. The LBT houses two 8.4\,m mirrors on a single mount. Each mirror contains a nearly identical set of facility instruments. The observations were made using only the left (or SX) mirror and the the LBT Utility Camera in the Infrared No.\ 1 \citep[LUCI-1;][]{seifert03}. LUCI-1 contains a 2048$\times$2048 pixel Teledyne HAWAII-2RG HgCdTe detector and was configured using the N3.75 camera, which yields a plate scale of $0\farcs$1178 pixel$^{-1}$. The observations were obtained using the Single conjugate adaptive Optics Upgrade for LBT \citep[SOUL;][]{pinna21}, the second-generation adaptive optics (AO) system at LBT. The AO observations were made using Enhanced Seeing Mode (ESM). Unlike full AO, ESM only uses 11 modes of corrections and a frequency of 100--400~Hz (depending on the brightness of the reference star used to determine corrections), but is used in conjunction with the wider N3.75 FOV to achieve significant improvements over seeing-limited observations over a $4\arcmin\times4\arcmin$ FOV \citep[see][]{rothberg19,rothberg20}. Observations were dithered, both on-target (to remove bad or hot pixels) and to blank sky, to remove thermal contributions from the sky and the telescope/instrument.  The data were reduced using standard \texttt{IRAF} procedures.  The final flat-fielded and sky-subtracted individual frames were aligned and shifted using the \texttt{IRAF} tasks \texttt{GEOMAP} and \texttt{GEOTRAN} into a final mosaic of $4\farcm03 \times 4\farcm4$. The observations were taken under varying nonphotometric conditions. A set of stars or clusters were selected to track the flux changes for each of the exposures. An optimal image (based on the flux and FWHM of the PSF of the AO reference star) was selected and any exposures which decreased by more than 0.3 mags were rejected for the final combined image.  The total exposure time for the LUCI-1 observations was 1174.68~s.

With the large field of view (FOV) of the MMIRS imager ($6\farcm9 \times 6\farcm9$), we were able to calibrate the photometric zero-points using aperture photometry of relatively isolated stars in the MMIRS images with cataloged $JHK_\mathrm{s}$-band magnitudes in the Two Micron All Sky Survey (2MASS; \citealp{skrutskie06}). We then derived a model of the ePSF for each image by fitting bright, isolated stars using the \texttt{EPSFBuilder} tool of the \texttt{photutils} package in \texttt{Astropy}. We performed PSF-fitting photometry at the location of M51-DS1 as well as for a set of approximately 60 stars spread across the images with varying degrees of crowding and galaxy-background emission. We include a low-order, two-dimensional polynomial in the fit to account for the spatially varying background for each star, taking care to avoid overfitting the data. We adopt the rms error of the fit residuals, scaled by a factor of the square root of the reduced $\chi^2$ (typically $\gtrsim 1$) for the fit, as the nominal statistical uncertainty per pixel, and multiply by the effective footprint, or number of ``noise pixels,'' of the ePSF\footnote{A derivation of this quantity is provided by F.\ Masci here: \url{http://web.ipac.caltech.edu/staff/fmasci/home/mystats/noisepix_specs.pdf}} to obtain an estimate of the statistical uncertainty for each flux measurement. We used the set of 2MASS calibration stars to derive aperture corrections ($\lesssim0.1$\,mag in all three filters) to place the PSF-fitting magnitudes on the scale of the image photometric zeropoints. We adopt the statistical flux uncertainty, summed in quadrature with the rms error of the stars used in estimations of the zeropoint and ePSF aperture correction, as the total uncertainty in our final magnitudes. Owing to the limited number of isolated 2MASS stars, even with the large field-of-view of MMIRS, the zeropoint rms (typically $\approx$0.1\,mag) dominates the error budget. 

The FOVs of the NIRI ($\approx$2$\arcmin\times$2$\arcmin$) and LUCI-1 ($4\arcmin\times4\arcmin$) images are smaller, and there were not enough isolated 2MASS stars in the central regions of M51 to do a direct calibration. We instead cross-calibrated our PSF photometry of stars in these images, performed in the same manner as described above, to a set of $\approx$15--20 common stars with the corresponding MMIRS image in the same filter. For our measurements of M51-DS1, we adopted the statistical uncertainty from the PSF-fitting (as above), summed in quadrature with the zero-point uncertainty (from the standard deviation of the individual stars used in the cross-calibration) as our measurement uncertainty. 

The image quality of all the ground-based near-IR data was between $\approx$0$\farcs5$ and $0\farcs8$ FWHM---a factor of $\approx$2--3 times that of WFC3/IR---generally with better seeing in the redder $H$ and $K$ bands. As noted in Section~\ref{sec:imaging}, M51-DS1 is relatively isolated in the near-IR, with the nearest comparably bright object in the WFC3/IR images at a separation of $\approx$1$\arcsec$. Still, our ground-based photometry may suffer some contamination from nearby unrelated sources, particularly in the $J$ band. The 2005 and 2021 $J$-band fluxes are a bit higher than those from the comparable $F110W$ images, though the discrepancies are $\lesssim1.7\sigma$. The 2021 $H$-band flux is fully consistent with the comparable $F160W$ measurement, and we also expect any contamination to be small or negligble in the $K$ band. Our final photometry of M51-DS1 from the ground-based near-IR images is provided in Table~\ref{table:GB_nearIR}.

\begin{deluxetable*}{ccccc}
\tablecaption{Ground-based Near-IR Photometry \label{table:GB_nearIR}}
\tablehead{\colhead{UT Date} & \colhead{MJD} & \colhead{Tel./Inst.} & \colhead{Band} & \colhead{App.\ Magnitude\tablenotemark{a}} \\ 
\colhead{} & \colhead{} & \colhead{} & \colhead{} & \colhead{(mag)} }
\startdata
2005 Jun 27.34 & 53548.34 & Gemini/NIRI & $J$   & 19.61 (0.14) \\
2005 Jun 27.36 & 53548.36 & Gemini/NIRI & $H$   & 19.14 (0.16) \\
2005 Jun 27.38 & 53548.38 & Gemini/NIRI & $K$   & 18.54 (0.21) \\
2021 Feb 23.36 & 59268.36 & MMT/MMIRS   & $J$   & 20.04 (0.11) \\
2021 Feb 23.49 & 59268.49 & MMT/MMIRS   & $H$   & 19.26 (0.09) \\
2021 Feb 24.40 & 59269.40 & MMT/MMIRS   & $K_\mathrm{s}$ & 18.73 (0.13) \\
2021 Apr 01.45 & 59305.45 & Gemini/NIRI & $J$   & 20.04 (0.08) \\
2021 Apr 02.41 & 59306.41 & Gemini/NIRI & $K_\mathrm{s}$ & 18.83 (0.12) \\
2022 Jan 22.48 & 59601.48 & LBT/LUCI-1   & $K_\mathrm{s}$ & 18.69 (0.16) \\
\enddata
\tablenotetext{a}{Vega magnitudes on the 2MASS system. 1$\sigma$ uncertainties are given in parentheses.}
\end{deluxetable*}

\section{Analysis of M51-DS1} \label{sec:analysis}

\subsection{Extinction, Distance, Host Galaxy Environment, and Metallicity}\label{sec:M51_env}
M51-DS1 is located at an R.A. and decl.\ of
$13^{\mathrm{h}}29^{\mathrm{m}}56\fs16, +47\degr11\arcmin47\farcs8$ (J2000.0), near the inner regions of the nearby, star-forming galaxy M51. Throughout this work, we assume a distance modulus for M51 from \citet{mcquinn16} of $m-M = 29.67 \pm 0.02$ (statistical) $\pm$ 0.07 (systematic; \citealp{rizzi07})~mag ($D = 8.58$\,Mpc) based on the luminosity of the tip of the red giant branch (TRGB) method, and that the systematic uncertainties associated with calibrating this method dominate over the statistical measurement uncertainties. We adopt the value from the NASA/IPAC Infrared Science Archive (IRSA) for the Galactic extinction toward the position of M51-DS1 of $E(B-V) = 0.03$~mag, based on the \citet{schlafly11} recalibration of the \citet{schlegel98} dust maps, and assuming a standard \citep{fitzpatrick99} reddening law with $R_V = 3.1$.

The star is $\approx$36$^{\arcsec}$ from the galaxy nucleus along the outer edge of a densely populated spiral arm and prominent dust lane (see Figure~\ref{fig:hst_im}). We thus expect the region may also be subject to substantial and spatially variable extinction from the host. \citet{croxall15} presented spectroscopic, gas-phase chemical abundances for 29 \ion{H}{2} regions in M51 including extinction estimates derived from Balmer-line ratios as part of the CHemical Abundances of Spirals (CHAOS) program. The two nearest \ion{H}{2} regions to M51-DS1 from this study are labeled in Figure~\ref{fig:hst_im}. One is centered on the dark dust lane with $E(B-V) = 1.0$\,mag, while the other is along the inside edge of the spiral arm in a region that appears visually similar to that of M51-DS1 with $E(B-V) = 0.4$\,mag. \citet{wei21} recently derived extinction maps of the M51 system from UV/optical and IR photometry, from which we again find M51-DS1 to be associated with a region of higher extinction with $A_V \approx 1.0$--$1.2$\,mag, or $E(B-V) \approx 0.3$--0.4\,mag for $R_V = 3.1$. These estimates are based on spatial averages in seeing-limited spectra ($\approx$1$\arcsec$; for CHAOS) or lower-resolution space-based imaging (e.g., $\approx$3$\arcsec$ with \textit{Spitzer} in \citealp{wei21}) and, while they are useful guideposts, the appearance of the dust lanes in the higher-resolution \textit{HST} images indicates that there is likely substantial variation in the foreground host extinction on smaller spatial scales. In our analysis below, we therefore consider the implications of various amounts of additional reddening from dust in the interstellar medium of M51, ranging from $E(B-V)$ = 0 mag to 1.6 mag.

From their set of \ion{H}{2} regions, \citet{croxall15} found O/H abundances in the central regions of M51 to be approximately solar or somewhat supersolar (assuming solar abundances of $12+\log(\mathrm{O/H})_{\odot} = 8.69$--8.78, \citealp{asplund09,ayres13}). Similarly, the metallicity map derived by \citet{wei21} suggests somewhat supersolar metallicities of $\log(Z/Z_{\odot}) \approx 0.0$--0.3\,dex in the vicinity of M51-DS1 along the inner spiral arm. 

\subsubsection{Membership in M51}
Given its red optical colors (see Section~\ref{sec:lcs} below), there is some possibility that M51-DS1 is a foreground cool dwarf in the Milky Way and not a true member of M51. We estimated the density of foreground cool dwarfs toward M51 using the TRIdimensional modeL of thE GALaxy \citep[TRILEGAL;][]{girardi12}, a simulator for stellar populations in the Milky Way.\footnote{TRILEGAL simulations can be run with a webform, here: \url{behttp://stev.oapd.inaf.it/cgi-bin/trilegal}} Based on our photometry catalogs (Section~\ref{sec:dolphot}), there are four stars within $2\farcs4$ of M51-DS1 (100\,pc at the distance of M51) that have broadly similar properties, i.e., that are brighter than $F814W\leq23.5$\,mag and with red colors $F606W-F814W\geq2.0$\,mag. Comparing this to the density of stars with the same properties from TRILEGAL simulation, we find a chance probability of foreground contamination of about 0.04\%. The density of foreground variables with these properties is likely much lower, and we note that M51-DS1 has no appreciable proper motion in HST imaging over 26~yr. Altogether, it is therefore very likely that M51-DS1 is a bona fide member of M51.

\subsection{Optical/Near-IR Light Curves and Colors}\label{sec:lcs}
\begin{figure*}
\centering
\includegraphics[width=\textwidth]{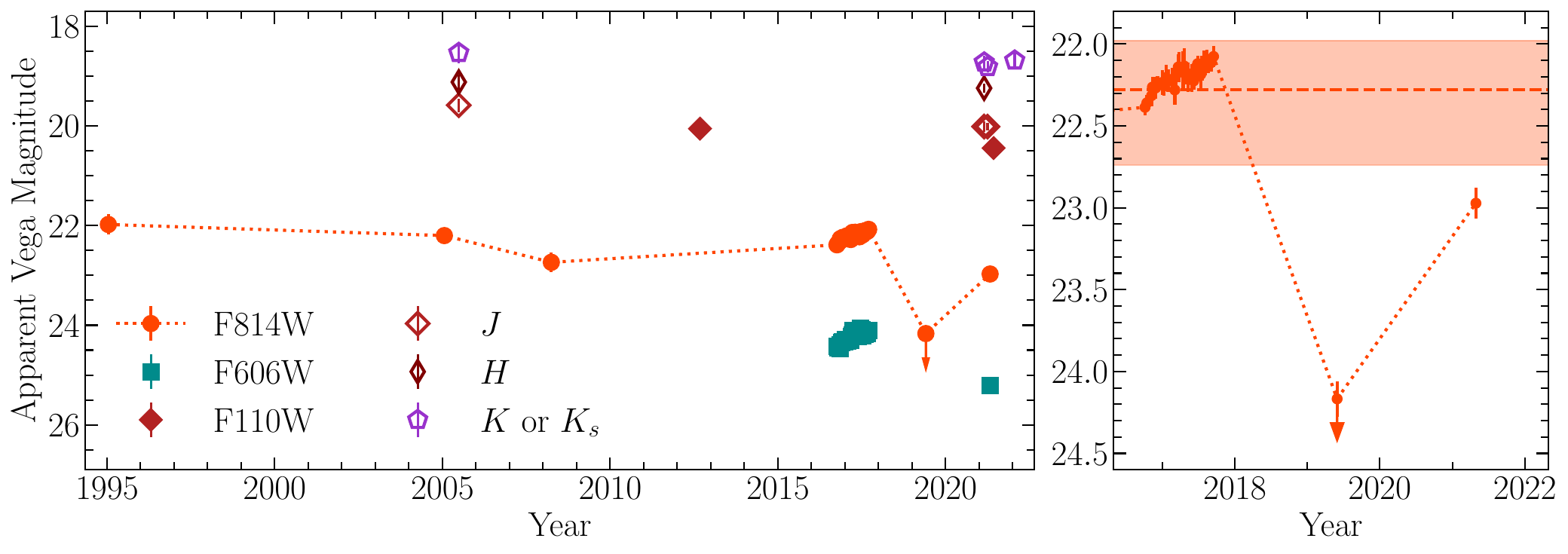}
\caption{\label{fig:hst_lcs}
Left: the light curves of M51-DS1 in $F606W$, $F814W$, and $F110W$ from archival and new \textit{HST} images (filled symbols) going back more than 26\,yr to 1995. Ground-based near-IR measurements from 2005 and new 2021/2022 images are shown as open symbols. All photometry has been corrected for Galactic extinction to M51. Right: zoom-in to the 2016--2021 $F814W$ light curve showing the pre-dimming brightening, the dimming, and the recovery of the star in more detail. The mean magnitude (dashed orange line) and magnitude range (orange shaded bar) of the pre-2019 $F814W$ photometry are also indicated.
}
\end{figure*}

The light curves of M51-DS1, dating back to 1995, are shown in Figure~\ref{fig:hst_lcs}. The magnitudes displayed and discussed here have been corrected only for Galactic extinction. The star was detected dozens of times at $F814W$ between 1995 and 2017, exhibiting some variability at the level of $\approx$0.7\,mag between $F814W = 22.74 \pm 0.20$ and $22.08 \pm 0.06$\,mag ($M_{F814W} = -6.9$ to $-7.6$\,mag). During 2016--2017, the evolution of the source was tracked at high cadence in both $F814W$ and $F606W$ with ACS/WFC (see inset in Figure~\ref{fig:hst_lcs}). The source displays a smooth, slow rise in both bands at a relatively constant, red color between $F606W - F814W = 1.6$ and 1.9\,mag, where the spread in the color measurements is comparable to their uncertainties. 

Subsequently, the star underwent a dramatic fading in the 2019 ACS/WFC $F814W$ observations, dropping to at least $F814W > 24.2$ ($>-5.5$\,mag) in our \texttt{DOLPHOT} measurement, $>2.2$\,mag below the brightest prior detection. Though \texttt{DOLPHOT} nominally obtained a detection (Section~\ref{sec:dolphot}), the source looks to have nearly completely disappeared (Figure~\ref{fig:hst_lcs}, bottom-center panel) at $F814W$, and, given the high degree of crowding at this location, we treat this measurement as an upper limit. Then, in our 2021 follow-up observations approximately two years later, the source has rebrightened partially to $F814W = 22.97 \pm 0.09$\,mag ($-6.6$\,mag, similar to the prior $F814W$ observed in 2008), and displays a slightly redder color at $F606W - F814W = 1.97 \pm 0.16$\,mag compared to the 2016--2019 measurements.

As expected given its red optical color, the source was brighter in the near-IR at $F110W = 20.05 \pm 0.17$\,mag ($M_{F110W} = -9.6$\,mag) in the 2012 WFC3/IR image. In our recent 2021 \textit{HST} follow-up imaging, around the time of the partial recovery seen in the optical, the star is also fainter at $F110W = 20.44 \pm 0.18$ ($M_{F110W} = -9.2$\,mag) with a red near-IR color of $F110W - F160W =  1.0 \pm 0.2$\,mag. The ground-based imaging with NIRI and MMIRS tells a similar story. In the 2005 NIRI imaging, the star is detected at $J = 19.58 \pm 0.14$ ($M_J = -10.1$\,mag), with $J-K = 1.05 \pm 0.25$\,mag. Subsequently, in our 2021 ground-based follow-up imaging, the star is again fainter in the near-IR filters with NIRI (MMIRS) compared to the pre-2019 levels at $J = 20.01 \pm 0.08$\,mag ($20.01 \pm 0.11$\,mag) and $J - K_\mathrm{s} = 1.20 \pm 0.14$\,mag ($1.29 \pm 0.17$\,mag). In our most recent 2022 $K_\mathrm{s}$-band image, the star remains at a similar brightness at $K_s = 18.69 \pm 0.16$\,mag ($M_{K_\mathrm{s}} = -11.0$\,mag).

\subsubsection{Near-IR Photometric Classification and Bolometric Luminosity}\label{sec:nearIR_Lbol}
At $M_K = -11.1$\,mag (2005; Galactic-extinction correction only), the star is well above the TRGB, and, moreover, is brighter than any asymptotic giant branch (AGB) stars identified in nearby galaxies including the Large and Small Magellanic Clouds (L/SMC), M31, and M33 \citep[see, e.g.,][and references therein]{cioni06,boyer11,massey21}. Its near-IR color of $J-K_\mathrm{s} \approx 1.0$--1.2\,mag is also consistent with the range often used to identify RSGs, though it may be somewhat bluer given the likelihood of significant foreground host extinction (see Section~\ref{sec:M51_env}). The $K$ band is particularly useful as a luminosity indicator for RSGs, both because the effects of extinction are reduced compared to the optical and bluer near-IR bands, and because the bolometric correction, $BC_K$, is found empirically by \citet{davies18a} to be constant across early-to-late M spectral types for cool supergiants in Milky Way and LMC star clusters. Thus, assuming an M-type spectrum ($T_{\rm eff} \lesssim 3700$\,K) for M51-DS1 and adopting their value of $BC_K = 3.0$\,mag, we obtain bolometric luminosities in the range $\log (L/L_{\odot}) \approx 5.15$--5.37 depending on the amount of extinction assumed in excess of the Milky Way foreground from $E(B-V) = 0.0$ up to 1.6\,mag (the highest value found to produce good fits in our modeling of the SED in Section~\ref{sec:SED_models}). The value for zero extinction, $\log (L/L_{\odot}) = 5.15$, can likely be viewed as a robust lower limit, as any foreground extinction or an earlier intrinsic spectral type ($<$M0 with $BC_K < 3.0$), will both increase the inferred bolometric luminosity of the star. We discuss the possible location of the star in a Hertzsprung--Russell diagram (HRD) and inferred evolutionary state in comparison with the results of our SED fitting in Section~\ref{sec:HRD}. 

\subsection{Color-Magnitude Diagrams}\label{sec:CMDs}
\begin{figure*}
\centering
\includegraphics[width=0.48\textwidth]{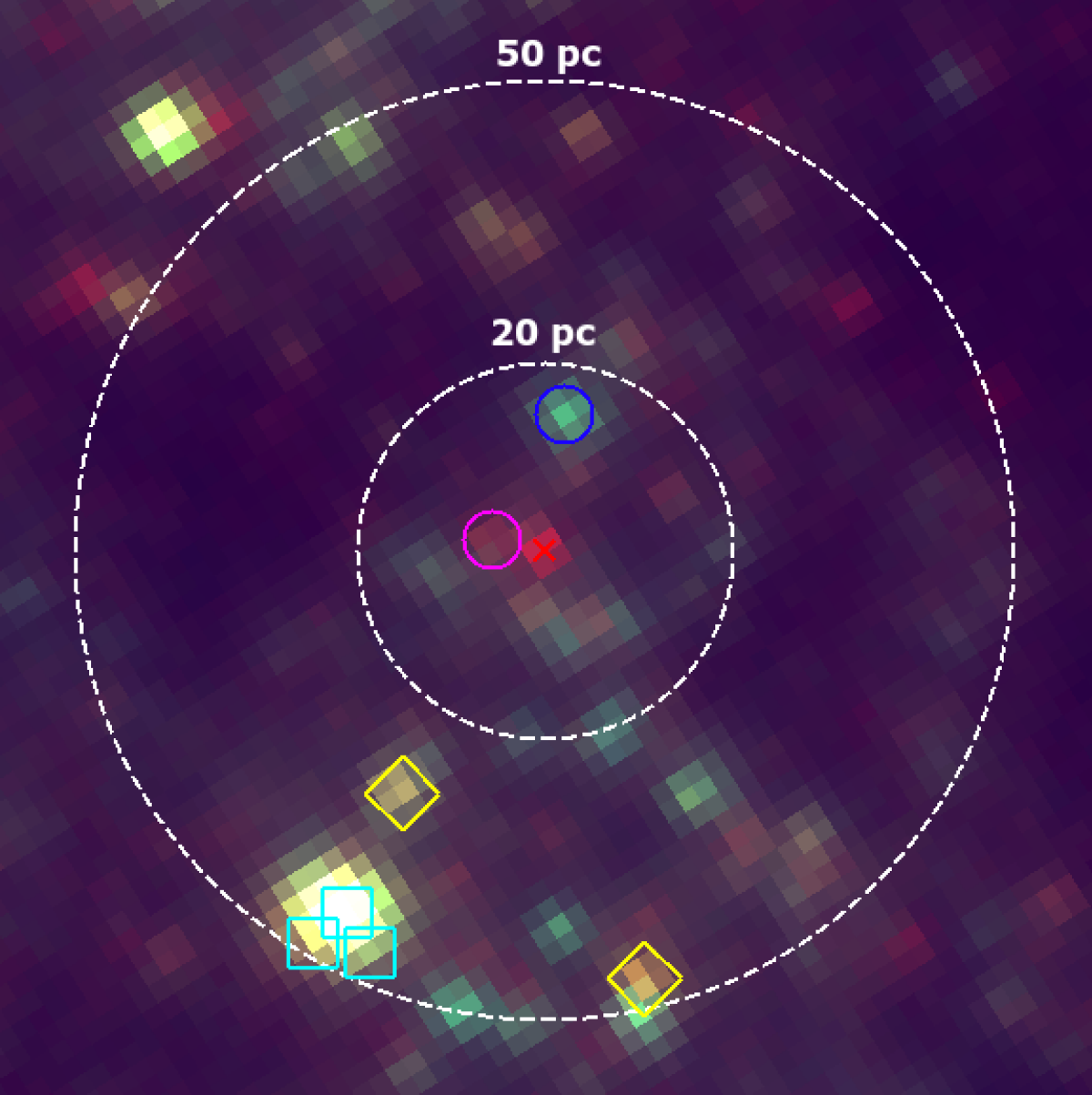}
\includegraphics[width=0.51\textwidth]{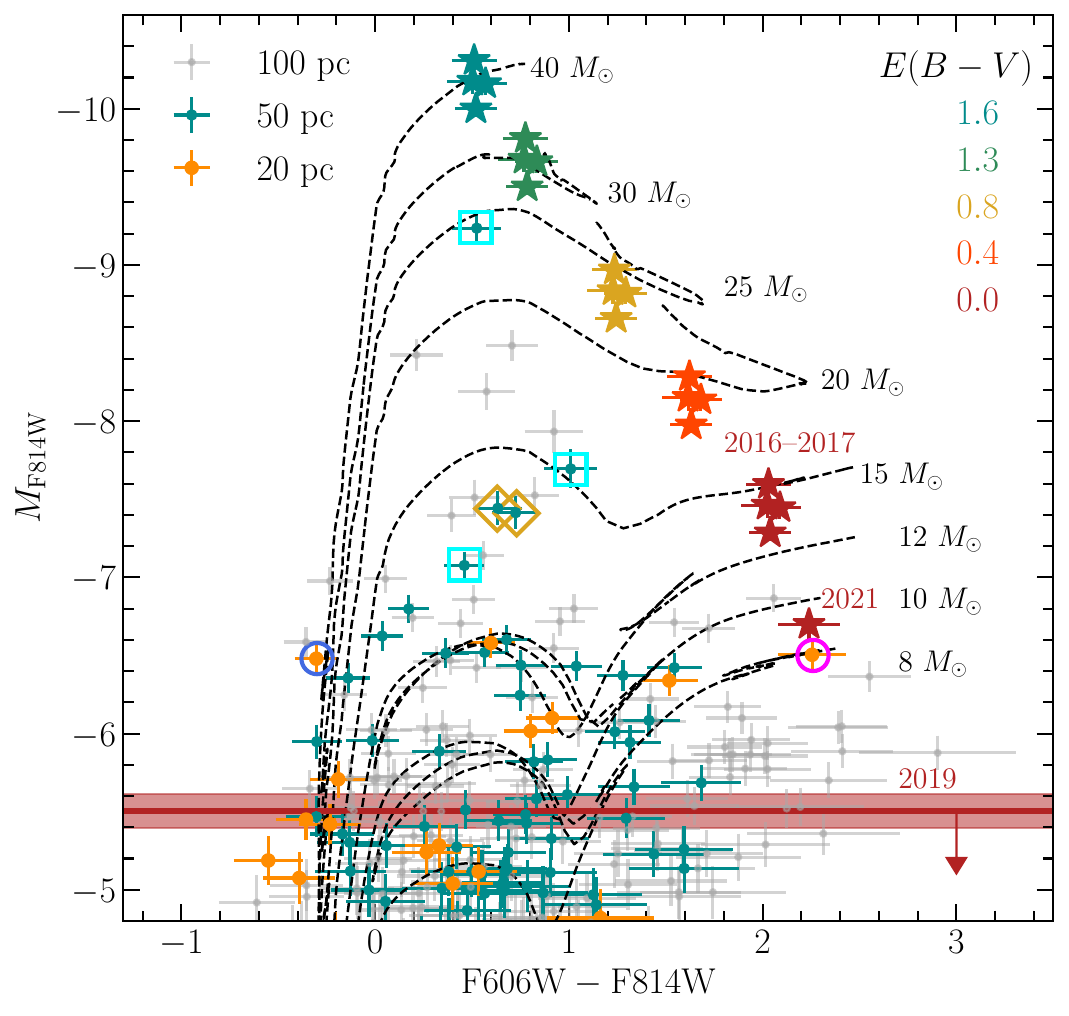}
\caption{\label{fig:HST_CMDs}
Left: immediate environment of M51-DS1 (red ``$\times$'' symbol) in the 2005 ACS/WFC images (color-composite of $F435W$, $F555W$, and $F814W$, as in Figure~\ref{fig:hst_im}) showing a mix of red and blue stars and dark regions that are likely obscured by dust. Radii corresponding to 20 ($0\farcs48$) and 50 pc ($1\farcs2$) projected distances from M51-DS1 are indicated. 
Sources indicated by multicolor symbols correspond to those objects marked in the CMD in the right panel. Right: the CMD of the region around M51-DS1 from the 2021 ACS/WFC $F606W$ and $F814W$ images. Stars within 20, 50, and 100 pc (projected; $0\farcs48$, $1\farcs2$, and $2\farcs4$, respectively) from the location of M51-DS1 are shown as orange, blue, and gray points, respectively. Additional multicolor markers correspond to those objects indicated in the left panel. We also show single-star, solar-metallicity, nonrotating stellar evolutionary tracks from MIST as the black and gray dashed curves for a range of masses between 8 and 40~$M_\odot$. Measurements for M51-DS1 are shown as the large red stars, including four epochs during 2016--2017 and the 2021 measurement. The 2016--2017 points are also shown with varying degrees of foreground host extinction, $E(B-V)$, indicated by different colors as labeled. The nominal 2019 $F814W$ \texttt{DOLPHOT} measurement (uncertainty) is indicated as the red horizontal line (shaded bar), though we consider this as an upper limit (see main text) as indicated by the downward arrow. 
}
\end{figure*}

In Figure~\ref{fig:HST_CMDs}, we present an F606W--F814W color-magnitude diagram (CMD) derived from our \texttt{DOLPHOT} photometry on the 2021 ACS/WFC images to place M51-DS1 in the context of the nearby stellar populations. Correcting only for Milky Way extinction, M51-DS1 appears as the most luminous red star ($F606W - F814W \gtrsim 2$\,mag) at $F814W$ within a projected distance of 100\,pc ($2\farcs4$) at its 2016--2017 level. It is one of only two stars within 50\,pc that appears so red; the other is indicated by the magenta circle in both figure panels. Compared to the single-star, nonrotating, solar-metallcity stellar tracks from the Mesa Isochrones and Stellar Tracks models (MIST; \citealp{choi16,choi17}), its location in the CMD would correspond to an RSG with an initial mass $\approx$12--15~$M_{\odot}$ (2016--2017) or $\approx$8--10~$M_{\odot}$ (2021), with an age younger than $\approx$15 or 30~Myr, respectively. As noted in Section~\ref{sec:M51_env}, however, there is likely to be significant and spatially variable foreground extinction in the region. For values of $E(B-V)$ between $0.4$ and $1.0$\,mag---in the range inferred from previous analysis of nearby \ion{H}{2} regions by \citep{croxall15}---the position of the star would correspond to higher masses up to $\approx$20--25~$M_\odot$, and younger ages $\lesssim$7--10~Myr. Our analysis of the full SED of M51-DS1 in Section~\ref{sec:SED_models} suggests even higher extinction values of $E(B-V) = 0.8$--1.6\,mag, possibly pointing to significant circumstellar extinction, corresponding to initial masses as high as $\approx$25--40~$M_\odot$ and implying an age as young as $\approx$5~Myr in the CMD. 

As is evident from the tricolor image shown in Figure~\ref{fig:HST_CMDs}, the immediate region surrounding M51-DS1 is populated by a range of blue-, yellow-, and red-appearing stars. This is reflected as a notable spread in the $F606W - F814W$ color between $\approx -0.3$ and $1.6$\,mag for the bulk of the stars in the CMD within the 20 and 50~pc radii. If one assumes a coeval population of stars undergoing isolated evolution, this could be attributed to effects of foreground dust, possibly implying variable extinction in the range $E(B-V) \approx 0.0$--$2.0$\,mag. Interpretation of the CMD as a whole, however, is complicated by additional factors on top of the uncertain extinction, including the possibility of multiple, overlapping stellar populations of different ages, and the effects of binary interaction---known to be increasingly common among more massive stars \citep[e.g.,][]{sana12,moe17}---on the location of a given source in the diagram. 

Despite these difficulties, we look briefly at a few individual sources that provide important context on the age of the population hosting M51-DS1 \citep[as has been done for several SN progenitors; e.g.,][]{vandyk99,maund05,maund17}. Within 50~pc, the apparently brightest object at $F814W$ in the CMD, along with two somewhat fainter objects, are associated with what appears to be a partially resolved star cluster (indicated with cyan squares). They are thus not likely to be secure detections of individual stars and should be viewed skeptically in the CMD. After this, there are two bright, yellowish stars (indicated with yellow diamonds), at $M_{F814W} \approx -7.4$\,mag and $F606W - F814W = 0.7$\,mag, which, in the absence of additional foreground extinction, would correspond to stars of at least $\approx$12--15~$M_{\odot}$ evolving across the Hertzsprung gap. Finally, within a smaller radius of 20~pc ($0\farcs48$), the most luminous object after M51-DS1 is a blue (presumably unreddened) source at $F606W - F814W = -0.3$\,mag (indicated in both the image and CMD by a blue circle). This would correspond to a $\approx$30~$M_{\odot}$ star at the end of its main-sequence lifetime and implying an age $\lesssim$6~Myr. This is consistent with the higher mass and younger age inferred for M51-DS1 above when one allows for significant foreground and/or circumstellar extinction ($E[B-V] \gtrsim 0.8$\,mag), assuming it is part of the same, coeval population, lending additional, contextual support to this interpretation.

\subsubsection{Comparison to Luminous, Red, Large-amplitude Variable Stars}
Regardless of the magnitude of the foreground extinction to M51-DS1, the track of possible locations it occupies in the CMD in Figure~\ref{fig:HST_CMDs} essentially aligns with the red-end points of the MIST evolutionary tracks for stars of initial masses $\gtrsim 15~M_{\odot}$. RSGs in this region are expected to be susceptible to pulsational instabilities driven by partial ionization of hydrogen in their envelopes (e.g., \citealp{li94,yoon10}, and see region of ``supergiant instabilities'' in Figures 13 and 16 of C18). 

\begin{figure*}
\centering
\includegraphics[width=0.49\textwidth]{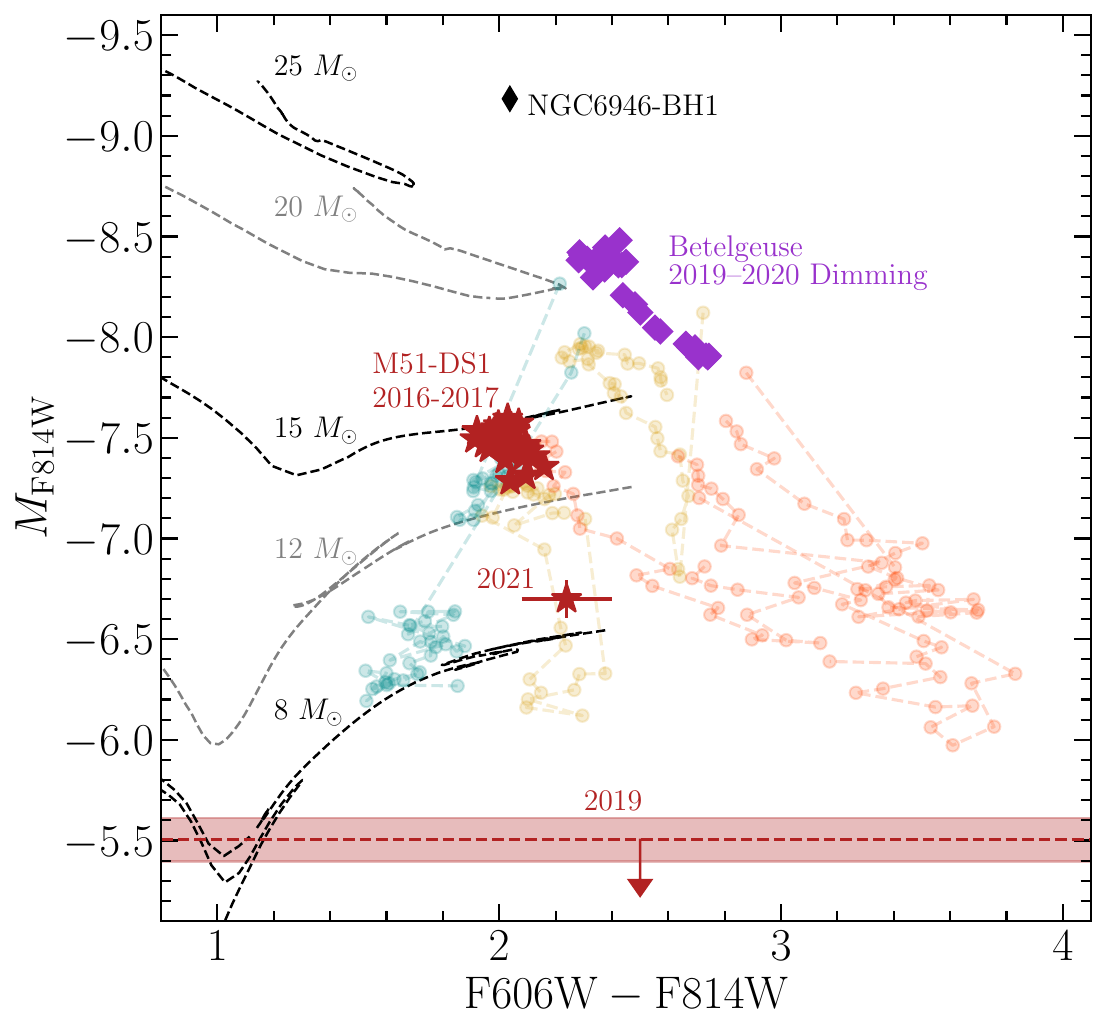}
\includegraphics[width=0.49\textwidth]{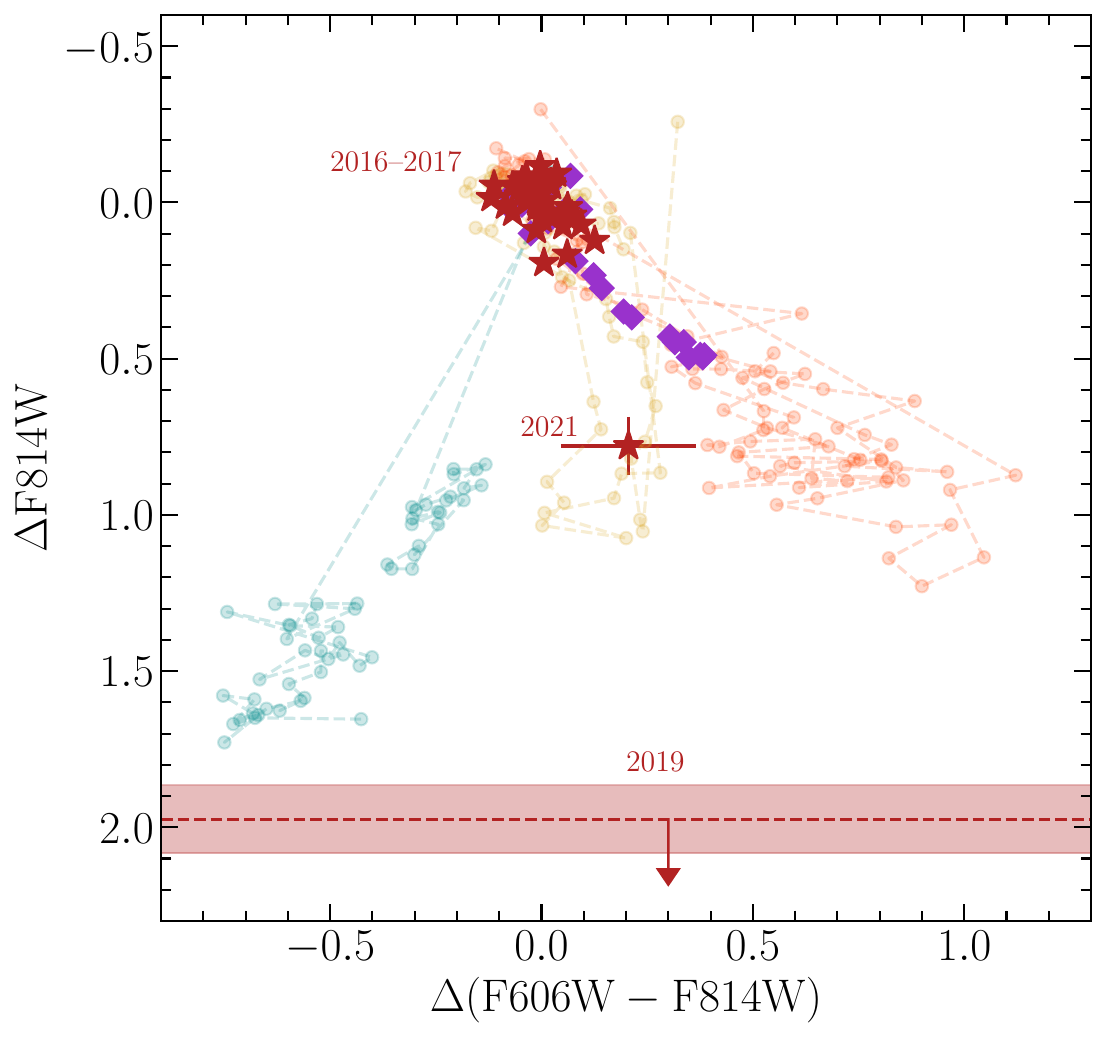}
\caption{\label{fig:LRVs_CMD}
Left: CMD of luminous ($M_{F814W} \leq -7.0$\,mag), red ($F606W - F814W \geq 1.5$), high-amplitude ($\Delta F814W \geq 1.0$\,mag) variable stars in M51 from the catalog of C18 (small circles connected by dashed lines) compared to the 2016--2017, 2019, and 2021 data of M51-DS1 (red stars). The plotting colors of the catalog variable points correspond to their photometric color evolution during fading events: to the blue (blue), approximately constant (yellow), and to the red (dark orange). We also show the $M_I, V-I$ evolution of Betelgeuse during its 2019 dimming event from the AAVSO, and the location of the progenitor of NGC\,6946-BH1 from \citep{adams17b}. As in Figure~\ref{fig:HST_CMDs}, we show single star tracks of various masses from MIST. Right: differential CMD showing the same data as in the left panel, but with baseline colors and magnitudes subtracted out for each source as described in the text.
}
\end{figure*}

To compare M51-DS1 to other such luminous, red variable stars, we selected stars from the catalog of light curves of luminous variables in M51 presenteed in C18. Of the more than $\approx$72000 objects in their catalog (including only the Milky Way extinction corrections already applied), we pulled out the 90 objects having $M_{F814W} \leq -7.0$\,mag at any epoch (1995, 2005, and 2016--2017 data), average colors (including epochs within 0.5\,mag of the $F814W$ peak) of $F606W - F814W \geq 1.5$\,mag, and maximum $F814W$ amplitudes $\Delta F814W \geq 1.0$\,mag. We examined both the images and light curves of these objects by eye to remove false positives such as foreground proper-motion stars and stars with large amplitude variability in the light curves that is not apparent in visual inspection of the images, likely indicating erroneous photometry. We also cull the sample by rejecting photometric points with large uncertainties ($\sigma_m > 0.2$\,mag) and stars where the color evolution is dominated by light-curve scatter, usually in the fainter $F606W$ measurements. We thus select a clean sample of eight luminous, red, large-amplitude variables, and show their evolution in the CMD in Figure~\ref{fig:LRVs_CMD}. 

These eight stars can then be separated according to their color evolution: those that evolve to the red as they fade, those that remain at a relatively constant color, and those that evolve to the blue (shown as orange, yellow, and blue circles in Figure~\ref{fig:LRVs_CMD}, respectively). The distinctions between these groups are seen more clearly in the differential CMD shown in the right panel of the figure, in which the average $F814W$ magnitudes and $F606W - F814W$ colors (using points within 0.5\,mag of the $F814W$ peak) have been subtracted out for each source. Interestingly, the ``yellow''-class objects appear near the end of the RSG evolutionary tracks at the Hayashi limit (though we note that only foreground Milky Way extinction corrections have been applied), possibly indicating that their near-vertical evolution in the CMD may indeed be attributed to large-amplitude, radial pulsations that are theoretically expected for such stars. Between the 2016--2017 and 2021 data, the evolution of M51-DS1 appears most similar to this group. The 2019 dimming of M51-DS1 is exceptional as the largest amplitude variation seen among these luminous variables in M51. Unfortunately, we lack any information on its color evolution during the 2019 minimum.

The red-evolving objects notably become much redder than the end of the MIST tracks at $F606W - F814W > 2.5$\,mag. The colors and large amplitudes of these stars may indicate they are very luminous AGB stars or Mira variables \citep[e.g.,][]{boyer11,yang19,neugent20}. At their peaks, they are brighter than any known AGB stars in the LMC ($M_{I} \gtrsim -7$\,mag; \citealp{soszynski09}), possibly pointing to extreme objects like super-AGBs---the proposed late-stage evolutionary phases of intermediate-mass stars in the range $\approx5$--$12~M_{\odot}$ \citep{doherty17,ogrady20}. The LMC may lack these rare, bright stars on account of its low mass compared to M51.

As shown in Figure~\ref{fig:LRVs_CMD}, the color evolution of these stars is also strikingly similar to that of the Galactic RSG Betelgeuse from the American Association of Variable Star Observers\footnote{\url{https://www.aavso.org/}} (AAVSO) during its historic photometric minimum in 2019--2020, known as the ``Great Dimming'' \citep{guinan19,guinan20,dupree20}. Multiepoch, high-resolution imaging revealed the appearance of an optically dark patch over part of the stellar surface during the dimming \citep{montarges21}. UV observations of chromospheric variability also indicate the formation of a dense, outflowing structure just before the event \citep{dupree20}, indicating an episode of enhanced mass loss. Using the method of TiO-band monitoring and tomography, \citet{harper20} and \citet{kravchenko21} have suggested that an increased molecular opacity in the outflowing material as the primary cause of the dimming event. \citet{levesque20}, however, find that optical spectra during the dimming event do not show significant increases in the depths of molecular (e.g., TiO) absorption features, and argue instead for the formation of large-grain dust. Based on the similarity of their color evolution, it is possible that some of these objects represent similar or even more extreme instances of episodic, enhanced mass loss in cool supergiants, though we reserve a full analysis for future investigations.

We  briefly comment on the two blue-evolving sources shown in Figure~\ref{fig:LRVs_CMD}. Their evolution is reminiscent of that typical of bluer massive stars, namely the characteristic S Doradus variations between a hot (visibly fainter), quiescent state and a cool (visibly brighter), outburst state \citep[e.g.,][]{humphreys94}.  Theoretically, massive stars in the bluer portion of the CMD are expected to be susceptible to instabilities arising in radiation-dominated envelopes that approach the local Eddington limit (\citealp[e.g.,][]{paxton13,jiang15,owocki15}; and see the ``radiation-dominated instabilities'' region in Figures~13 and 16 of C18). We suggest that the ``blue''-evolving variables could be intrinsically bluer stars, possibly LBVs or YSG experiencing LBV-like episodes \citep[as in][]{smith04a}, that are now reddened by the dust in the foreground ISM or formed in their own circumstellar environments during prior eruptive mass-loss events. A detailed examination of these objects is outside the scope of the present work.

Finally, having laid out the modes of variability observed (and expected) for luminous, red-appearing stars, we return to the variability of M51-DS1. As noted above, where color information is available for the 2016--2017 and 2021 epochs, M51-DS1 tracks with the mostly vertical evolution of the yellow-colored variables in Figure~\ref{fig:LRVs_CMD}, consistent with expectations for pulsational instabilities occuring in massive ($\gtrsim 15 M_{\odot}$) RSGs, though large-grain circumstellar dust may also produce extinction that is relatively gray \citep[see, e.g.,][]{scicluna15,massey06,haubois19}. The long period of brightening seen between 2016--2017 may correspond at most to one-half of a full pulsation cycle, requiring that any periodicity be $\gtrsim$2\,yr in duration. Given the star's high luminosity ($M_K = -11.1$\,mag), this is consistent with expectations from the RSG period--luminosity relation observed in the Milky Way, L/SMC, M33, and M31 \citep{kiss06,yang12,soraisam18}. The maximum amplitude of this variability (excluding the 2019 minimum), is $\approx$0.9\,mag, only slightly larger than the pre-2019 variability observed at the level of $\Delta F814W \approx 0.7$\,mag. At an exceptional $\Delta F814W \gtrsim 2.1$\,mag, it is unlikely, though not strongly excluded, that the 2019 minimum is part of the inferred (semi)regular pulsation cycle ($\lesssim10$\% chance of finding the source in the bottom portion of a pulsation only once in the six light-curve samples over 26~yr). We therefore suggest that the 2019 dimming event of M51-DS1 represents a relatively rare occurrence, possibly an exceptional mass-loss event and more extreme cousin of the ``Great Dimming'' of Betelgeuse. 

\subsection{Spectral Energy Distribution Analysis} \label{sec:SEDs}
\begin{figure*}
\centering
\includegraphics[width=0.32\textwidth]{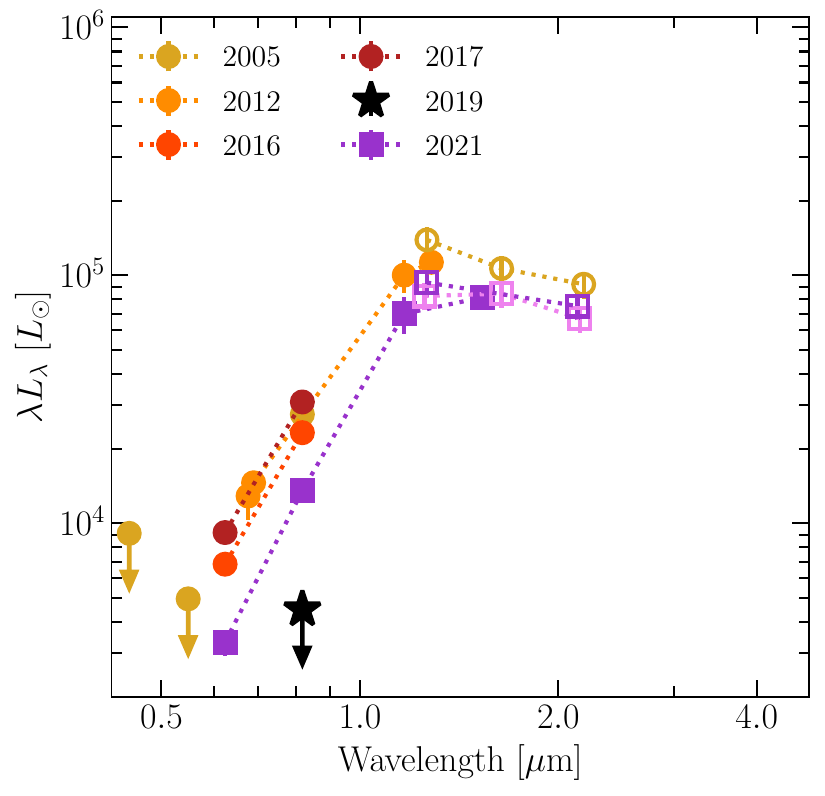}
\includegraphics[width=0.32\textwidth]{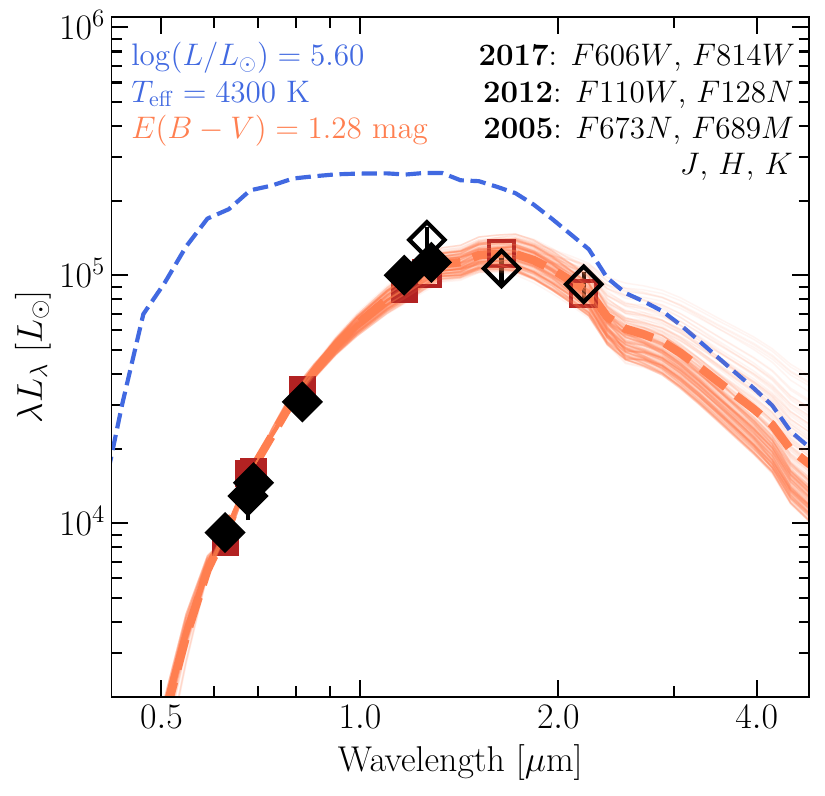}
\includegraphics[width=0.32\textwidth]{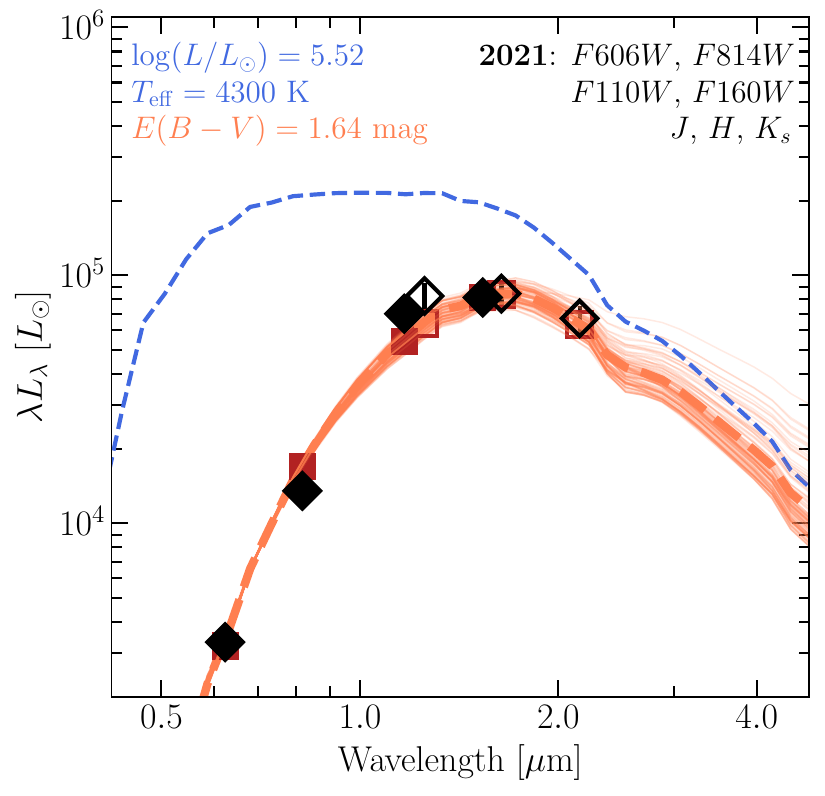}
\caption{\label{fig:SEDs}
Left: multiepoch SEDs of M51-DS1 constructed from visible and near-IR photometry. Space-based measurements from \textit{HST} are shown as filled symbols, while ground-based near-IR measurements are open symbols. Upper limits are indicated with downward arrows. Center: the pre-2019 SED constructed from the 2017 ACS/WFC, 2012 WFC3/UVIS and IR, and 2005 NIRI photometry (black diamonds) is shown along with GRAMS RSG models that provide good fits ($\chi^2 \leq \chi^2_{\mathrm{min}} + \Delta \chi^2_{90}$; see text) to the data. The best-fitting model, described by the parameters listed in the upper-left corner of the panel, is shown as the orange, thick-dashed curve, with the corresponding synthetic photometry points in the available bands shown as red squares. The input (unreddened) stellar photosphere to the model is shown as the blue dashed curve. Right: same as the center panel, but for the 2021 SED constructed from our ACS/WFC, WFC3/IR, and MMIRS photometry. 
}
\end{figure*}

Figure~\ref{fig:SEDs} shows the multiepoch SEDs of M51-DS1, constructed from the available \textit{HST} and ground-based photometry. The photometric magnitudes were converted to band luminosities, $\lambda L_{\lambda}$, using the zero-point fluxes and effective wavelengths available with the \texttt{pysynphot} package \citep{pysynphot} for \textit{HST} instrument and filter setups, and those compiled by the Spanish Virtual Observatory (SVO) Filter Profile Service\footnote{Documentation for the SVO Filter Profile Service is available at \url{http://ivoa.net/documents/Notes/SVOFPSDAL/index.html} and \url{http://ivoa.net/documents/Notes/SVOFPSDAL/index.html}} for the ground-based observations \citep{svo1,svo2}. 

Prior to its dramatic dimming in 2019, the source was characterized by a very red SED peaking in the near-IR between 1 and 2\,$\mu$m at $\lambda L_{\lambda} \approx 10^{5}~L_{\odot}$. The variability seen in the light curves (Section~\ref{sec:lcs}) is also reflected in the SEDs constructed from the 2005, 2012, and 2016--2017 data, though the overall color and shape of the SED remains largely constant. In 2019, the source faded by $\Delta \lambda L_{\lambda} \gtrsim 2.6\times10^4~L_{\odot}$ at $F814W$, a factor of $\gtrsim 7$ or $\gtrsim 85$\% of the flux in that band. Without multiband observations in 2019, we are unable to constrain the shape of the SED or total drop in bolometric luminosity at that epoch. During our follow-up 2021 observations, the source has partially recovered in flux at $F814W$. The 2021 SED also peaks in the near-IR at $\lambda L_{\lambda} \approx 8.2\times10^4~L_{\odot}$ in $F160W$. While overall similar in color and shape, the source is dimmer across all the available optical and near-IR bands compared to the pre-2019 SEDs.

\subsubsection{Spectral Energy Distribution Modeling}\label{sec:SED_models}
To estimate the physical parameters of the star, we attempted to fit the SEDs with the Grid of Red supergiant and Asymptotic Giant Branch ModelS (GRAMS; \citealp{sargent11,srinivasan11}). This suite of radiative transfer models consists of a base grid of 1225 spectra from spherically symmetric shells of varying amounts of silicate dust \citep[][appropriate for RSGs]{ossenkopf92} around stars of constant mass-loss rates computed using the dust radiative transfer code \texttt{2-Dust} \citep{ueta03}. The published grid uses input PHOENIX model photospheres \citep{kucinskas05,kucinskas06} for 1\,$M_{\odot}$ stars (model spectra can be scaled for more luminous and massive, i.e., supergiant, stars) with effective temperatures, $T_{\mathrm{eff}}$, between 2100 and 4700\,K, and at a fixed subsolar metallicity $\log(Z/Z_{\odot}) = -0.5$  and a fixed surface gravity $\log g = -0.5$. The amount of circumstellar dust is characterized in terms of the optical depth at 1~$\mu$m, $\tau_1$, from which a dust mass-loss rate, $\dot M_{\mathrm{d}}$ can be inferred assuming a wind speed of $v_w = 10$\,km\,s$^{-1}$. An additional parameter in the GRAMS grid is the inner radius of the dust shell, $R_{\mathrm{in}}$. We found that our results were largely insensitive to this parameter and we chose to fix it at $R_{\mathrm{in}} = 11.0 R_*$, where $R_*$ is the stellar radius, to reduce the number of free parameters.

To model the source prior to its near-disappearance in 2019, we combined \textit{HST} measurements from the 2017 ACS/WFC observations in $F606W$ and $F814W$ (MJD 58011) and the 2012 WFC3/UVIS observations in $F673N$ and $F689M$ (MJD 56020--56027) and IR observations in $F110W$ and $F128N$ (MJD 56174), together with the 2005 ground-based NIRI measurements in $J$, $H$, and $K$ (MJD 53548). Given the observed variability of the source during this time period, we experimented with fitting different combinations of the available pre-2019 data, but found that these epochs together gave the best results. We also included additional reddening (\citealp{fitzpatrick99} law with $R_V =3.1$) for a range of $E(B-V)$ values between 0.0 and 3.0\,mag to account for the likely significant and uncertain host extinction (see Section~\ref{sec:M51_env}). We then fit each (reddened) model spectrum in the expanded grid (99,225 models in total) by computing $\chi^2$ values between the data points and synthetic photometry on the spectra, weighted by data-point uncertainties and allowing for an overall scaling factor of the flux as a free parameter. In our analysis, we will consider ``good'' models as those with $\chi^2 \leq \chi^2_{\mathrm{min}} + \Delta \chi^2_{90}$, where $\chi^2_{\mathrm{min}}$ is the minimum value of $\chi^2$ found for all the models, and $\Delta \chi^2_{90}$ is the 90\%-tile of the $\chi^2$ distribution with degrees of freedom $\nu$. 

The results of this fitting procedure for the pre-2019 SED are shown in the center panel of Figure~\ref{fig:SEDs}. The best-fitting model ($\chi^2/\nu = 3.12$) has $T_{\mathrm{eff}} = 4300$, $E(B-V) = 1.28$, and $\log(L/L_{\odot}) = 5.60$. With $\nu = 5$ degrees of freedom, $\Delta \chi^2 = 9.24$, and we find a set of 257 good models with $T_{\mathrm{eff}} = 3700$--4700~K, $E(B-V) = 0.56$--1.56\,mag, and $\log(L/L_{\odot}) = 5.35$--5.75. Unsurprisingly, $T_{\mathrm{eff}}$ and $E(B-V)$ are strongly positively correlated for the set of good models, and we note that the range of $T_{\mathrm{eff}}$ they span extends to the maximum value in the grid. The $\chi^2$ distribution is well behaved around the minimum at $T_{\mathrm{eff}} = 4300$~K in the range we tested, although highly reddened, warmer models could, in principle, also provide acceptable fits to the SED.  These results point to a very luminous, cool ($\approx$K or early M spectral type) supergiant star, near the empirical Humphreys--Davidson limit ($\log[L/L_{\odot}] = 5.5$--5.8; \citealp{humphreys79,davies18b}). In comparison to single-star evolutionary tracks from the MIST models \citep{choi16,choi17}, this would correspond to a star with an initial mass $\approx$24--40~$M_{\odot}$  with terminal age $\lesssim 8$~Myr (see Figure~\ref{fig:HRD} and further discussion in Section~\ref{sec:HRD}). 

We fit the post-fading SED, constructed from the 2021 ACS/WFC ($F606W$, $F814W$), WFC3/IR ($F110W$, $F160W$), and MMIRS ($J$, $H$, $K_\mathrm{s}$) imaging, using the same procedure and show the results in the rightmost panel of Figure~\ref{fig:SEDs}. In this case, the best-fitting model ($\chi^2/\nu = 4.78$) has $T_{\mathrm{eff}} = 4300$, $E(B-V) = 1.64$, and $\log(L/L_{\odot}) = 5.52$. The set of 125 good models ($\nu = 3$; $\Delta \chi^2_{90} = 6.25$) now has $T_{\mathrm{eff}} = 3900$ -- 4700~K, $E(B-V) = 1.0$ -- 1.9\,mag, and $\log(L/L_{\odot}) = 5.31$--5.65. Overall, results for the 2021 SED suggest a slightly dimmer star than the pre-fading data, though there is substantial overlap in the allowed parameter ranges. 

While the GRAMS grid includes models with high values of circumstellar extinction up to $\tau_1 \approx 60$, our fitting results for both the pre- and post-fading data favor $\tau_1 \lesssim 0.9$ and corresponding to dust mass-loss rates below $\dot M_{\mathrm{d}} \lesssim 10^{-7}~M_{\odot}$~yr$^{-1}$. This is in accord with modern estimates of mass-loss rates for normal RSG winds in this luminosity range \citep{beasor20}. At the same time, the models favor relatively high values of foreground host (i.e., ISM-like) extinction. Importantly, longer-wavelength data ($\gtrsim 3$--5\,$\mu$m, and especially at the 10\,$\mu$m silicate feature) that are sensitive to emission from warm dust would be required to directly constrain the amount of CSM, so our ability to distinguish circum- and interstellar extinction should be viewed cautiously. Notably too, the modeling results for the 2021 data favor higher values of the foreground extinction to account for the redder colors of the source compared to the pre-fading source. An increase in the amount of intervening ISM material is presumably unphysical, indicating that other effects not captured by the models may be at play in the observed SED evolution. These may include the effects of increased molecular opacity during episodes of enhanced mass loss \citep[see, e.g.,][]{davies21}, changes in composition or the grain-size distribution of circumstellar dust, and/or the effects of nonspherical geometry in the stellar surface (e.g., large convective bubbles and cold spots) or the CSM.  

Lastly, the input PHOENIX model photospheres of the published GRAMS grid are at a metallicity of $\log(Z/Z_{\odot}) = -0.5$, originally chosen to be similar to that of the LMC, while the local environment of M51-DS1 indicates that models at higher, somewhat supersolar, metallicities ($\log[Z/Z_{\odot}] = 0.0$--0.3; see Section~\ref{sec:M51_env}) would be more appropriate. Though metallicity will affect the photospheric spectrum in the optical/near-IR, we do not expect this to be significant compared to the effects of of an external dusty wind or the possible effects of a complex geometry mentioned above. We examined this by also fitting the pre-fading SED with bare PHOENIX photospheres at low, solar, and high metallcities ($\log[Z/Z_{\odot}] = -0.5, 0.0, +0.5$), again including the amount of foreground extinction, $E(B-V)$, as a free parameter. In general, we find similar results regardless of metallicity. The best-fitting models have $T_{\mathrm{eff}} = 4300$ at low and solar metallicities and $T_{\mathrm{eff}} = 4500$ at high metallicity, all well within the preferred parameter ranges found for the GRAMS models.

\subsection{Location in the Hertzsprung--Russell Diagram and Evolutionary State}\label{sec:HRD}
\begin{figure}
\centering
\includegraphics[width=0.5\textwidth]{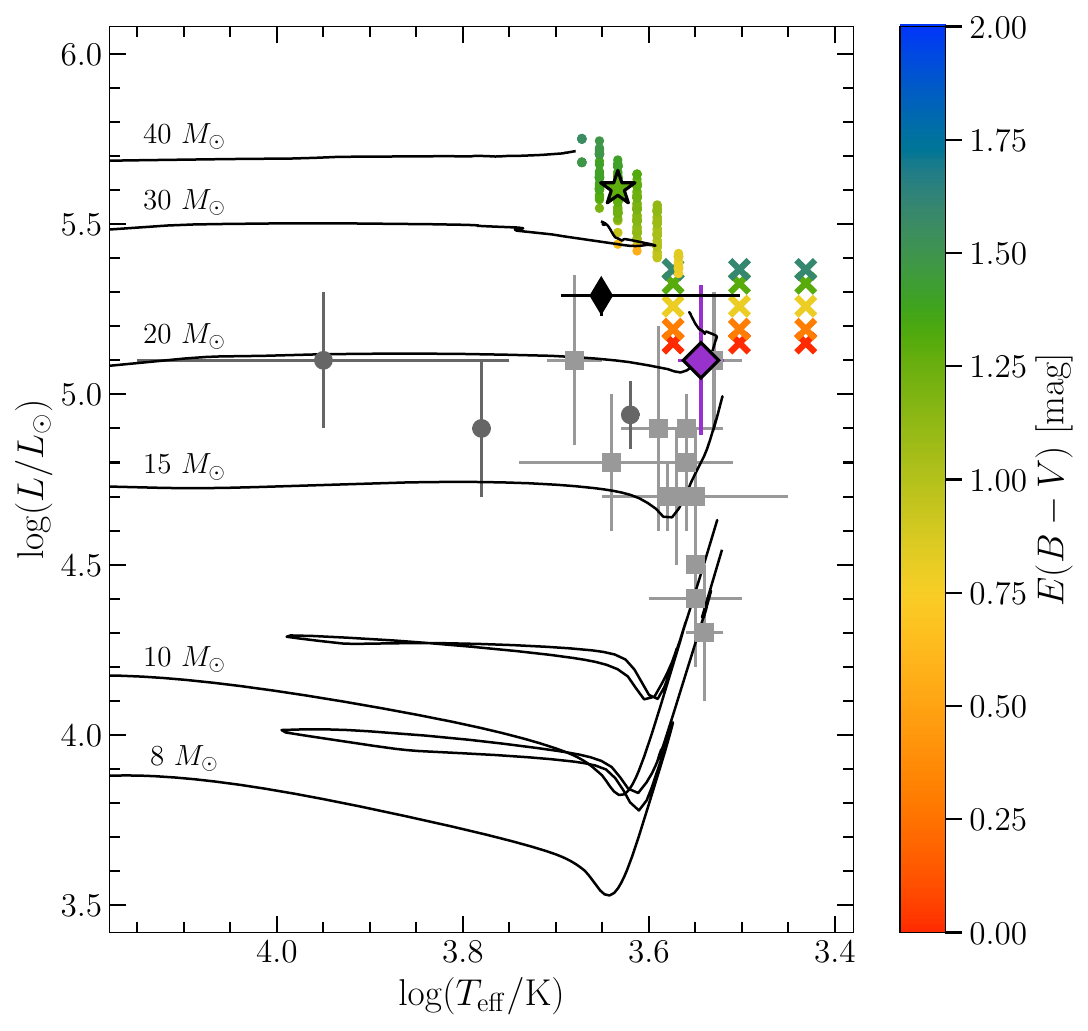}
\caption{\label{fig:HRD}
HRD showing possible locations of M51-DS1. The result for the best-fitting model for the pre-2019 SED is shown as the star symbol, while the range of good models are represented by small circles. The color of each point corresponds to the value of the extinction (in excess of the Galactic foreground), $E(B-V)$, for that model as indicated by the color bar. The ``$\times$'' symbols show the locations of M51-DS1 assuming early, mid, and late M-type spectra and applying the $K$-band bolometric corrections of \citet{davies18a} to the 2005 NIRI measurement, and with the same mapping of color to $E(B-V)$. We show stellar evolutionary tracks from MIST (nonrotating, solar metallicity) for a set of massive stars in the range $M = 8$--40\,$M_{\odot}$ as black curves for comparison. We also indicate the locations of Betelgeuse (purple diamond; \citealp{dolan16}), the RSG progenitor of NGC\,6949-BH1 (thin black diamond; model P5 from \citealp{adams17a}), and the collection of directly detected SN progenitors of types II (light gray squares) and IIb (dark gray circles) from \citet{smartt15}. 
}
\end{figure}

We examine possible locations M51-DS1 in an HRD (Figure~\ref{fig:HRD}) and discuss implications for the inferred initial mass and evolutionary state of the star. Specifically, we directly compare the results of applying a simple $K$-band bolometric correction \citep[][assuming an M-type RSG; see Section~\ref{sec:nearIR_Lbol}]{davies18a} to those of our SED model fitting (Section~\ref{sec:SED_models}). Overall, the luminosity of the star inferred from applying $BC_K$ is lower than that from the SED fitting, falling in the range $\log (L/L_{\odot}) \approx 5.15$--5.37, notably above the range inferred for the growing collection of SN progenitors and comparable to that inferred for the failed SN candidate NGC\,6946-BH1, depending on the value assumed for the foreground extinction between $E(B-V) = 0.0$ and 1.6\,mag. The SED modeling prefers somewhat warmer ($T_{\mathrm{eff}} = 3700$--4700~K) photospheric models at relatively high extinction ($E[B-V] = 0.56$--1.56\,mag) and luminosities in the range $\log(L/L_{\odot}) = 5.35$--5.75. Given the constraints on the ISM environment discussed in Section~\ref{sec:M51_env}, the high $E(B-V)$ values $\gtrsim 1$\,mag would likely point to significant circumstellar extinction.

Even for an early M-type star ($T_{\mathrm{eff}} \approx 3700$~K), where there is overlap between the two methods, the inferred luminosity from SED fitting is about a factor of 1.3 higher than from applying $BC_K$ for the same extinction of $E(B-V) = 0.8$\,mag. This modest discrepancy could be resolved if the effective extinction law were grayer---either from a shallower ISM law (i.e., $R_V > 3.1$) for the central regions of M51 or perhaps related to properties of the stellar wind or circumstellar dust not captured by the models (e.g., nonspherical geometry or nonisotropic scattering by dust grains)---resulting in a smaller correction at bluer wavelengths and a lower bolometric luminosity for a given model. Alternatively, using a smaller $BC_K$ at earlier spectral types (as found for cool supergiants in the Magellanic Clouds; 
\citealp[e.g.,][]{elias85,levesque06,davies18b}) would also bring the luminosity estimates closer to agreement. 

Altogether, we infer that M51-DS1 is likely a luminous RSG at $\log(L/L_{\odot}) \approx 5.2$--$5.3$ (for $T_{\mathrm{eff}} \lesssim 3700$~K and reasonable values of $E(B-V)=0.4$--1.3), corresponding to a highly evolved star of initial mass $M \approx 19$--22\,$M_{\odot}$, or somewhat warmer ($T_{\mathrm{eff}} = 3700$--4700~K)---possibly a YSG on a post-RSG track back to higher temperatures---at higher luminosities up to $\log(L/L_{\odot}) \approx 5.4$--$5.7$. This more extreme case would suggest an initial mass as high as $M \approx 26$--40\,$M_{\odot}$ for a star near the empirical Humphreys--Davidson limit in an exceptionally short-lived and rare state of stellar evolution.

\section{Summary and Conclusions} \label{sec:summary}
We have presented a detailed analysis and characterization of a remarkable dimming event of a luminous, cool supergiant star in the nearby, star-forming galaxy M51, found as part of a new search for failed SNe as ``disappearing'' massive stars with the \textit{HST}. The object, which we call M51-DS1, was detected dozens of times in the wealth of archival \textit{HST} imaging for more than two decades since 1995 before undergoing a dramatic, near-total disappearance in the 2019 ACS/WFC $F814W$ images obtained as part of our search. Thus meeting our criteria for a strong failed SN candidate (outlined in Section~\ref{sec:candidates}), we conducted follow-up \textit{HST} imaging observations with a Cycle 28 mid-cycle program using ACS/WFC and WFC3/IR, supplemented with ground-based, near-IR imaging from Gemini-N/NIRI and MMT/MMIRS, to attempt to confirm the disappearance of the star. The star was found to have partially rebrightened to near its pre-disappearance levels across the optical and near-IR, ruling out the terminal collapse of the star in a failed SN. As summarized below, our analysis indicates instead that an isolated episode of enhanced mass loss on a very massive, cool Y/RSG star---potentially a rare, more extreme cousin of the recent ``Great Dimming'' of Betelgeuse---can explain its observed properties. 

M51-DS1 is located in the inner regions of M51 along the edge of a spiral arm. Previous studies of the host galaxy suggest an environment of approximately solar or somewhat supersolar metallicity ($\log[Z/Z_{\odot}] \approx 0.0$--0.3) and significant foreground extinction ($E[B-V] \approx 0.4$--1.0\,mag). Photometry of stars in the immediate vicinity indicates the presence of a young population ($\lesssim 6$\,Myr old) within a 20~pc radius. M51-DS1 itself is the brightest red source in this region, implying an initial mass of at least $\approx$30\,$M_{\odot}$ if it is part of the same population. It is difficult, however, to disentangle possible confounding effects, including a mix of stellar populations of different ages and high spatial variation in the foreground extinction toward individual sources up to $E(B-V) \approx 2.0$\,mag. At $M_K = -11.1$\,mag in 2005 and with $J-K\approx1.0$--1.2\,mag (Galactic-extinction correction only), the star is consistent with a luminous RSG, and applying the $K$-band bolometric correction of \citet{davies18a} yields $\log(L/L_{\odot}) = 5.2$--5.3, implying an initial mass $M = 19$--22\,$M_\odot$ (assuming an M-type supergiant with $T_{\mathrm{eff}}\lesssim3700$~K and foreground extinction $E[B-V]=0.4$--1.3\,mag). Modeling of the SED suggests higher temperatures ($T_{\mathrm{eff}}\approx3700$--$4700$~K) and luminosities ($\log[L/L_{\odot}] = 5.4$--5.7), pointing to a more massive (initial mass $M = 26$--40\,$M_\odot$) YSG or post-RSG star that would be consistent with the inferred age of the very young stellar population in the immediate vicinity, and possibly requiring circumstellar extinction. 

The extended, 26~yr light curve of M51-DS1 indicates a history of variability with $\Delta F814W \approx 0.7$--0.9\,mag, consistent with semiregular variations typical of Y/RSGs and expected theoretically for massive cool stars above $\gtrsim$15$\,M_{\odot}$. At $\Delta F814W \gtrsim 2.1$\,mag, we find that the 2019 minimum is unlikely to be part of the typical variation cycle for this star, but rather an uncommon and exceptionally deep dimming event---notably, more than $\gtrsim$1.5\,mag deeper than the 2019--2020 historic minimum of Betelgeuse in the comparable $I$ band. In the context of other large-amplitude modes of variation seen in luminous red stars in M51, we suggest that the 2019 event of M51-DS1 could be associated with an enhanced episode of mass loss from the star, in which increased opacity from the dense, molecular wind or the formation of new dust grains temporarily obscured the star.

This discovery highlights a central challenge for the definitive identification of failed SNe, namely, that the massive stars in question exhibit optical variability that can mimic a disappearing star. In particular, episodic mass loss in massive, cool supergiants remains poorly understood \citep[e.g.,][]{smith14}, both in terms of the physical mechanisms that drive it and critical properties such as the frequency, possible duration, and mass-loss rates of the most extreme events; an extended mass-loss event could plausibly obscure a massive star in the optical for years. While Galactic examples of episodic mass loss from cool supergiants, like Betelgeuse's Great Dimming, can be studied in real time and in exquisite detail, such events will be few and far between. Still, significant progress can be made in mining the ever-growing trove of archival data from both space-based (e.g., \textit{HST}, \textit{Spitzer}) and numerous ground-based time-domain surveys to uncover and characterize the population of such sources in nearby galaxies, both as possible contaminants in ongoing failed SN searches and as direct probes of episodic mass-loss in massive, evolved stars. At the same time, a continued, concerted effort to monitor existing and newly uncovered failed SN candidates is vital, especially observations in the near-IR and mid-IR, e.g., soon with the \textit{James Webb Space Telescope} and later with the \textit{Nancy Grace Roman Space Telescope}, to test for surviving obscured stars and direct signatures of a potential obscuring wind or dust. 

\vspace{0.5cm}
We thank the anonymous referee for their helpful comments that improved the paper. We thank K.\ Paterson for help with MMIRS imaging data reduction. We also thank the observing support staffs of the MMT, LBT, Gemini Observatory, and the \textit{HST} for their help in planning, obtaining, and analyzing the observations presented in this work. 

Some of the data presented in this paper were obtained from the Mikulski Archive for Space Telescopes (MAST) at the Space Telescope Science Institute. The specific observations analyzed can be accessed via \dataset[10.17909/t9-z1c2-ye93]{https://doi.org/10.17909/t9-z1c2-ye93}.

Time-domain research by D.J.S.\ is also supported by NSF grant Nos.\ AST-1821987, 1813466, 1908972, \& 2108032, and by the Heising-Simons Foundation under grant \#2020-1864. 
J.S.\ acknowledges support from NASA grant No.\ HST-GO-15645.003-A and the Packard Foundation.
Research by S.V.\ is supported by NSF grant Nos.\ AST-1813176 and AST-2008108. 
E.R.B.\ is supported by NASA through a Hubble Fellowship grant No.\ HST-HF2-51428 awarded by the Space Telescope Science Institute, which is operated by the Association of Universities for Research in Astronomy, Inc., for NASA, under contract NAS5-26555. 

Based on observations made with the NASA/ESA Hubble Space Telescope, obtained at the Space Telescope Science Institute, which is operated by the Association of Universities for Research in Astronomy, Inc., under NASA contract NAS5-26555. These observations are associated with programs \#HST-GO-15645, 16508, 10452, 14704, 12490, and 12762. Support for program \#HST-GO-15645 and \#HST-GO-16508 was provided by NASA through a grant from the Space Telescope Science Institute, which is operated by the Association of Universities for Research in Astronomy, Inc., under NASA contract NAS5-26555.

Observations reported here were obtained at the MMT Observatory, a joint facility of the University of Arizona and the Smithsonian Institution.

Based on observations obtained at the Gemini Observatory (Programs GN-2005A-Q-49 and GN-2021A-DD-101), which is operated by the Association of Universities for Research in Astronomy, Inc., under a cooperative agreement with the NSF on behalf of the Gemini partnership: the National Science Foundation (United States), National Research Council (Canada), CONICYT (Chile), Ministerio de Ciencia, Tecnolog\'{i}a e Innovaci\'{o}n Productiva (Argentina), Minist\'{e}rio da Ci\^{e}ncia, Tecnologia e Inova\c{c}\~{a}o (Brazil), and Korea Astronomy and Space Science Institute (Republic of Korea). 

We acknowledge with thanks the variable star observations from the AAVSO International Database contributed by observers worldwide and used in this research.

\facilities{HST (ACS, WFC3, WFPC2), Gemini:Gillett (NIRI), MMT (MMIRS), LBT (LUCI-1), AAVSO}

\software{\texttt{AstroDrizzle}, \texttt{TweakReg} (\url{http://drizzlepac.stsci.edu/}; \citealp{hack12}), \texttt{SExtractor}, \texttt{PSFEx} (\url{https://www.astromatic.net/software/};
\citealp{bertin96,bertin11}), \texttt{DOLPHOT} (\url{http://americano.dolphinsim.com/dolphot/}; \citealp{dolphin00,dolphin16}, \texttt{DRAGONS} (\url{https://dragons.readthedocs.io/en/v2.1.1/index.html}), \texttt{Astropy} (\url{https://www.astropy.org/}; \citealp{2013A&A...558A..33A,astropy}), \texttt{photutils}, \texttt{EPSFBuilder} (\url{https://photutils.readthedocs.io/en/stable/}; \citealp{bradley20_photutils})} 



\bibliography{jencson}

\begin{thebibliography}{}
\expandafter\ifx\csname natexlab\endcsname\relax\def\natexlab#1{#1}\fi
\providecommand{\url}[1]{\href{#1}{#1}}
\providecommand{\dodoi}[1]{doi:~\href{http://doi.org/#1}{\nolinkurl{#1}}}
\providecommand{\doeprint}[1]{\href{http://ascl.net/#1}{\nolinkurl{http://ascl.net/#1}}}
\providecommand{\doarXiv}[1]{\href{https://arxiv.org/abs/#1}{\nolinkurl{https://arxiv.org/abs/#1}}}

\bibitem[{{Abbott} {et~al.}(2019){Abbott}, {Abbott}, {Abbott}, {Abraham},
  {Acernese}, {Ackley}, {Adams}, {Adhikari}, {Adya}, {Affeldt}, {Agathos},
  {Agatsuma}, {Aggarwal}, {Aguiar}, {Aiello}, {Ain}, {Ajith}, {Allen},
  {Allocca}, {Aloy}, {Altin}, {Amato}, {Ananyeva}, {Anderson}, {Anderson},
  {Angelova}, {Antier}, {Appert}, {Arai}, {Araya}, {Areeda}, {Ar{\`e}ne},
  {Arnaud}, {Arun}, {Ascenzi}, {Ashton}, {Aston}, {Astone}, {Aubin}, {Aufmuth},
  {AultONeal}, {Austin}, {Avendano}, {Avila-Alvarez}, {Babak}, {Bacon},
  {Badaracco}, {Bader}, {Bae}, {Baker}, {Baldaccini}, {Ballardin}, {Ballmer},
  {Banagiri}, {Barayoga}, {Barclay}, {Barish}, {Barker}, {Barkett}, {Barnum},
  {Barone}, {Barr}, {Barsotti}, {Barsuglia}, {Barta}, {Bartlett}, {Bartos},
  {Bassiri}, {Basti}, {Bawaj}, {Bayley}, {Bazzan}, {B{\'e}csy}, {Bejger},
  {Belahcene}, {Bell}, {Beniwal}, {Berger}, {Bergmann}, {Bernuzzi}, {Bero},
  {Berry}, {Bersanetti}, {Bertolini}, {Betzwieser}, {Bhandare}, {Bidler},
  {Bilenko}, {Bilgili}, {Billingsley}, {Birch}, {Birney}, {Birnholtz},
  {Biscans}, {Biscoveanu}, {Bisht}, {Bitossi}, {Bizouard}, {Blackburn},
  {Blackman}, {Blair}, {Blair}, {Blair}, {Bloemen}, {Bode}, {Boer}, {Boetzel},
  {Bogaert}, {Bondu}, {Bonilla}, {Bonnand}, {Booker}, {Boom}, {Booth}, {Bork},
  {Boschi}, {Bose}, {Bossie}, {Bossilkov}, {Bosveld}, {Bouffanais}, {Bozzi},
  {Bradaschia}, {Brady}, {Bramley}, {Branchesi}, {Brau}, {Briant}, {Briggs},
  {Brighenti}, {Brillet}, {Brinkmann}, {Brisson}, {Brockill}, {Brooks},
  {Brown}, {Brunett}, {Buikema}, {Bulik}, {Bulten}, {Buonanno}, {Buskulic},
  {Bustamante Rosell}, {Buy}, {Byer}, {Cabero}, {Cadonati}, {Cagnoli},
  {Cahillane}, {Calder{\'o}n Bustillo}, {Callister}, {Calloni}, {Camp},
  {Campbell}, {Canepa}, {Cannon}, {Cao}, {Cao}, {Capocasa}, {Carbognani},
  {Caride}, {Carney}, {Carullo}, {Casanueva Diaz}, {Casentini}, {Caudill},
  {Cavagli{\`a}}, {Cavalier}, {Cavalieri}, {Cella}, {Cerd{\'a}-Dur{\'a}n},
  {Cerretani}, {Cesarini}, {Chaibi}, {Chakravarti}, {Chamberlin}, {Chan},
  {Chao}, {Charlton}, {Chase}, {Chassande-Mottin}, {Chatterjee}, {Chaturvedi},
  {Chatziioannou}, {Cheeseboro}, {Chen}, {Chen}, {Chen}, {Cheng}, {Cheong},
  {Chia}, {Chincarini}, {Chiummo}, {Cho}, {Cho}, {Cho}, {Christensen}, {Chu},
  {Chua}, {Chung}, {Chung}, {Ciani}, {Ciobanu}, {Ciolfi}, {Cipriano}, {Cirone},
  {Clara}, {Clark}, {Clearwater}, {Cleva}, {Cocchieri}, {Coccia}, {Cohadon},
  {Cohen}, {Colgan}, {Colleoni}, {Collette}, {Collins}, {Cominsky},
  {Constancio}, {Conti}, {Cooper}, {Corban}, {Corbitt}, {Cordero-Carri{\'o}n},
  {Corley}, {Cornish}, {Corsi}, {Cortese}, {Costa}, {Cotesta}, {Coughlin},
  {Coughlin}, {Coulon}, {Countryman}, {Couvares}, {Covas}, {Cowan}, {Coward},
  {Cowart}, {Coyne}, {Coyne}, {Creighton}, {Creighton}, {Cripe}, {Croquette},
  {Crowder}, {Cullen}, {Cumming}, {Cunningham}, {Cuoco}, {Canton}, {D{\'a}lya},
  {Danilishin}, {D'Antonio}, {Danzmann}, {Dasgupta}, {Da Silva Costa},
  {Datrier}, {Dattilo}, {Dave}, {Davier}, {Davis}, {Daw}, {DeBra},
  {Deenadayalan}, {Degallaix}, {De Laurentis}, {Del{\'e}glise}, {Del Pozzo},
  {DeMarchi}, {Demos}, {Dent}, {De Pietri}, {Derby}, {De Rosa}, {De Rossi},
  {DeSalvo}, {de Varona}, {Dhurandhar}, {D{\'\i}az}, {Dietrich}, {Di Fiore},
  {Di Giovanni}, {Di Girolamo}, {Di Lieto}, {Ding}, {Di Pace}, {Di Palma}, {Di
  Renzo}, {Dmitriev}, {Doctor}, {Donovan}, {Dooley}, {Doravari}, {Dorrington},
  {Downes}, {Drago}, {Driggers}, {Du}, {Ducoin}, {Dupej}, {Dwyer}, {Easter},
  {Edo}, {Edwards}, {Effler}, {Ehrens}, {Eichholz}, {Eikenberry}, {Eisenmann},
  {Eisenstein}, {Essick}, {Estelles}, {Estevez}, {Etienne}, {Etzel}, {Evans},
  {Evans}, {Fafone}, {Fair}, {Fairhurst}, {Fan}, {Farinon}, {Farr}, {Farr},
  {Fauchon-Jones}, {Favata}, {Fays}, {Fazio}, {Fee}, {Feicht}, {Fejer}, {Feng},
  {Fernandez-Galiana}, {Ferrante}, {Ferreira}, {Ferreira}, {Ferrini},
  {Fidecaro}, {Fiori}, {Fiorucci}, {Fishbach}, {Fisher}, {Fishner},
  {Fitz-Axen}, {Flaminio}, {Fletcher}, {Flynn}, {Fong}, {Font}, {Forsyth},
  {Fournier}, {Frasca}, {Frasconi}, {Frei}, {Freise}, {Frey}, {Frey},
  {Fritschel}, {Frolov}, {Fulda}, {Fyffe}, {Gabbard}, {Gadre}, {Gaebel},
  {Gair}, {Gammaitoni}, {Ganija}, {Gaonkar}, {Garcia},
  {Garc{\'\i}a-Quir{\'o}s}, {Garufi}, {Gateley}, {Gaudio}, {Gaur}, {Gayathri},
  {Gemme}, {Genin}, {Gennai}, {George}, {George}, {Gergely}, {Germain},
  {Ghonge}, {Ghosh}, {Ghosh}, {Ghosh}, {Giacomazzo}, {Giaime}, {Giardina},
  {Giazotto}, {Gill}, {Giordano}, {Glover}, {Godwin}, {Goetz}, {Goetz},
  {Goncharov}, {Gonz{\'a}lez}, {Gonzalez Castro}, {Gopakumar}, {Gorodetsky},
  {Gossan}, {Gosselin}, {Gouaty}, {Grado}, {Graef}, {Granata}, {Grant}, {Gras},
  {Grassia}, {Gray}, {Gray}, {Greco}, {Green}, {Green}, {Gretarsson}, {Groot},
  {Grote}, {Grunewald}, {Gruning}, {Guidi}, {Gulati}, {Guo}, {Gupta}, {Gupta},
  {Gustafson}, {Gustafson}, {Haegel}, {Halim}, {Hall}, {Hall}, {Hamilton},
  {Hammond}, {Haney}, {Hanke}, {Hanks}, {Hanna}, {Hannam}, {Hannuksela},
  {Hanson}, {Hardwick}, {Haris}, {Harms}, {Harry}, {Harry}, {Haster},
  {Haughian}, {Hayes}, {Healy}, {Heidmann}, {Heintze}, {Heitmann}, {Hello},
  {Hemming}, {Hendry}, {Heng}, {Hennig}, {Heptonstall}, {Hernandez Vivanco},
  {Heurs}, {Hild}, {Hinderer}, {Hoak}, {Hochheim}, {Hofman}, {Holgado},
  {Holland}, {Holt}, {Holz}, {Hopkins}, {Horst}, {Hough}, {Howell}, {Hoy},
  {Hreibi}, {Huang}, {Huerta}, {Huet}, {Hughey}, {Hulko}, {Husa}, {Huttner},
  {Huynh-Dinh}, {Idzkowski}, {Iess}, {Ingram}, {Inta}, {Intini}, {Irwin},
  {Isa}, {Isac}, {Isi}, {Iyer}, {Izumi}, {Jacqmin}, {Jadhav}, {Jani},
  {Janthalur}, {Jaranowski}, {Jenkins}, {Jiang}, {Johnson}, {Johnson-McDaniel},
  {Jones}, {Jones}, {Jones}, {Jonker}, {Ju}, {Junker}, {Kalaghatgi},
  {Kalogera}, {Kamai}, {Kandhasamy}, {Kang}, {Kanner}, {Kapadia}, {Karki},
  {Karvinen}, {Kashyap}, {Kasprzack}, {Katsanevas}, {Katsavounidis}, {Katzman},
  {Kaufer}, {Kawabe}, {Keerthana}, {K{\'e}f{\'e}lian}, {Keitel}, {Kennedy},
  {Key}, {Khalili}, {Khan}, {Khan}, {Khan}, {Khan}, {Khazanov}, {Khursheed},
  {Kijbunchoo}, {Kim}, {Kim}, {Kim}, {Kim}, {Kim}, {Kim}, {Kimball}, {King},
  {King}, {Kinley-Hanlon}, {Kirchhoff}, {Kissel}, {Kleybolte}, {Klika},
  {Klimenko}, {Knowles}, {Koch}, {Koehlenbeck}, {Koekoek}, {Koley},
  {Kondrashov}, {Kontos}, {Koper}, {Korobko}, {Korth}, {Kowalska}, {Kozak},
  {Kringel}, {Krishnendu}, {Kr{\'o}lak}, {Kuehn}, {Kumar}, {Kumar}, {Kumar},
  {Kumar}, {Kuo}, {Kutynia}, {Kwang}, {Lackey}, {Lai}, {Lam}, {Landry}, {Lane},
  {Lang}, {Lange}, {Lantz}, {Lanza}, {Lartaux-Vollard}, {Lasky}, {Laxen},
  {Lazzarini}, {Lazzaro}, {Leaci}, {Leavey}, {Lecoeuche}, {Lee}, {Lee}, {Lee},
  {Lee}, {Lee}, {Lee}, {Lehmann}, {Lenon}, {Leroy}, {Letendre}, {Levin}, {Li},
  {Li}, {Li}, {Li}, {Lin}, {Linde}, {Linker}, {Littenberg}, {Liu}, {Liu}, {Lo},
  {Lockerbie}, {London}, {Longo}, {Lorenzini}, {Loriette}, {Lormand},
  {Losurdo}, {Lough}, {Lousto}, {Lovelace}, {Lower}, {L{\"u}ck}, {Lumaca},
  {Lundgren}, {Lynch}, {Ma}, {Macas}, {Macfoy}, {MacInnis}, {Macleod},
  {Macquet}, {Maga{\~n}a-Sandoval}, {Maga{\~n}a Zertuche}, {Magee}, {Majorana},
  {Maksimovic}, {Malik}, {Man}, {Mandic}, {Mangano}, {Mansell}, {Manske},
  {Mantovani}, {Marchesoni}, {Marion}, {M{\'a}rka}, {M{\'a}rka}, {Markakis},
  {Markosyan}, {Markowitz}, {Maros}, {Marquina}, {Marsat}, {Martelli},
  {Martin}, {Martin}, {Martynov}, {Mason}, {Massera}, {Masserot}, {Massinger},
  {Masso-Reid}, {Mastrogiovanni}, {Matas}, {Matichard}, {Matone}, {Mavalvala},
  {Mazumder}, {McCann}, {McCarthy}, {McClelland}, {McCormick}, {McCuller},
  {McGuire}, {McIver}, {McManus}, {McRae}, {McWilliams}, {Meacher}, {Meadors},
  {Mehmet}, {Mehta}, {Meidam}, {Melatos}, {Mendell}, {Mercer}, {Mereni},
  {Merilh}, {Merzougui}, {Meshkov}, {Messenger}, {Messick}, {Metzdorff},
  {Meyers}, {Miao}, {Michel}, {Middleton}, {Mikhailov}, {Milano}, {Miller},
  {Miller}, {Millhouse}, {Mills}, {Milovich-Goff}, {Minazzoli}, {Minenkov},
  {Mishkin}, {Mishra}, {Mistry}, {Mitra}, {Mitrofanov}, {Mitselmakher},
  {Mittleman}, {Mo}, {Moffa}, {Mogushi}, {Mohapatra}, {Montani}, {Moore},
  {Moraru}, {Moreno}, {Morisaki}, {Mours}, {Mow-Lowry}, {Mukherjee},
  {Mukherjee}, {Mukherjee}, {Mukund}, {Mullavey}, {Munch}, {Mu{\~n}iz},
  {Muratore}, {Murray}, {Nagar}, {Nardecchia}, {Naticchioni}, {Nayak},
  {Neilson}, {Nelemans}, {Nelson}, {Nery}, {Neunzert}, {Ng}, {Ng}, {Nguyen},
  {Nichols}, {Nielsen}, {Nissanke}, {Nitz}, {Nocera}, {North}, {Nuttall},
  {Obergaulinger}, {Oberling}, {O'Brien}, {O'Dea}, {Ogin}, {Oh}, {Oh}, {Ohme},
  {Ohta}, {Okada}, {Oliver}, {Oppermann}, {Oram}, {O'Reilly}, {Ormiston},
  {Ortega}, {O'Shaughnessy}, {Ossokine}, {Ottaway}, {Overmier}, {Owen}, {Pace},
  {Pagano}, {Page}, {Pai}, {Pai}, {Palamos}, {Palashov}, {Palomba},
  {Pal-Singh}, {Pan}, {Pang}, {Pang}, {Pankow}, {Pannarale}, {Pant},
  {Paoletti}, {Paoli}, {Papa}, {Parida}, {Parker}, {Pascucci}, {Pasqualetti},
  {Passaquieti}, {Passuello}, {Patil}, {Patricelli}, {Pearlstone}, {Pedersen},
  {Pedraza}, {Pedurand}, {Pele}, {Penn}, {Perego}, {Perez}, {Perreca},
  {Pfeiffer}, {Phelps}, {Phukon}, {Piccinni}, {Pichot}, {Piergiovanni},
  {Pillant}, {Pinard}, {Pirello}, {Pitkin}, {Poggiani}, {Pong}, {Ponrathnam},
  {Popolizio}, {Porter}, {Powell}, {Prajapati}, {Prasad}, {Prasai}, {Prasanna},
  {Pratten}, {Prestegard}, {Privitera}, {Prodi}, {Prokhorov}, {Puncken},
  {Punturo}, {Puppo}, {P{\"u}rrer}, {Qi}, {Quetschke}, {Quinonez}, {Quintero},
  {Quitzow-James}, {Raab}, {Radkins}, {Radulescu}, {Raffai}, {Raja}, {Rajan},
  {Rajbhandari}, {Rakhmanov}, {Ramirez}, {Ramos-Buades}, {Rana}, {Rao},
  {Rapagnani}, {Raymond}, {Razzano}, {Read}, {Regimbau}, {Rei}, {Reid},
  {Reitze}, {Ren}, {Ricci}, {Richardson}, {Richardson}, {Ricker},
  {Riemenschneider}, {Riles}, {Rizzo}, {Robertson}, {Robie}, {Robinet},
  {Rocchi}, {Rolland}, {Rollins}, {Roma}, {Romanelli}, {Romano}, {Romel},
  {Romie}, {Rose}, {Rosi{\'n}ska}, {Rosofsky}, {Ross}, {Rowan}, {R{\"u}diger},
  {Ruggi}, {Rutins}, {Ryan}, {Sachdev}, {Sadecki}, {Sakellariadou}, {Salafia},
  {Salconi}, {Saleem}, {Salemi}, {Samajdar}, {Sammut}, {Sanchez}, {Sanchez},
  {Sanchis-Gual}, {Sandberg}, {Sanders}, {Santiago}, {Sarin}, {Sassolas},
  {Sathyaprakash}, {Saulson}, {Sauter}, {Savage}, {Schale}, {Scheel},
  {Scheuer}, {Schmidt}, {Schnabel}, {Schofield}, {Sch{\"o}nbeck}, {Schreiber},
  {Schulte}, {Schutz}, {Schwalbe}, {Scott}, {Scott}, {Seidel}, {Sellers},
  {Sengupta}, {Sennett}, {Sentenac}, {Sequino}, {Sergeev}, {Setyawati},
  {Shaddock}, {Shaffer}, {Shahriar}, {Shaner}, {Shao}, {Sharma}, {Shawhan},
  {Shen}, {Shink}, {Shoemaker}, {Shoemaker}, {ShyamSundar}, {Siellez},
  {Sieniawska}, {Sigg}, {Silva}, {Singer}, {Singh}, {Singhal}, {Sintes},
  {Sitmukhambetov}, {Skliris}, {Slagmolen}, {Slaven-Blair}, {Smith}, {Smith},
  {Somala}, {Son}, {Sorazu}, {Sorrentino}, {Souradeep}, {Sowell}, {Spencer},
  {Srivastava}, {Srivastava}, {Staats}, {Stachie}, {Standke}, {Steer},
  {Steinke}, {Steinlechner}, {Steinlechner}, {Steinmeyer}, {Stevenson},
  {Stocks}, {Stone}, {Stops}, {Strain}, {Stratta}, {Strigin}, {Strunk},
  {Sturani}, {Stuver}, {Sudhir}, {Summerscales}, {Sun}, {Sunil}, {Suresh},
  {Sutton}, {Swinkels}, {Szczepa{\'n}czyk}, {Tacca}, {Tait}, {Talbot},
  {Talukder}, {Tanner}, {T{\'a}pai}, {Taracchini}, {Tasson}, {Taylor}, {Thies},
  {Thomas}, {Thomas}, {Thondapu}, {Thorne}, {Thrane}, {Tiwari}, {Tiwari},
  {Tiwari}, {Toland}, {Tonelli}, {Tornasi}, {Torres-Forn{\'e}}, {Torrie},
  {T{\"o}yr{\"a}}, {Travasso}, {Traylor}, {Tringali}, {Trovato}, {Trozzo},
  {Trudeau}, {Tsang}, {Tse}, {Tso}, {Tsukada}, {Tsuna}, {Tuyenbayev}, {Ueno},
  {Ugolini}, {Unnikrishnan}, {Urban}, {Usman}, {Vahlbruch}, {Vajente},
  {Valdes}, {van Bakel}, {van Beuzekom}, {van den Brand}, {Van Den Broeck},
  {Vander-Hyde}, {van Heijningen}, {van der Schaaf}, {van Veggel}, {Vardaro},
  {Varma}, {Vass}, {Vas{\'u}th}, {Vecchio}, {Vedovato}, {Veitch}, {Veitch},
  {Venkateswara}, {Venugopalan}, {Verkindt}, {Vetrano}, {Vicer{\'e}}, {Viets},
  {Vine}, {Vinet}, {Vitale}, {Vo}, {Vocca}, {Vorvick}, {Vyatchanin}, {Wade},
  {Wade}, {Wade}, {Walet}, {Walker}, {Wallace}, {Walsh}, {Wang}, {Wang},
  {Wang}, {Wang}, {Wang}, {Ward}, {Warden}, {Warner}, {Was}, {Watchi},
  {Weaver}, {Wei}, {Weinert}, {Weinstein}, {Weiss}, {Wellmann}, {Wen},
  {Wessel}, {We{\ss}els}, {Westhouse}, {Wette}, {Whelan}, {White}, {Whiting},
  {Whittle}, {Wilken}, {Williams}, {Williamson}, {Willis}, {Willke}, {Wimmer},
  {Winkler}, {Wipf}, {Wittel}, {Woan}, {Woehler}, {Wofford}, {Worden},
  {Wright}, {Wu}, {Wysocki}, {Xiao}, {Yamamoto}, {Yancey}, {Yang}, {Yap},
  {Yazback}, {Yeeles}, {Yu}, {Yu}, {Yuen}, {Yvert}, {Zadro{\.Z}ny}, {Zanolin},
  {Zappa}, {Zelenova}, {Zendri}, {Zevin}, {Zhang}, {Zhang}, {Zhang}, {Zhao},
  {Zhou}, {Zhou}, {Zhu}, {Zimmerman}, {Zlochower}, {Zucker}, {Zweizig}, {LIGO
  Scientific Collaboration}, \& {Virgo Collaboration}}]{GWTC-1}
{Abbott}, B.~P., {Abbott}, R., {Abbott}, T.~D., {et~al.} 2019, Physical Review
  X, 9, 031040, \dodoi{10.1103/PhysRevX.9.031040}

\bibitem[{{Abbott} {et~al.}(2021){Abbott}, {Abbott}, {Abraham}, {Acernese},
  {Ackley}, {Adams}, {Adams}, {Adhikari}, {Adya}, {Affeldt}, {Agathos},
  {Agatsuma}, {Aggarwal}, {Aguiar}, {Aiello}, {Ain}, {Ajith}, {Akcay}, {Allen},
  {Allocca}, {Altin}, {Amato}, {Anand}, {Ananyeva}, {Anderson}, {Anderson},
  {Angelova}, {Ansoldi}, {Antelis}, {Antier}, {Appert}, {Arai}, {Araya},
  {Areeda}, {Ar{\`e}ne}, {Arnaud}, {Aronson}, {Arun}, {Asali}, {Ascenzi},
  {Ashton}, {Aston}, {Astone}, {Aubin}, {Aufmuth}, {AultONeal}, {Austin},
  {Avendano}, {Babak}, {Badaracco}, {Bader}, {Bae}, {Baer}, {Bagnasco},
  {Baird}, {Ball}, {Ballardin}, {Ballmer}, {Bals}, {Balsamo}, {Baltus},
  {Banagiri}, {Bankar}, {Bankar}, {Barayoga}, {Barbieri}, {Barish}, {Barker},
  {Barneo}, {Barnum}, {Barone}, {Barr}, {Barsotti}, {Barsuglia}, {Barta},
  {Bartlett}, {Bartos}, {Bassiri}, {Basti}, {Bawaj}, {Bayley}, {Bazzan},
  {Becher}, {B{\'e}csy}, {Bedakihale}, {Bejger}, {Belahcene}, {Beniwal},
  {Benjamin}, {Bennett}, {Bentley}, {Bergamin}, {Berger}, {Bergmann},
  {Bernuzzi}, {Berry}, {Bersanetti}, {Bertolini}, {Betzwieser}, {Bhandare},
  {Bhandari}, {Bhattacharjee}, {Bidler}, {Bilenko}, {Billingsley}, {Birney},
  {Birnholtz}, {Biscans}, {Bischi}, {Biscoveanu}, {Bisht}, {Bitossi},
  {Bizouard}, {Blackburn}, {Blackman}, {Blair}, {Blair}, {Blair}, {Blanch},
  {Bobba}, {Bode}, {Boer}, {Boetzel}, {Bogaert}, {Boldrini}, {Bondu},
  {Bonilla}, {Bonnand}, {Booker}, {Boom}, {Bork}, {Boschi}, {Bose},
  {Bossilkov}, {Boudart}, {Bouffanais}, {Bozzi}, {Bradaschia}, {Brady},
  {Bramley}, {Branchesi}, {Brau}, {Breschi}, {Briant}, {Briggs}, {Brighenti},
  {Brillet}, {Brinkmann}, {Brockill}, {Brooks}, {Brooks}, {Brown}, {Brunett},
  {Bruno}, {Bruntz}, {Buikema}, {Bulik}, {Bulten}, {Buonanno}, {Buscicchio},
  {Buskulic}, {Byer}, {Cabero}, {Cadonati}, {Caesar}, {Cagnoli}, {Cahillane},
  {Calder{\'o}n Bustillo}, {Callaghan}, {Callister}, {Calloni}, {Camp},
  {Canepa}, {Cannon}, {Cao}, {Cao}, {Carapella}, {Carbognani}, {Carney},
  {Carpinelli}, {Carullo}, {Carver}, {Casanueva Diaz}, {Casentini}, {Caudill},
  {Cavagli{\`a}}, {Cavalier}, {Cavalieri}, {Cella}, {Cerd{\'a}-Dur{\'a}n},
  {Cesarini}, {Chaibi}, {Chakravarti}, {Chan}, {Chan}, {Chandra}, {Chanial},
  {Chao}, {Charlton}, {Chase}, {Chassande-Mottin}, {Chatterjee},
  {Chattopadhyay}, {Chaturvedi}, {Chatziioannou}, {Chen}, {Chen}, {Chen},
  {Chen}, {Cheng}, {Cheong}, {Chia}, {Chiadini}, {Chierici}, {Chincarini},
  {Chiummo}, {Cho}, {Cho}, {Cho}, {Choate}, {Christensen}, {Chu}, {Chua},
  {Chung}, {Chung}, {Ciani}, {Ciecielag}, {Cie{\'s}lar}, {Cifaldi}, {Ciobanu},
  {Ciolfi}, {Cipriano}, {Cirone}, {Clara}, {Clark}, {Clark}, {Clarke},
  {Clearwater}, {Clesse}, {Cleva}, {Coccia}, {Cohadon}, {Cohen}, {Colleoni},
  {Collette}, {Collins}, {Colpi}, {Constancio}, {Conti}, {Cooper}, {Corban},
  {Corbitt}, {Cordero-Carri{\'o}n}, {Corezzi}, {Corley}, {Cornish}, {Corre},
  {Corsi}, {Cortese}, {Costa}, {Cotesta}, {Coughlin}, {Coughlin}, {Coulon},
  {Countryman}, {Cousins}, {Couvares}, {Covas}, {Coward}, {Cowart}, {Coyne},
  {Coyne}, {Creighton}, {Creighton}, {Croquette}, {Crowder}, {Cudell},
  {Cullen}, {Cumming}, {Cummings}, {Cunningham}, {Cuoco}, {Cury{\l}o},
  {Canton}, {D{\'a}lya}, {Dana}, {DaneshgaranBajastani}, {D'Angelo}, {Danila},
  {Danilishin}, {D'Antonio}, {Danzmann}, {Darsow-Fromm}, {Dasgupta}, {Datrier},
  {Dattilo}, {Dave}, {Davier}, {Davies}, {Davis}, {Daw}, {Dean}, {DeBra},
  {Deenadayalan}, {Degallaix}, {De Laurentis}, {Del{\'e}glise}, {Del Favero},
  {De Lillo}, {De Lillo}, {Del Pozzo}, {DeMarchi}, {De Matteis}, {D'Emilio},
  {Demos}, {Denker}, {Dent}, {Depasse}, {De Pietri}, {De Rosa}, {De Rossi},
  {DeSalvo}, {de Varona}, {Dhurandhar}, {D{\'\i}az}, {Diaz-Ortiz}, {Didio},
  {Dietrich}, {Di Fiore}, {DiFronzo}, {Di Giorgio}, {Di Giovanni}, {Di
  Giovanni}, {Di Girolamo}, {Di Lieto}, {Ding}, {Di Pace}, {Di Palma}, {Di
  Renzo}, {Divakarla}, {Dmitriev}, {Doctor}, {D'Onofrio}, {Donovan}, {Dooley},
  {Doravari}, {Dorrington}, {Downes}, {Drago}, {Driggers}, {Du}, {Ducoin},
  {Dupej}, {Durante}, {D'Urso}, {Duverne}, {Dwyer}, {Easter}, {Eddolls},
  {Edelman}, {Edo}, {Edy}, {Effler}, {Eichholz}, {Eikenberry}, {Eisenmann},
  {Eisenstein}, {Ejlli}, {Errico}, {Essick}, {Estell{\'e}s}, {Estevez},
  {Etienne}, {Etzel}, {Evans}, {Evans}, {Ewing}, {Fafone}, {Fair}, {Fairhurst},
  {Fan}, {Farah}, {Farinon}, {Farr}, {Farr}, {Fauchon-Jones}, {Favata}, {Fays},
  {Fazio}, {Feicht}, {Fejer}, {Feng}, {Fenyvesi}, {Ferguson},
  {Fernandez-Galiana}, {Ferrante}, {Ferreira}, {Fidecaro}, {Figura}, {Fiori},
  {Fiorucci}, {Fishbach}, {Fisher}, {Fishner}, {Fittipaldi}, {Fitz-Axen},
  {Fiumara}, {Flaminio}, {Floden}, {Flynn}, {Fong}, {Font}, {Forsyth},
  {Fournier}, {Frasca}, {Frasconi}, {Frei}, {Freise}, {Frey}, {Frey},
  {Fritschel}, {Frolov}, {Fronz{\'e}}, {Fulda}, {Fyffe}, {Gabbard}, {Gadre},
  {Gaebel}, {Gair}, {Gais}, {Galaudage}, {Gamba}, {Ganapathy}, {Ganguly},
  {Gaonkar}, {Garaventa}, {Garc{\'\i}a-Quir{\'o}s}, {Garufi}, {Gateley},
  {Gaudio}, {Gayathri}, {Gemme}, {Gennai}, {George}, {George}, {George},
  {Gergely}, {Ghonge}, {Ghosh}, {Ghosh}, {Ghosh}, {Giacomazzo}, {Giacoppo},
  {Giaime}, {Giardina}, {Gibson}, {Gier}, {Gill}, {Giri}, {Glanzer}, {Gleckl},
  {Godwin}, {Goetz}, {Goetz}, {Gohlke}, {Goncharov}, {Gonz{\'a}lez},
  {Gopakumar}, {Gossan}, {Gosselin}, {Gouaty}, {Grace}, {Grado}, {Granata},
  {Granata}, {Grant}, {Gras}, {Grassia}, {Gray}, {Gray}, {Greco}, {Green},
  {Green}, {Gretarsson}, {Griggs}, {Grignani}, {Grimaldi}, {Grimes}, {Grimm},
  {Grote}, {Grunewald}, {Gruning}, {Guerrero}, {Guidi}, {Guimaraes},
  {Guix{\'e}}, {Gulati}, {Guo}, {Gupta}, {Gupta}, {Gupta}, {Gustafson},
  {Gustafson}, {Guzman}, {Haegel}, {Halim}, {Hall}, {Hamilton}, {Hammond},
  {Haney}, {Hanke}, {Hanks}, {Hanna}, {Hannam}, {Hannuksela}, {Hannuksela},
  {Hansen}, {Hansen}, {Hanson}, {Harder}, {Hardwick}, {Haris}, {Harms},
  {Harry}, {Harry}, {Hartwig}, {Hasskew}, {Haster}, {Haughian}, {Hayes},
  {Healy}, {Heidmann}, {Heintze}, {Heinze}, {Heinzel}, {Heitmann}, {Hellman},
  {Hello}, {Helmling-Cornell}, {Hemming}, {Hendry}, {Heng}, {Hennes}, {Hennig},
  {Hennig}, {Hernandez Vivanco}, {Heurs}, {Hild}, {Hill}, {Hines}, {Hochheim},
  {Hofgard}, {Hofman}, {Hohmann}, {Holgado}, {Holland}, {Hollows}, {Holmes},
  {Holt}, {Holz}, {Hopkins}, {Horst}, {Hough}, {Howell}, {Hoy}, {Hoyland},
  {Huang}, {H{\"u}bner}, {Huddart}, {Huerta}, {Hughey}, {Hui}, {Husa},
  {Huttner}, {Hutzler}, {Huxford}, {Huynh-Dinh}, {Idzkowski}, {Iess},
  {Imperato}, {Inchauspe}, {Ingram}, {Intini}, {Isi}, {Iyer},
  {JaberianHamedan}, {Jacqmin}, {Jadhav}, {Jadhav}, {James}, {Jani},
  {Janssens}, {Janthalur}, {Jaranowski}, {Jariwala}, {Jaume}, {Jenkins},
  {Jeunon}, {Jiang}, {Johns}, {Johnson-McDaniel}, {Jones}, {Jones}, {Jones},
  {Jones}, {Jones}, {Jonker}, {Ju}, {Junker}, {Kalaghatgi}, {Kalogera},
  {Kamai}, {Kandhasamy}, {Kang}, {Kanner}, {Kapadia}, {Kapasi}, {Karathanasis},
  {Karki}, {Kashyap}, {Kasprzack}, {Kastaun}, {Katsanevas}, {Katsavounidis},
  {Katzman}, {Kawabe}, {K{\'e}f{\'e}lian}, {Keitel}, {Key}, {Khadka},
  {Khalili}, {Khan}, {Khan}, {Khazanov}, {Khetan}, {Khursheed}, {Kijbunchoo},
  {Kim}, {Kim}, {Kim}, {Kim}, {Kim}, {Kim}, {Kimball}, {King}, {Kinley-Hanlon},
  {Kirchhoff}, {Kissel}, {Kleybolte}, {Klimenko}, {Knowles}, {Knyazev}, {Koch},
  {Koehlenbeck}, {Koekoek}, {Koley}, {Kolstein}, {Komori}, {Kondrashov},
  {Kontos}, {Koper}, {Korobko}, {Korth}, {Kovalam}, {Kozak}, {Kr{\"a}mer},
  {Kringel}, {Krishnendu}, {Kr{\'o}lak}, {Kuehn}, {Kumar}, {Kumar}, {Kumar},
  {Kumar}, {Kuns}, {Kwang}, {Lackey}, {Laghi}, {Lalande}, {Lam}, {Lamberts},
  {Landry}, {Lane}, {Lang}, {Lange}, {Lantz}, {Lanza}, {La Rosa},
  {Lartaux-Vollard}, {Lasky}, {Laxen}, {Lazzarini}, {Lazzaro}, {Leaci},
  {Leavey}, {Lecoeuche}, {Lee}, {Lee}, {Lee}, {Lee}, {Lehmann}, {Leon},
  {Leroy}, {Letendre}, {Levin}, {Li}, {Li}, {Li}, {Li}, {Li}, {Linde},
  {Linker}, {Linley}, {Littenberg}, {Liu}, {Liu}, {Llorens-Monteagudo}, {Lo},
  {Lockwood}, {London}, {Longo}, {Lorenzini}, {Loriette}, {Lormand}, {Losurdo},
  {Lough}, {Lousto}, {Lovelace}, {L{\"u}ck}, {Lumaca}, {Lundgren}, {Ma},
  {Macas}, {MacInnis}, {Macleod}, {MacMillan}, {Macquet}, {Maga{\~n}a
  Hernandez}, {Maga{\~n}a-Sandoval}, {Magazz{\`u}}, {Magee}, {Majorana},
  {Maksimovic}, {Maliakal}, {Malik}, {Man}, {Mandic}, {Mangano}, {Mansell},
  {Manske}, {Mantovani}, {Mapelli}, {Marchesoni}, {Marion}, {M{\'a}rka},
  {M{\'a}rka}, {Markakis}, {Markosyan}, {Markowitz}, {Maros}, {Marquina},
  {Marsat}, {Martelli}, {Martin}, {Martin}, {Martinez}, {Martinez}, {Martynov},
  {Masalehdan}, {Mason}, {Massera}, {Masserot}, {Massinger}, {Masso-Reid},
  {Mastrogiovanni}, {Matas}, {Mateu-Lucena}, {Matichard}, {Matiushechkina},
  {Mavalvala}, {Maynard}, {McCann}, {McCarthy}, {McClelland}, {McCormick},
  {McCuller}, {McGuire}, {McIsaac}, {McIver}, {McManus}, {McRae}, {McWilliams},
  {Meacher}, {Meadors}, {Mehmet}, {Mehta}, {Melatos}, {Melchor}, {Mendell},
  {Menendez-Vazquez}, {Mercer}, {Mereni}, {Merfeld}, {Merilh}, {Merritt},
  {Merzougui}, {Meshkov}, {Messenger}, {Messick}, {Metzdorff}, {Meyers},
  {Meylahn}, {Mhaske}, {Miani}, {Miao}, {Michaloliakos}, {Michel}, {Middleton},
  {Milano}, {Miller}, {Millhouse}, {Mills}, {Milotti}, {Milovich-Goff},
  {Minazzoli}, {Minenkov}, {Mir}, {Mishkin}, {Mishra}, {Mistry}, {Mitra},
  {Mitrofanov}, {Mitselmakher}, {Mittleman}, {Mo}, {Mogushi}, {Mohapatra},
  {Mohite}, {Molina}, {Molina-Ruiz}, {Mondin}, {Montani}, {Moore}, {Moraru},
  {Morawski}, {Moreno}, {Morisaki}, {Mours}, {Mow-Lowry}, {Mozzon},
  {Muciaccia}, {Mukherjee}, {Mukherjee}, {Mukherjee}, {Mukherjee}, {Mukund},
  {Mullavey}, {Munch}, {Mu{\~n}iz}, {Murray}, {Nadji}, {Nagar}, {Nardecchia},
  {Naticchioni}, {Nayak}, {Neil}, {Neilson}, {Nelemans}, {Nelson}, {Nery},
  {Neunzert}, {Nitz}, {Ng}, {Ng}, {Nguyen}, {Nguyen}, {Nguyen}, {Nichols},
  {Nissanke}, {Nocera}, {Noh}, {North}, {Nothard}, {Nuttall}, {Oberling},
  {O'Brien}, {O'Dell}, {Oganesyan}, {Ogin}, {Oh}, {Oh}, {Ohme}, {Ohta},
  {Okada}, {Olivetto}, {Oppermann}, {Oram}, {O'Reilly}, {Ormiston}, {Ortega},
  {O'Shaughnessy}, {Ossokine}, {Osthelder}, {Ottaway}, {Overmier}, {Owen},
  {Pace}, {Pagano}, {Page}, {Pagliaroli}, {Pai}, {Pai}, {Palamos}, {Palashov},
  {Palomba}, {Pan}, {Panda}, {Pang}, {Pankow}, {Pannarale}, {Pant}, {Paoletti},
  {Paoli}, {Paolone}, {Parker}, {Pascucci}, {Pasqualetti}, {Passaquieti},
  {Passuello}, {Patel}, {Patricelli}, {Payne}, {Pechsiri}, {Pedraza},
  {Pegoraro}, {Pele}, {Penn}, {Perego}, {Perez}, {P{\'e}rigois}, {Perreca},
  {Perri{\`e}s}, {Petermann}, {Petterson}, {Pfeiffer}, {Pham}, {Phukon},
  {Piccinni}, {Pichot}, {Piendibene}, {Piergiovanni}, {Pierini}, {Pierro},
  {Pillant}, {Pilo}, {Pinard}, {Pinto}, {Piotrzkowski}, {Pirello}, {Pitkin},
  {Placidi}, {Plastino}, {Pluchar}, {Poggiani}, {Polini}, {Pong}, {Ponrathnam},
  {Popolizio}, {Porter}, {Poverman}, {Powell}, {Pracchia}, {Prajapati},
  {Prasai}, {Prasanna}, {Pratten}, {Prestegard}, {Principe}, {Prodi},
  {Prokhorov}, {Prosposito}, {Prudenzi}, {Puecher}, {Punturo}, {Puosi},
  {Puppo}, {P{\"u}rrer}, {Qi}, {Quetschke}, {Quinonez}, {Quitzow-James},
  {Raab}, {Raaijmakers}, {Radkins}, {Radulesco}, {Raffai}, {Rafferty}, {Rail},
  {Raja}, {Rajan}, {Rajbhandari}, {Rakhmanov}, {Ramirez}, {Ramirez},
  {Ramos-Buades}, {Rana}, {Rao}, {Rapagnani}, {Rapol}, {Ratto}, {Raymond},
  {Razzano}, {Read}, {Regimbau}, {Rei}, {Reid}, {Reitze}, {Rettegno}, {Ricci},
  {Richardson}, {Richardson}, {Richardson}, {Ricker}, {Riemenschneider},
  {Riles}, {Rizzo}, {Robertson}, {Robinet}, {Rocchi}, {Rocha}, {Rodriguez},
  {Rodriguez-Soto}, {Rolland}, {Rollins}, {Roma}, {Romanelli}, {Romano},
  {Romel}, {Romero}, {Romero-Shaw}, {Romie}, {Ronchini}, {Rose}, {Rose},
  {Rose}, {Rosell}, {Rosi{\'n}ska}, {Rosofsky}, {Ross}, {Rowan}, {Rowlinson},
  {Roy}, {Roy}, {Ruggi}, {Ryan}, {Sachdev}, {Sadecki}, {Sadiq},
  {Sakellariadou}, {Salafia}, {Salconi}, {Saleem}, {Samajdar}, {Sanchez},
  {Sanchez}, {Sanchez}, {Sanchis-Gual}, {Sanders}, {Sandles}, {Santiago},
  {Santos}, {Saravanan}, {Sarin}, {Sassolas}, {Sathyaprakash}, {Sauter},
  {Savage}, {Savant}, {Sawant}, {Sayah}, {Schaetzl}, {Schale}, {Scheel},
  {Scheuer}, {Schindler-Tyka}, {Schmidt}, {Schnabel}, {Schofield},
  {Sch{\"o}nbeck}, {Schreiber}, {Schulte}, {Schutz}, {Schwarm}, {Schwartz},
  {Scott}, {Scott}, {Seglar-Arroyo}, {Seidel}, {Sellers}, {Sengupta},
  {Sennett}, {Sentenac}, {Sequino}, {Sergeev}, {Setyawati}, {Shaffer},
  {Shahriar}, {Sharifi}, {Sharma}, {Sharma}, {Shawhan}, {Shen}, {Shikauchi},
  {Shink}, {Shoemaker}, {Shoemaker}, {Shukla}, {ShyamSundar}, {Sieniawska},
  {Sigg}, {Singer}, {Singh}, {Singh}, {Singha}, {Singhal}, {Sintes}, {Sipala},
  {Skliris}, {Slagmolen}, {Slaven-Blair}, {Smetana}, {Smith}, {Smith},
  {Somala}, {Son}, {Soni}, {Soni}, {Sorazu}, {Sordini}, {Sorrentino},
  {Sorrentino}, {Soulard}, {Souradeep}, {Sowell}, {Spencer}, {Spera},
  {Srivastava}, {Srivastava}, {Staats}, {Stachie}, {Steer}, {Steinhoff},
  {Steinke}, {Steinlechner}, {Steinlechner}, {Steinmeyer}, {Stevenson},
  {Stolle-McAllister}, {Stops}, {Stover}, {Strain}, {Stratta}, {Strunk},
  {Sturani}, {Stuver}, {S{\"u}dbeck}, {Sudhagar}, {Sudhir}, {Suh},
  {Summerscales}, {Sun}, {Sun}, {Sunil}, {Sur}, {Suresh}, {Sutton}, {Swinkels},
  {Szczepa{\'n}czyk}, {Tacca}, {Tait}, {Talbot}, {Tanasijczuk}, {Tanner},
  {Tao}, {Tapia}, {Tapia San Martin}, {Tasson}, {Taylor}, {Tenorio},
  {Terkowski}, {Thirugnanasambandam}, {Thomas}, {Thomas}, {Thomas}, {Thompson},
  {Thondapu}, {Thorne}, {Thrane}, {Tiwari}, {Tiwari}, {Tiwari}, {Toland},
  {Tolley}, {Tonelli}, {Tornasi}, {Torres-Forn{\'e}}, {Torrie}, {e Melo},
  {T{\"o}yr{\"a}}, {Tran}, {Trapananti}, {Travasso}, {Traylor}, {Tringali},
  {Tripathee}, {Trovato}, {Trudeau}, {Tsai}, {Tsang}, {Tse}, {Tso}, {Tsukada},
  {Tsuna}, {Tsutsui}, {Turconi}, {Ubhi}, {Udall}, {Ueno}, {Ugolini},
  {Unnikrishnan}, {Urban}, {Usman}, {Utina}, {Vahlbruch}, {Vajente}, {Vajpeyi},
  {Valdes}, {Valentini}, {Valsan}, {van Bakel}, {van Beuzekom}, {van den
  Brand}, {Van Den Broeck}, {Vander-Hyde}, {van der Schaaf}, {van Heijningen},
  {Vardaro}, {Vargas}, {Varma}, {Vass}, {Vas{\'u}th}, {Vecchio}, {Vedovato},
  {Veitch}, {Veitch}, {Venkateswara}, {Venneberg}, {Venugopalan}, {Verkindt},
  {Verma}, {Veske}, {Vetrano}, {Vicer{\'e}}, {Viets}, {Vijaykumar},
  {Villa-Ortega}, {Vinet}, {Vitale}, {Vo}, {Vocca}, {Vorvick}, {Vyatchanin},
  {Wade}, {Wade}, {Wade}, {Walet}, {Walker}, {Wallace}, {Wallace}, {Walsh},
  {Wang}, {Wang}, {Wang}, {Wang}, {Ward}, {Warner}, {Was}, {Washington},
  {Watchi}, {Weaver}, {Wei}, {Weinert}, {Weinstein}, {Weiss}, {Wellmann},
  {Wen}, {We{\ss}els}, {Westhouse}, {Wette}, {Whelan}, {White}, {White},
  {Whiting}, {Whittle}, {Wilken}, {Williams}, {Williams}, {Williamson},
  {Willis}, {Willke}, {Wilson}, {Wimmer}, {Winkler}, {Wipf}, {Woan}, {Woehler},
  {Wofford}, {Wong}, {Wrangel}, {Wright}, {Wu}, {Wysocki}, {Xiao}, {Yamamoto},
  {Yang}, {Yang}, {Yang}, {Yap}, {Yeeles}, {Yoon}, {Yu}, {Yu}, {Yuen},
  {Zadro{\.Z}ny}, {Zanolin}, {Zelenova}, {Zendri}, {Zevin}, {Zhang}, {Zhang},
  {Zhang}, {Zhang}, {Zhao}, {Zhao}, {Zheng}, {Zhou}, {Zhou}, {Zhu},
  {Zimmerman}, {Zlochower}, {Zucker}, {Zweizig}, {LIGO Scientific
  Collaboration}, \& {Virgo Collaboration}}]{GWTC-2}
{Abbott}, R., {Abbott}, T.~D., {Abraham}, S., {et~al.} 2021, Physical Review X,
  11, 021053, \dodoi{10.1103/PhysRevX.11.021053}

\bibitem[{{Adams} {et~al.}(2017{\natexlab{a}}){Adams}, {Kochanek}, {Gerke}, \&
  {Stanek}}]{adams17a}
{Adams}, S.~M., {Kochanek}, C.~S., {Gerke}, J.~R., \& {Stanek}, K.~Z.
  2017{\natexlab{a}}, \mnras, 469, 1445, \dodoi{10.1093/mnras/stx898}

\bibitem[{{Adams} {et~al.}(2017{\natexlab{b}}){Adams}, {Kochanek}, {Gerke},
  {Stanek}, \& {Dai}}]{adams17b}
{Adams}, S.~M., {Kochanek}, C.~S., {Gerke}, J.~R., {Stanek}, K.~Z., \& {Dai},
  X. 2017{\natexlab{b}}, \mnras, 468, 4968, \dodoi{10.1093/mnras/stx816}

\bibitem[{{Asplund} {et~al.}(2009){Asplund}, {Grevesse}, {Sauval}, \&
  {Scott}}]{asplund09}
{Asplund}, M., {Grevesse}, N., {Sauval}, A.~J., \& {Scott}, P. 2009, \araa, 47,
  481, \dodoi{10.1146/annurev.astro.46.060407.145222}

\bibitem[{{Astropy Collaboration} {et~al.}(2013){Astropy Collaboration},
  {Robitaille}, {Tollerud}, {Greenfield}, {Droettboom}, {Bray}, {Aldcroft},
  {Davis}, {Ginsburg}, {Price-Whelan}, {Kerzendorf}, {Conley}, {Crighton},
  {Barbary}, {Muna}, {Ferguson}, {Grollier}, {Parikh}, {Nair}, {Unther},
  {Deil}, {Woillez}, {Conseil}, {Kramer}, {Turner}, {Singer}, {Fox}, {Weaver},
  {Zabalza}, {Edwards}, {Azalee Bostroem}, {Burke}, {Casey}, {Crawford},
  {Dencheva}, {Ely}, {Jenness}, {Labrie}, {Lim}, {Pierfederici}, {Pontzen},
  {Ptak}, {Refsdal}, {Servillat}, \& {Streicher}}]{2013A&A...558A..33A}
{Astropy Collaboration}, {Robitaille}, T.~P., {Tollerud}, E.~J., {et~al.} 2013,
  \aap, 558, A33, \dodoi{10.1051/0004-6361/201322068}

\bibitem[{{Ayres} {et~al.}(2013){Ayres}, {Lyons}, {Ludwig}, {Caffau}, \&
  {Wedemeyer-B{\"o}hm}}]{ayres13}
{Ayres}, T.~R., {Lyons}, J.~R., {Ludwig}, H.~G., {Caffau}, E., \&
  {Wedemeyer-B{\"o}hm}, S. 2013, \apj, 765, 46,
  \dodoi{10.1088/0004-637X/765/1/46}

\bibitem[{{Barbon} {et~al.}(1999){Barbon}, {Buond{\'\i}}, {Cappellaro}, \&
  {Turatto}}]{asiago}
{Barbon}, R., {Buond{\'\i}}, V., {Cappellaro}, E., \& {Turatto}, M. 1999,
  \aaps, 139, 531, \dodoi{10.1051/aas:1999404}

\bibitem[{{Basinger} {et~al.}(2021){Basinger}, {Kochanek}, {Adams}, {Dai}, \&
  {Stanek}}]{basinger21}
{Basinger}, C.~M., {Kochanek}, C.~S., {Adams}, S.~M., {Dai}, X., \& {Stanek},
  K.~Z. 2021, \mnras, 508, 1156, \dodoi{10.1093/mnras/stab2620}

\bibitem[{{Beasor} {et~al.}(2020){Beasor}, {Davies}, {Smith}, {van Loon},
  {Gehrz}, \& {Figer}}]{beasor20}
{Beasor}, E.~R., {Davies}, B., {Smith}, N., {et~al.} 2020, \mnras, 492, 5994,
  \dodoi{10.1093/mnras/staa255}

\bibitem[{{Bertin}(2011)}]{bertin11}
{Bertin}, E. 2011, in Astronomical Society of the Pacific Conference Series,
  Vol. 442, Astronomical Data Analysis Software and Systems XX, ed. I.~N.
  {Evans}, A.~{Accomazzi}, D.~J. {Mink}, \& A.~H. {Rots}, 435

\bibitem[{{Bertin} \& {Arnouts}(1996)}]{bertin96}
{Bertin}, E., \& {Arnouts}, S. 1996, \aaps, 117, 393,
  \dodoi{10.1051/aas:1996164}

\bibitem[{{Blagorodnova} {et~al.}(2017){Blagorodnova}, {Kotak}, {Polshaw},
  {Kasliwal}, {Cao}, {Cody}, {Doran}, {Elias-Rosa}, {Fraser}, {Fremling},
  {Gonzalez-Fernandez}, {Harmanen}, {Jencson}, {Kankare}, {Kudritzki},
  {Kulkarni}, {Magnier}, {Manulis}, {Masci}, {Mattila}, {Nugent}, {Ochner},
  {Pastorello}, {Reynolds}, {Smith}, {Sollerman}, {Taddia}, {Terreran},
  {Tomasella}, {Turatto}, {Vreeswijk}, {Wozniak}, \& {Zaggia}}]{blagorodnova17}
{Blagorodnova}, N., {Kotak}, R., {Polshaw}, J., {et~al.} 2017, \apj, 834, 107,
  \dodoi{10.3847/1538-4357/834/2/107}

\bibitem[{{Bohlin}(2016)}]{bohlin16}
{Bohlin}, R.~C. 2016, \aj, 152, 60, \dodoi{10.3847/0004-6256/152/3/60}

\bibitem[{{Boyer} {et~al.}(2011){Boyer}, {Srinivasan}, {van Loon}, {McDonald},
  {Meixner}, {Zaritsky}, {Gordon}, {Kemper}, {Babler}, {Block}, {Bracker},
  {Engelbracht}, {Hora}, {Indebetouw}, {Meade}, {Misselt}, {Robitaille},
  {Sewi{\l}o}, {Shiao}, \& {Whitney}}]{boyer11}
{Boyer}, M.~L., {Srinivasan}, S., {van Loon}, J.~T., {et~al.} 2011, \aj, 142,
  103, \dodoi{10.1088/0004-6256/142/4/103}

\bibitem[{Bradley {et~al.}(2020)Bradley, Sipőcz, Robitaille, Tollerud,
  Vinícius, Deil, Barbary, Wilson, Busko, Günther, Cara, Conseil, Bostroem,
  Droettboom, Bray, Bratholm, Lim, Barentsen, Craig, Pascual, Perren, Greco,
  Donath, de~Val-Borro, Kerzendorf, Bach, Weaver, D'Eugenio, Souchereau, \&
  Ferreira}]{bradley20_photutils}
Bradley, L., Sipőcz, B., Robitaille, T., {et~al.} 2020, astropy/photutils:
  1.0.0, 1.0.0,  Zenodo, \dodoi{10.5281/zenodo.4044744}

\bibitem[{{Choi} {et~al.}(2017){Choi}, {Conroy}, \& {Byler}}]{choi17}
{Choi}, J., {Conroy}, C., \& {Byler}, N. 2017, \apj, 838, 159,
  \dodoi{10.3847/1538-4357/aa679f}

\bibitem[{{Choi} {et~al.}(2016){Choi}, {Dotter}, {Conroy}, {Cantiello},
  {Paxton}, \& {Johnson}}]{choi16}
{Choi}, J., {Dotter}, A., {Conroy}, C., {et~al.} 2016, \apj, 823, 102,
  \dodoi{10.3847/0004-637X/823/2/102}

\bibitem[{{Cioni} {et~al.}(2006){Cioni}, {Girardi}, {Marigo}, \&
  {Habing}}]{cioni06}
{Cioni}, M. R.~L., {Girardi}, L., {Marigo}, P., \& {Habing}, H.~J. 2006, \aap,
  448, 77, \dodoi{10.1051/0004-6361:20053933}

\bibitem[{{Conroy} {et~al.}(2018){Conroy}, {Strader}, {van Dokkum}, {Dolphin},
  {Weisz}, {Murphy}, {Dotter}, {Johnson}, \& {Cargile}}]{conroy18}
{Conroy}, C., {Strader}, J., {van Dokkum}, P., {et~al.} 2018, \apj, 864, 111,
  \dodoi{10.3847/1538-4357/aad460}

\bibitem[{{Croxall} {et~al.}(2015){Croxall}, {Pogge}, {Berg}, {Skillman}, \&
  {Moustakas}}]{croxall15}
{Croxall}, K.~V., {Pogge}, R.~W., {Berg}, D.~A., {Skillman}, E.~D., \&
  {Moustakas}, J. 2015, \apj, 808, 42, \dodoi{10.1088/0004-637X/808/1/42}

\bibitem[{{Dalcanton} {et~al.}(2012){Dalcanton}, {Williams}, {Lang}, {Lauer},
  {Kalirai}, {Seth}, {Dolphin}, {Rosenfield}, {Weisz}, {Bell}, {Bianchi},
  {Boyer}, {Caldwell}, {Dong}, {Dorman}, {Gilbert}, {Girardi}, {Gogarten},
  {Gordon}, {Guhathakurta}, {Hodge}, {Holtzman}, {Johnson}, {Larsen}, {Lewis},
  {Melbourne}, {Olsen}, {Rix}, {Rosema}, {Saha}, {Sarajedini}, {Skillman}, \&
  {Stanek}}]{dalcanton12}
{Dalcanton}, J.~J., {Williams}, B.~F., {Lang}, D., {et~al.} 2012, \apjs, 200,
  18, \dodoi{10.1088/0067-0049/200/2/18}

\bibitem[{{Davies} \& {Beasor}(2018)}]{davies18a}
{Davies}, B., \& {Beasor}, E.~R. 2018, \mnras, 474, 2116,
  \dodoi{10.1093/mnras/stx2734}

\bibitem[{{Davies} \& {Beasor}(2020{\natexlab{a}})}]{davies20a}
---. 2020{\natexlab{a}}, \mnras, 493, 468, \dodoi{10.1093/mnras/staa174}

\bibitem[{{Davies} \& {Beasor}(2020{\natexlab{b}})}]{davies20b}
---. 2020{\natexlab{b}}, \mnras, 496, L142, \dodoi{10.1093/mnrasl/slaa102}

\bibitem[{{Davies} {et~al.}(2018){Davies}, {Crowther}, \& {Beasor}}]{davies18b}
{Davies}, B., {Crowther}, P.~A., \& {Beasor}, E.~R. 2018, \mnras, 478, 3138,
  \dodoi{10.1093/mnras/sty1302}

\bibitem[{{Davies} {et~al.}(2007){Davies}, {Figer}, {Kudritzki}, {MacKenty},
  {Najarro}, \& {Herrero}}]{davies07}
{Davies}, B., {Figer}, D.~F., {Kudritzki}, R.-P., {et~al.} 2007, \apj, 671,
  781, \dodoi{10.1086/522224}

\bibitem[{{Davies} \& {Plez}(2021)}]{davies21}
{Davies}, B., \& {Plez}, B. 2021, \mnras, 508, 5757,
  \dodoi{10.1093/mnras/stab2645}

\bibitem[{{Doherty} {et~al.}(2017){Doherty}, {Gil-Pons}, {Siess}, \&
  {Lattanzio}}]{doherty17}
{Doherty}, C.~L., {Gil-Pons}, P., {Siess}, L., \& {Lattanzio}, J.~C. 2017,
  \pasa, 34, e056, \dodoi{10.1017/pasa.2017.52}

\bibitem[{{Dolan} {et~al.}(2016){Dolan}, {Mathews}, {Lam}, {Quynh Lan},
  {Herczeg}, \& {Dearborn}}]{dolan16}
{Dolan}, M.~M., {Mathews}, G.~J., {Lam}, D.~D., {et~al.} 2016, \apj, 819, 7,
  \dodoi{10.3847/0004-637X/819/1/7}

\bibitem[{{Dolphin}(2016)}]{dolphin16}
{Dolphin}, A. 2016, {DOLPHOT: Stellar photometry}, Astrophysics Source Code
  Library.
\newblock \doeprint{1608.013}

\bibitem[{{Dolphin}(2000)}]{dolphin00}
{Dolphin}, A.~E. 2000, \pasp, 112, 1383, \dodoi{10.1086/316630}

\bibitem[{{Dupree} {et~al.}(2020){Dupree}, {Strassmeier}, {Matthews},
  {Uitenbroek}, {Calderwood}, {Granzer}, {Guinan}, {Leike}, {Montarg{\`e}s},
  {Richards}, {Wasatonic}, \& {Weber}}]{dupree20}
{Dupree}, A.~K., {Strassmeier}, K.~G., {Matthews}, L.~D., {et~al.} 2020, \apj,
  899, 68, \dodoi{10.3847/1538-4357/aba516}

\bibitem[{{Elias} {et~al.}(1985){Elias}, {Frogel}, \& {Humphreys}}]{elias85}
{Elias}, J.~H., {Frogel}, J.~A., \& {Humphreys}, R.~M. 1985, \apjs, 57, 91,
  \dodoi{10.1086/190997}

\bibitem[{{Ertl} {et~al.}(2016){Ertl}, {Janka}, {Woosley}, {Sukhbold}, \&
  {Ugliano}}]{ertl16}
{Ertl}, T., {Janka}, H.~T., {Woosley}, S.~E., {Sukhbold}, T., \& {Ugliano}, M.
  2016, \apj, 818, 124, \dodoi{10.3847/0004-637X/818/2/124}

\bibitem[{{Fitzpatrick}(1999)}]{fitzpatrick99}
{Fitzpatrick}, E.~L. 1999, \pasp, 111, 63, \dodoi{10.1086/316293}

\bibitem[{{Gerke} {et~al.}(2015){Gerke}, {Kochanek}, \& {Stanek}}]{gerke15}
{Gerke}, J.~R., {Kochanek}, C.~S., \& {Stanek}, K.~Z. 2015, \mnras, 450, 3289,
  \dodoi{10.1093/mnras/stv776}

\bibitem[{{Girardi} {et~al.}(2012){Girardi}, {Barbieri}, {Groenewegen},
  {Marigo}, {Bressan}, {Rocha-Pinto}, {Santiago}, {Camargo}, \& {da
  Costa}}]{girardi12}
{Girardi}, L., {Barbieri}, M., {Groenewegen}, M. A.~T., {et~al.} 2012, in
  Astrophysics and Space Science Proceedings, Vol.~26, Red Giants as Probes of
  the Structure and Evolution of the Milky Way, 165,
  \dodoi{10.1007/978-3-642-18418-5\_17}

\bibitem[{{Guinan} {et~al.}(2020){Guinan}, {Wasatonic}, {Calderwood}, \&
  {Carona}}]{guinan20}
{Guinan}, E., {Wasatonic}, R., {Calderwood}, T., \& {Carona}, D. 2020, The
  Astronomer's Telegram, 13512, 1

\bibitem[{{Guinan} {et~al.}(2019){Guinan}, {Wasatonic}, \&
  {Calderwood}}]{guinan19}
{Guinan}, E.~F., {Wasatonic}, R.~J., \& {Calderwood}, T.~J. 2019, The
  Astronomer's Telegram, 13341, 1

\bibitem[{{Hack} {et~al.}(2012){Hack}, {Dencheva}, {Fruchter}, {Armstrong},
  {Avila}, {Baggett}, {Bray}, {Droettboom}, {Dulude}, {Gonzaga}, {Grogin},
  {Kozhurina-Platais}, {Lucas}, {Mack}, {MacKenty}, {Petro}, {Pirzkal},
  {Rajan}, {Smith}, {Sontag}, \& {Ubeda}}]{hack12}
{Hack}, W.~J., {Dencheva}, N., {Fruchter}, A.~S., {et~al.} 2012, in American
  Astronomical Society Meeting Abstracts, Vol. 220, American Astronomical
  Society Meeting Abstracts \#220, 135.15

\bibitem[{{Harper} {et~al.}(2020){Harper}, {Guinan}, {Wasatonic}, \&
  {Ryde}}]{harper20}
{Harper}, G.~M., {Guinan}, E.~F., {Wasatonic}, R., \& {Ryde}, N. 2020, \apj,
  905, 34, \dodoi{10.3847/1538-4357/abc1f0}

\bibitem[{{Haubois} {et~al.}(2019){Haubois}, {Norris}, {Tuthill}, {Pinte},
  {Kervella}, {Girard}, {Kostogryz}, {Berdyugina}, {Perrin}, {Lacour},
  {Chiavassa}, \& {Ridgway}}]{haubois19}
{Haubois}, X., {Norris}, B., {Tuthill}, P.~G., {et~al.} 2019, \aap, 628, A101,
  \dodoi{10.1051/0004-6361/201833258}

\bibitem[{{Heger} {et~al.}(1997){Heger}, {Jeannin}, {Langer}, \&
  {Baraffe}}]{heger97}
{Heger}, A., {Jeannin}, L., {Langer}, N., \& {Baraffe}, I. 1997, \aap, 327,
  224.
\newblock \doarXiv{astro-ph/9705097}

\bibitem[{{Humphreys} \& {Davidson}(1979)}]{humphreys79}
{Humphreys}, R.~M., \& {Davidson}, K. 1979, \apj, 232, 409,
  \dodoi{10.1086/157301}

\bibitem[{{Humphreys} \& {Davidson}(1994)}]{humphreys94}
---. 1994, \pasp, 106, 1025, \dodoi{10.1086/133478}

\bibitem[{{Humphreys} {et~al.}(1999){Humphreys}, {Davidson}, \&
  {Smith}}]{humphreys99}
{Humphreys}, R.~M., {Davidson}, K., \& {Smith}, N. 1999, \pasp, 111, 1124,
  \dodoi{10.1086/316420}

\bibitem[{{Humphreys} {et~al.}(1997){Humphreys}, {Smith}, {Davidson}, {Jones},
  {Gehrz}, {Mason}, {Hayward}, {Houck}, \& {Krautter}}]{humphreys97}
{Humphreys}, R.~M., {Smith}, N., {Davidson}, K., {et~al.} 1997, \aj, 114, 2778,
  \dodoi{10.1086/118686}

\bibitem[{{Jencson} {et~al.}(2019){Jencson}, {Kasliwal}, {Adams}, {Bond}, {De},
  {Johansson}, {Karambelkar}, {Lau}, {Tinyanont}, {Ryder}, {Cody}, {Masci},
  {Bally}, {Blagorodnova}, {Castell{\'o}n}, {Fremling}, {Gehrz}, {Helou},
  {Kilpatrick}, {Milne}, {Morrell}, {Perley}, {Phillips}, {Smith}, {van Dyk},
  \& {Williams}}]{jencson19}
{Jencson}, J.~E., {Kasliwal}, M.~M., {Adams}, S.~M., {et~al.} 2019, \apj, 886,
  40, \dodoi{10.3847/1538-4357/ab4a01}

\bibitem[{{Jiang} {et~al.}(2015){Jiang}, {Cantiello}, {Bildsten}, {Quataert},
  \& {Blaes}}]{jiang15}
{Jiang}, Y.-F., {Cantiello}, M., {Bildsten}, L., {Quataert}, E., \& {Blaes}, O.
  2015, \apj, 813, 74, \dodoi{10.1088/0004-637X/813/1/74}

\bibitem[{{Kashiyama} \& {Quataert}(2015)}]{kashiyama15}
{Kashiyama}, K., \& {Quataert}, E. 2015, \mnras, 451, 2656,
  \dodoi{10.1093/mnras/stv1164}

\bibitem[{{Kiss} {et~al.}(2006){Kiss}, {Szab{\'o}}, \& {Bedding}}]{kiss06}
{Kiss}, L.~L., {Szab{\'o}}, G.~M., \& {Bedding}, T.~R. 2006, \mnras, 372, 1721,
  \dodoi{10.1111/j.1365-2966.2006.10973.x}

\bibitem[{{Kochanek}(2020)}]{kochanek20}
{Kochanek}, C.~S. 2020, \mnras, 493, 4945, \dodoi{10.1093/mnras/staa605}

\bibitem[{{Kochanek} {et~al.}(2014){Kochanek}, {Adams}, \&
  {Belczynski}}]{kochanek14}
{Kochanek}, C.~S., {Adams}, S.~M., \& {Belczynski}, K. 2014, \mnras, 443, 1319,
  \dodoi{10.1093/mnras/stu1226}

\bibitem[{{Kochanek} {et~al.}(2008){Kochanek}, {Beacom}, {Kistler}, {Prieto},
  {Stanek}, {Thompson}, \& {Y{\"u}ksel}}]{kochanek08}
{Kochanek}, C.~S., {Beacom}, J.~F., {Kistler}, M.~D., {et~al.} 2008, \apj, 684,
  1336, \dodoi{10.1086/590053}

\bibitem[{{Kochanek} {et~al.}(2012){Kochanek}, {Khan}, \& {Dai}}]{kochanek12}
{Kochanek}, C.~S., {Khan}, R., \& {Dai}, X. 2012, \apj, 759, 20,
  \dodoi{10.1088/0004-637X/759/1/20}

\bibitem[{{Kravchenko} {et~al.}(2021){Kravchenko}, {Jorissen}, {Van Eck},
  {Merle}, {Chiavassa}, {Paladini}, {Freytag}, {Plez}, {Montarg{\`e}s}, \& {Van
  Winckel}}]{kravchenko21}
{Kravchenko}, K., {Jorissen}, A., {Van Eck}, S., {et~al.} 2021, \aap, 650, L17,
  \dodoi{10.1051/0004-6361/202039801}

\bibitem[{{Ku{\v{c}}inskas} {et~al.}(2006){Ku{\v{c}}inskas}, {Hauschildt},
  {Brott}, {Vansevi{\v{c}}ius}, {Lindegren}, {Tanab{\'e}}, \&
  {Allard}}]{kucinskas06}
{Ku{\v{c}}inskas}, A., {Hauschildt}, P.~H., {Brott}, I., {et~al.} 2006, \aap,
  452, 1021, \dodoi{10.1051/0004-6361:20054431}

\bibitem[{{Ku{\v{c}}inskas} {et~al.}(2005){Ku{\v{c}}inskas}, {Hauschildt},
  {Ludwig}, {Brott}, {Vansevi{\v{c}}ius}, {Lindegren}, {Tanab{\'e}}, \&
  {Allard}}]{kucinskas05}
{Ku{\v{c}}inskas}, A., {Hauschildt}, P.~H., {Ludwig}, H.~G., {et~al.} 2005,
  \aap, 442, 281, \dodoi{10.1051/0004-6361:20053028}

\bibitem[{{Levesque} \& {Massey}(2020)}]{levesque20}
{Levesque}, E.~M., \& {Massey}, P. 2020, \apjl, 891, L37,
  \dodoi{10.3847/2041-8213/ab7935}

\bibitem[{{Levesque} {et~al.}(2006){Levesque}, {Massey}, {Olsen}, {Plez},
  {Meynet}, \& {Maeder}}]{levesque06}
{Levesque}, E.~M., {Massey}, P., {Olsen}, K.~A.~G., {et~al.} 2006, \apj, 645,
  1102, \dodoi{10.1086/504417}

\bibitem[{{Li} \& {Gong}(1994)}]{li94}
{Li}, Y., \& {Gong}, Z.~G. 1994, \aap, 289, 449

\bibitem[{{Lovegrove} \& {Woosley}(2013)}]{lovegrove13}
{Lovegrove}, E., \& {Woosley}, S.~E. 2013, \apj, 769, 109,
  \dodoi{10.1088/0004-637X/769/2/109}

\bibitem[{{Massey} {et~al.}(2006){Massey}, {Levesque}, \& {Plez}}]{massey06}
{Massey}, P., {Levesque}, E.~M., \& {Plez}, B. 2006, \apj, 646, 1203,
  \dodoi{10.1086/505025}

\bibitem[{{Massey} {et~al.}(2021){Massey}, {Neugent}, {Levesque}, {Drout}, \&
  {Courteau}}]{massey21}
{Massey}, P., {Neugent}, K.~F., {Levesque}, E.~M., {Drout}, M.~R., \&
  {Courteau}, S. 2021, \aj, 161, 79, \dodoi{10.3847/1538-3881/abd01f}

\bibitem[{{Maund}(2017)}]{maund17}
{Maund}, J.~R. 2017, \mnras, 469, 2202, \dodoi{10.1093/mnras/stx879}

\bibitem[{{Maund} \& {Smartt}(2005)}]{maund05}
{Maund}, J.~R., \& {Smartt}, S.~J. 2005, \mnras, 360, 288,
  \dodoi{10.1111/j.1365-2966.2005.09034.x}

\bibitem[{{McLeod} {et~al.}(2012){McLeod}, {Fabricant}, {Nystrom}, {McCracken},
  {Amato}, {Bergner}, {Brown}, {Burke}, {Chilingarian}, {Conroy}, {Curley},
  {Furesz}, {Geary}, {Hertz}, {Holwell}, {Matthews}, {Norton}, {Park}, {Roll},
  {Zajac}, {Epps}, \& {Martini}}]{mcleod12}
{McLeod}, B., {Fabricant}, D., {Nystrom}, G., {et~al.} 2012, \pasp, 124, 1318,
  \dodoi{10.1086/669044}

\bibitem[{{McQuinn} {et~al.}(2016){McQuinn}, {Skillman}, {Dolphin}, {Berg}, \&
  {Kennicutt}}]{mcquinn16}
{McQuinn}, K.~B.~W., {Skillman}, E.~D., {Dolphin}, A.~E., {Berg}, D., \&
  {Kennicutt}, R. 2016, \apj, 826, 21, \dodoi{10.3847/0004-637X/826/1/21}

\bibitem[{{Moe} \& {Di Stefano}(2017)}]{moe17}
{Moe}, M., \& {Di Stefano}, R. 2017, \apjs, 230, 15,
  \dodoi{10.3847/1538-4365/aa6fb6}

\bibitem[{{Monnier} {et~al.}(2004){Monnier}, {Millan-Gabet}, {Tuthill},
  {Traub}, {Carleton}, {Coud{\'e} du Foresto}, {Danchi}, {Lacasse}, {Morel},
  {Perrin}, {Porro}, {Schloerb}, \& {Townes}}]{monnier04}
{Monnier}, J.~D., {Millan-Gabet}, R., {Tuthill}, P.~G., {et~al.} 2004, \apj,
  605, 436, \dodoi{10.1086/382218}

\bibitem[{{Montarg{\`e}s} {et~al.}(2021){Montarg{\`e}s}, {Cannon}, {Lagadec},
  {de Koter}, {Kervella}, {Sanchez-Bermudez}, {Paladini}, {Cantalloube},
  {Decin}, {Scicluna}, {Kravchenko}, {Dupree}, {Ridgway}, {Wittkowski},
  {Anugu}, {Norris}, {Rau}, {Perrin}, {Chiavassa}, {Kraus}, {Monnier},
  {Millour}, {Le Bouquin}, {Haubois}, {Lopez}, {Stee}, \&
  {Danchi}}]{montarges21}
{Montarg{\`e}s}, M., {Cannon}, E., {Lagadec}, E., {et~al.} 2021, \nat, 594,
  365, \dodoi{10.1038/s41586-021-03546-8}

\bibitem[{{Neugent} {et~al.}(2020){Neugent}, {Massey}, {Georgy}, {Drout},
  {Mommert}, {Levesque}, {Meynet}, \& {Ekstr{\"o}m}}]{neugent20}
{Neugent}, K.~F., {Massey}, P., {Georgy}, C., {et~al.} 2020, \apj, 889, 44,
  \dodoi{10.3847/1538-4357/ab5ba0}

\bibitem[{{Neustadt} {et~al.}(2021){Neustadt}, {Kochanek}, {Stanek},
  {Basinger}, {Jayasinghe}, {Garling}, {Adams}, \& {Gerke}}]{neustadt21}
{Neustadt}, J.~M.~M., {Kochanek}, C.~S., {Stanek}, K.~Z., {et~al.} 2021,
  \mnras, 508, 516, \dodoi{10.1093/mnras/stab2605}

\bibitem[{{O'Connor} \& {Ott}(2011)}]{oconnor11}
{O'Connor}, E., \& {Ott}, C.~D. 2011, \apj, 730, 70,
  \dodoi{10.1088/0004-637X/730/2/70}

\bibitem[{{O'Grady} {et~al.}(2020){O'Grady}, {Drout}, {Shappee}, {Bauer},
  {Fuller}, {Kochanek}, {Jayasinghe}, {Gaensler}, {Stanek}, {Holoien},
  {Prieto}, \& {Thompson}}]{ogrady20}
{O'Grady}, A. J.~G., {Drout}, M.~R., {Shappee}, B.~J., {et~al.} 2020, \apj,
  901, 135, \dodoi{10.3847/1538-4357/abafad}

\bibitem[{{Ossenkopf} {et~al.}(1992){Ossenkopf}, {Henning}, \&
  {Mathis}}]{ossenkopf92}
{Ossenkopf}, V., {Henning}, T., \& {Mathis}, J.~S. 1992, \aap, 261, 567

\bibitem[{{Owocki}(2015)}]{owocki15}
{Owocki}, S.~P. 2015, {Instabilities in the Envelopes and Winds of Very Massive
  Stars}, ed. J.~S. {Vink}, Vol. 412, 113, \dodoi{10.1007/978-3-319-09596-7\_5}

\bibitem[{{Paxton} {et~al.}(2013){Paxton}, {Cantiello}, {Arras}, {Bildsten},
  {Brown}, {Dotter}, {Mankovich}, {Montgomery}, {Stello}, {Timmes}, \&
  {Townsend}}]{paxton13}
{Paxton}, B., {Cantiello}, M., {Arras}, P., {et~al.} 2013, \apjs, 208, 4,
  \dodoi{10.1088/0067-0049/208/1/4}

\bibitem[{{Pejcha} {et~al.}(2016{\natexlab{a}}){Pejcha}, {Metzger}, \&
  {Tomida}}]{pejcha16a}
{Pejcha}, O., {Metzger}, B.~D., \& {Tomida}, K. 2016{\natexlab{a}}, \mnras,
  455, 4351, \dodoi{10.1093/mnras/stv2592}

\bibitem[{{Pejcha} {et~al.}(2016{\natexlab{b}}){Pejcha}, {Metzger}, \&
  {Tomida}}]{pejcha16b}
---. 2016{\natexlab{b}}, \mnras, 461, 2527, \dodoi{10.1093/mnras/stw1481}

\bibitem[{{Pejcha} \& {Thompson}(2015)}]{pejcha15}
{Pejcha}, O., \& {Thompson}, T.~A. 2015, \apj, 801, 90,
  \dodoi{10.1088/0004-637X/801/2/90}

\bibitem[{{Perna} {et~al.}(2014){Perna}, {Duffell}, {Cantiello}, \&
  {MacFadyen}}]{perna14}
{Perna}, R., {Duffell}, P., {Cantiello}, M., \& {MacFadyen}, A.~I. 2014, \apj,
  781, 119, \dodoi{10.1088/0004-637X/781/2/119}

\bibitem[{{Pinna} {et~al.}(2021){Pinna}, {Rossi}, {Puglisi}, {Agapito},
  {Bonaglia}, {Plantet}, {Mazzoni}, {Briguglio}, {Carbonaro}, {Xompero},
  {Grani}, {Riccardi}, {Esposito}, {Hinz}, {Vaz}, {Ertel}, {Montoya}, {Durney},
  {Christou}, {Miller}, {Taylor}, {Cavallaro}, \& {Lefebvre}}]{pinna21}
{Pinna}, E., {Rossi}, F., {Puglisi}, A., {et~al.} 2021, arXiv e-prints,
  arXiv:2101.07091.
\newblock \doarXiv{2101.07091}

\bibitem[{{Piro}(2013)}]{piro13}
{Piro}, A.~L. 2013, \apjl, 768, L14, \dodoi{10.1088/2041-8205/768/1/L14}

\bibitem[{{Reid} \& {Goldston}(2002)}]{reid02}
{Reid}, M.~J., \& {Goldston}, J.~E. 2002, \apj, 568, 931,
  \dodoi{10.1086/338947}

\bibitem[{{Reynolds} {et~al.}(2015){Reynolds}, {Fraser}, \&
  {Gilmore}}]{reynolds15}
{Reynolds}, T.~M., {Fraser}, M., \& {Gilmore}, G. 2015, \mnras, 453, 2885,
  \dodoi{10.1093/mnras/stv1809}

\bibitem[{{Rizzi} {et~al.}(2007){Rizzi}, {Tully}, {Makarov}, {Makarova},
  {Dolphin}, {Sakai}, \& {Shaya}}]{rizzi07}
{Rizzi}, L., {Tully}, R.~B., {Makarov}, D., {et~al.} 2007, \apj, 661, 815,
  \dodoi{10.1086/516566}

\bibitem[{{Rodrigo} \& {Solano}(2020)}]{svo2}
{Rodrigo}, C., \& {Solano}, E. 2020, in Contributions to the XIV.0 Scientific
  Meeting (virtual) of the Spanish Astronomical Society, 182

\bibitem[{{Rodrigo} {et~al.}(2012){Rodrigo}, {Solano}, \& {Bayo}}]{svo1}
{Rodrigo}, C., {Solano}, E., \& {Bayo}, A. 2012, {SVO Filter Profile Service
  Version 1.0}, IVOA Working Draft 15 October 2012,
  \dodoi{10.5479/ADS/bib/2012ivoa.rept.1015R}

\bibitem[{{Rothberg} {et~al.}(2019){Rothberg}, {Christou}, {Miller},
  {Thompson}, {Taylor}, \& {Veillet}}]{rothberg19}
{Rothberg}, B., {Christou}, J.~C., {Miller}, D.~L., {et~al.} 2019, arXiv
  e-prints, arXiv:1911.00549.
\newblock \doarXiv{1911.00549}

\bibitem[{{Rothberg} {et~al.}(2020){Rothberg}, {Christou}, {Miller},
  {Thompson}, {Taylor}, {Veillet}, \& {Gatto}}]{rothberg20}
{Rothberg}, B., {Christou}, J.~C., {Miller}, D.~L., {et~al.} 2020, in Society
  of Photo-Optical Instrumentation Engineers (SPIE) Conference Series, Vol.
  11448, Society of Photo-Optical Instrumentation Engineers (SPIE) Conference
  Series, 114485I, \dodoi{10.1117/12.2563359}

\bibitem[{{Sana} {et~al.}(2012){Sana}, {de Mink}, {de Koter}, {Langer},
  {Evans}, {Gieles}, {Gosset}, {Izzard}, {Le Bouquin}, \& {Schneider}}]{sana12}
{Sana}, H., {de Mink}, S.~E., {de Koter}, A., {et~al.} 2012, Science, 337, 444,
  \dodoi{10.1126/science.1223344}

\bibitem[{{Sargent} {et~al.}(2011){Sargent}, {Srinivasan}, \&
  {Meixner}}]{sargent11}
{Sargent}, B.~A., {Srinivasan}, S., \& {Meixner}, M. 2011, \apj, 728, 93,
  \dodoi{10.1088/0004-637X/728/2/93}

\bibitem[{{Schlafly} \& {Finkbeiner}(2011)}]{schlafly11}
{Schlafly}, E.~F., \& {Finkbeiner}, D.~P. 2011, \apj, 737, 103,
  \dodoi{10.1088/0004-637X/737/2/103}

\bibitem[{{Schlegel} {et~al.}(1998){Schlegel}, {Finkbeiner}, \&
  {Davis}}]{schlegel98}
{Schlegel}, D.~J., {Finkbeiner}, D.~P., \& {Davis}, M. 1998, \apj, 500, 525,
  \dodoi{10.1086/305772}

\bibitem[{{Scicluna} {et~al.}(2015){Scicluna}, {Siebenmorgen}, {Wesson},
  {Blommaert}, {Kasper}, {Voshchinnikov}, \& {Wolf}}]{scicluna15}
{Scicluna}, P., {Siebenmorgen}, R., {Wesson}, R., {et~al.} 2015, \aap, 584,
  L10, \dodoi{10.1051/0004-6361/201527563}

\bibitem[{{Seifert} {et~al.}(2003){Seifert}, {Appenzeller}, {Baumeister},
  {Bizenberger}, {Bomans}, {Dettmar}, {Grimm}, {Herbst}, {Hofmann}, {Juette},
  {Laun}, {Lehmitz}, {Lemke}, {Lenzen}, {Mandel}, {Polsterer}, {Rohloff},
  {Schuetze}, {Seltmann}, {Thatte}, {Weiser}, \& {Xu}}]{seifert03}
{Seifert}, W., {Appenzeller}, I., {Baumeister}, H., {et~al.} 2003, in Society
  of Photo-Optical Instrumentation Engineers (SPIE) Conference Series, Vol.
  4841, Instrument Design and Performance for Optical/Infrared Ground-based
  Telescopes, ed. M.~{Iye} \& A.~F.~M. {Moorwood}, 962--973,
  \dodoi{10.1117/12.459494}

\bibitem[{{Skrutskie} {et~al.}(2006){Skrutskie}, {Cutri}, {Stiening},
  {Weinberg}, {Schneider}, {Carpenter}, {Beichman}, {Capps}, {Chester},
  {Elias}, {Huchra}, {Liebert}, {Lonsdale}, {Monet}, {Price}, {Seitzer},
  {Jarrett}, {Kirkpatrick}, {Gizis}, {Howard}, {Evans}, {Fowler}, {Fullmer},
  {Hurt}, {Light}, {Kopan}, {Marsh}, {McCallon}, {Tam}, {Van Dyk}, \&
  {Wheelock}}]{skrutskie06}
{Skrutskie}, M.~F., {Cutri}, R.~M., {Stiening}, R., {et~al.} 2006, \aj, 131,
  1163, \dodoi{10.1086/498708}

\bibitem[{{Smartt}(2015)}]{smartt15}
{Smartt}, S.~J. 2015, \pasa, 32, e016, \dodoi{10.1017/pasa.2015.17}

\bibitem[{{Smartt} {et~al.}(2009){Smartt}, {Eldridge}, {Crockett}, \&
  {Maund}}]{smartt09}
{Smartt}, S.~J., {Eldridge}, J.~J., {Crockett}, R.~M., \& {Maund}, J.~R. 2009,
  \mnras, 395, 1409, \dodoi{10.1111/j.1365-2966.2009.14506.x}

\bibitem[{{Smith}(2004)}]{smith04b}
{Smith}, N. 2004, \mnras, 349, L31, \dodoi{10.1111/j.1365-2966.2004.07718.x}

\bibitem[{{Smith}(2014)}]{smith14}
---. 2014, \araa, 52, 487, \dodoi{10.1146/annurev-astro-081913-040025}

\bibitem[{{Smith}(2017)}]{smith17}
---. 2017, Philosophical Transactions of the Royal Society of London Series A,
  375, 20160268, \dodoi{10.1098/rsta.2016.0268}

\bibitem[{{Smith} {et~al.}(2001{\natexlab{a}}){Smith}, {Humphreys}, {Davidson},
  {Gehrz}, {Schuster}, \& {Krautter}}]{smith01a}
{Smith}, N., {Humphreys}, R.~M., {Davidson}, K., {et~al.} 2001{\natexlab{a}},
  \aj, 121, 1111, \dodoi{10.1086/318748}

\bibitem[{{Smith} {et~al.}(2001{\natexlab{b}}){Smith}, {Humphreys}, \&
  {Gehrz}}]{smith01b}
{Smith}, N., {Humphreys}, R.~M., \& {Gehrz}, R.~D. 2001{\natexlab{b}}, \pasp,
  113, 692, \dodoi{10.1086/320812}

\bibitem[{{Smith} {et~al.}(2011){Smith}, {Li}, {Silverman}, {Ganeshalingam}, \&
  {Filippenko}}]{smith11}
{Smith}, N., {Li}, W., {Silverman}, J.~M., {Ganeshalingam}, M., \&
  {Filippenko}, A.~V. 2011, \mnras, 415, 773,
  \dodoi{10.1111/j.1365-2966.2011.18763.x}

\bibitem[{{Smith} {et~al.}(2004){Smith}, {Vink}, \& {de Koter}}]{smith04a}
{Smith}, N., {Vink}, J.~S., \& {de Koter}, A. 2004, \apj, 615, 475,
  \dodoi{10.1086/424030}

\bibitem[{{Smith} {et~al.}(2016){Smith}, {Andrews}, {Van Dyk}, {Mauerhan},
  {Kasliwal}, {Bond}, {Filippenko}, {Clubb}, {Graham}, {Perley}, {Jencson},
  {Bally}, {Ubeda}, \& {Sabbi}}]{smith16}
{Smith}, N., {Andrews}, J.~E., {Van Dyk}, S.~D., {et~al.} 2016, \mnras, 458,
  950, \dodoi{10.1093/mnras/stw219}

\bibitem[{{Smith} {et~al.}(2018){Smith}, {Andrews}, {Rest}, {Bianco}, {Prieto},
  {Matheson}, {James}, {Smith}, {Strampelli}, \& {Zenteno}}]{smith18}
{Smith}, N., {Andrews}, J.~E., {Rest}, A., {et~al.} 2018, \mnras, 480, 1466,
  \dodoi{10.1093/mnras/sty1500}

\bibitem[{{Soraisam} {et~al.}(2018){Soraisam}, {Bildsten}, {Drout}, {Bauer},
  {Gilfanov}, {Kupfer}, {Laher}, {Masci}, {Prince}, {Kulkarni}, {Matheson}, \&
  {Saha}}]{soraisam18}
{Soraisam}, M.~D., {Bildsten}, L., {Drout}, M.~R., {et~al.} 2018, \apj, 859,
  73, \dodoi{10.3847/1538-4357/aabc59}

\bibitem[{{Soszy{\'n}ski} {et~al.}(2009){Soszy{\'n}ski}, {Udalski},
  {Szyma{\'n}ski}, {Kubiak}, {Pietrzy{\'n}ski}, {Wyrzykowski}, {Szewczyk},
  {Ulaczyk}, \& {Poleski}}]{soszynski09}
{Soszy{\'n}ski}, I., {Udalski}, A., {Szyma{\'n}ski}, M.~K., {et~al.} 2009,
  \actaa, 59, 239.
\newblock \doarXiv{0910.1354}

\bibitem[{{Srinivasan} {et~al.}(2011){Srinivasan}, {Sargent}, \&
  {Meixner}}]{srinivasan11}
{Srinivasan}, S., {Sargent}, B.~A., \& {Meixner}, M. 2011, \aap, 532, A54,
  \dodoi{10.1051/0004-6361/201117033}

\bibitem[{{STScI Development Team}(2013)}]{pysynphot}
{STScI Development Team}. 2013, {pysynphot: Synthetic photometry software
  package}.
\newblock \doeprint{1303.023}

\bibitem[{{Sukhbold} \& {Adams}(2020)}]{sukhbold19}
{Sukhbold}, T., \& {Adams}, S. 2020, \mnras, 492, 2578,
  \dodoi{10.1093/mnras/staa059}

\bibitem[{{Sukhbold} {et~al.}(2016){Sukhbold}, {Ertl}, {Woosley}, {Brown}, \&
  {Janka}}]{sukhbold16}
{Sukhbold}, T., {Ertl}, T., {Woosley}, S.~E., {Brown}, J.~M., \& {Janka}, H.~T.
  2016, \apj, 821, 38, \dodoi{10.3847/0004-637X/821/1/38}

\bibitem[{{Sukhbold} \& {Woosley}(2014)}]{sukhbold14}
{Sukhbold}, T., \& {Woosley}, S.~E. 2014, \apj, 783, 10,
  \dodoi{10.1088/0004-637X/783/1/10}

\bibitem[{{The Astropy Collaboration} {et~al.}(2018){The Astropy
  Collaboration}, {Price-Whelan}, {Sip{\H o}cz}, {G{\"u}nther}, {Lim},
  {Crawford}, {Conseil}, {Shupe}, {Craig}, {Dencheva}, {Ginsburg},
  {VanderPlas}, {Bradley}, {P{\'e}rez-Su{\'a}rez}, {de Val-Borro}, {Aldcroft},
  {Cruz}, {Robitaille}, {Tollerud}, {Ardelean}, {Babej}, {Bachetti}, {Bakanov},
  {Bamford}, {Barentsen}, {Barmby}, {Baumbach}, {Berry}, {Biscani}, {Boquien},
  {Bostroem}, {Bouma}, {Brammer}, {Bray}, {Breytenbach}, {Buddelmeijer},
  {Burke}, {Calderone}, {Cano Rodr{\'{\i}}guez}, {Cara}, {Cardoso},
  {Cheedella}, {Copin}, {Crichton}, {D{\'A}vella}, {Deil}, {Depagne},
  {Dietrich}, {Donath}, {Droettboom}, {Earl}, {Erben}, {Fabbro}, {Ferreira},
  {Finethy}, {Fox}, {Garrison}, {Gibbons}, {Goldstein}, {Gommers}, {Greco},
  {Greenfield}, {Groener}, {Grollier}, {Hagen}, {Hirst}, {Homeier}, {Horton},
  {Hosseinzadeh}, {Hu}, {Hunkeler}, {Ivezi{\'c}}, {Jain}, {Jenness}, {Kanarek},
  {Kendrew}, {Kern}, {Kerzendorf}, {Khvalko}, {King}, {Kirkby}, {Kulkarni},
  {Kumar}, {Lee}, {Lenz}, {Littlefair}, {Ma}, {Macleod}, {Mastropietro},
  {McCully}, {Montagnac}, {Morris}, {Mueller}, {Mumford}, {Muna}, {Murphy},
  {Nelson}, {Nguyen}, {Ninan}, {N{\"o}the}, {Ogaz}, {Oh}, {Parejko}, {Parley},
  {Pascual}, {Patil}, {Patil}, {Plunkett}, {Prochaska}, {Rastogi}, {Reddy
  Janga}, {Sabater}, {Sakurikar}, {Seifert}, {Sherbert}, {Sherwood-Taylor},
  {Shih}, {Sick}, {Silbiger}, {Singanamalla}, {Singer}, {Sladen}, {Sooley},
  {Sornarajah}, {Streicher}, {Teuben}, {Thomas}, {Tremblay}, {Turner},
  {Terr{\'o}n}, {van Kerkwijk}, {de la Vega}, {Watkins}, {Weaver}, {Whitmore},
  {Woillez}, \& {Zabalza}}]{astropy}
{The Astropy Collaboration}, {Price-Whelan}, A.~M., {Sip{\H o}cz}, B.~M.,
  {et~al.} 2018, ArXiv e-prints.
\newblock \doarXiv{1801.02634}

\bibitem[{{Tisserand} {et~al.}(2009){Tisserand}, {Wood}, {Marquette}, {Afonso},
  {Albert}, {Andersen}, {Ansari}, {Aubourg}, {Bareyre}, {Beaulieu}, {Charlot},
  {Coutures}, {Ferlet}, {Fouqu{\'e}}, {Glicenstein}, {Goldman}, {Gould},
  {Gros}, {de Kat}, {Lesquoy}, {Loup}, {Magneville}, {Maurice}, {Maury},
  {Milsztajn}, {Moniez}, {Palanque-Delabrouille}, {Perdereau}, {Rich},
  {Schwemling}, {Spiro}, \& {Vidal-Madjar}}]{tisserand09}
{Tisserand}, P., {Wood}, P.~R., {Marquette}, J.~B., {et~al.} 2009, \aap, 501,
  985, \dodoi{10.1051/0004-6361/200911808}

\bibitem[{{Ueta} \& {Meixner}(2003)}]{ueta03}
{Ueta}, T., \& {Meixner}, M. 2003, \apj, 586, 1338, \dodoi{10.1086/367818}

\bibitem[{{Ugliano} {et~al.}(2012){Ugliano}, {Janka}, {Marek}, \&
  {Arcones}}]{ugliano12}
{Ugliano}, M., {Janka}, H.-T., {Marek}, A., \& {Arcones}, A. 2012, \apj, 757,
  69, \dodoi{10.1088/0004-637X/757/1/69}

\bibitem[{{Van Dyk} {et~al.}(1999){Van Dyk}, {Peng}, {Barth}, \&
  {Filippenko}}]{vandyk99}
{Van Dyk}, S.~D., {Peng}, C.~Y., {Barth}, A.~J., \& {Filippenko}, A.~V. 1999,
  \aj, 118, 2331, \dodoi{10.1086/301068}

\bibitem[{{Walmswell} \& {Eldridge}(2012)}]{walmswell12}
{Walmswell}, J.~J., \& {Eldridge}, J.~J. 2012, \mnras, 419, 2054,
  \dodoi{10.1111/j.1365-2966.2011.19860.x}

\bibitem[{{Wei} {et~al.}(2021){Wei}, {Zou}, {Lin}, {Zhou}, {Liu}, {Kong}, {Ma},
  \& {Ma}}]{wei21}
{Wei}, P., {Zou}, H., {Lin}, L., {et~al.} 2021, Research in Astronomy and
  Astrophysics, 21, 006, \dodoi{10.1088/1674-4527/21/1/6}

\bibitem[{{Williams} {et~al.}(2014){Williams}, {Lang}, {Dalcanton}, {Dolphin},
  {Weisz}, {Bell}, {Bianchi}, {Byler}, {Gilbert}, {Girardi}, {Gordon},
  {Gregersen}, {Johnson}, {Kalirai}, {Lauer}, {Monachesi}, {Rosenfield},
  {Seth}, \& {Skillman}}]{williams14}
{Williams}, B.~F., {Lang}, D., {Dalcanton}, J.~J., {et~al.} 2014, \apjs, 215,
  9, \dodoi{10.1088/0067-0049/215/1/9}

\bibitem[{{Yang} \& {Jiang}(2011)}]{yang11}
{Yang}, M., \& {Jiang}, B.~W. 2011, \apj, 727, 53,
  \dodoi{10.1088/0004-637X/727/1/53}

\bibitem[{{Yang} \& {Jiang}(2012)}]{yang12}
---. 2012, \apj, 754, 35, \dodoi{10.1088/0004-637X/754/1/35}

\bibitem[{{Yang} {et~al.}(2019){Yang}, {Bonanos}, {Jiang}, {Gao}, {Gavras},
  {Maravelias}, {Ren}, {Wang}, {Xue}, {Tramper}, {Spetsieri}, \&
  {Pouliasis}}]{yang19}
{Yang}, M., {Bonanos}, A.~Z., {Jiang}, B.-W., {et~al.} 2019, \aap, 629, A91,
  \dodoi{10.1051/0004-6361/201935916}

\bibitem[{{Yoon} \& {Cantiello}(2010)}]{yoon10}
{Yoon}, S.-C., \& {Cantiello}, M. 2010, \apjl, 717, L62,
  \dodoi{10.1088/2041-8205/717/1/L62}

\bibitem[{{Zackay} {et~al.}(2016){Zackay}, {Ofek}, \& {Gal-Yam}}]{zackay16}
{Zackay}, B., {Ofek}, E.~O., \& {Gal-Yam}, A. 2016, \apj, 830, 27,
  \dodoi{10.3847/0004-637X/830/1/27}

\end{thebibliography}
\bibliographystyle{aasjournal}

\end{document}